\documentclass[
  aps,
  pra,
  reprint,
  superscriptaddress,
  floatfix,
  longbibliography
]{revtex4-2}

\newcommand{\subcaptext}[1]{\par\vspace{2pt}{\footnotesize #1\par}}

\usepackage{graphicx}
\usepackage{amsmath,amssymb,amsthm}
\usepackage{booktabs}
\usepackage{multirow}
\usepackage[table]{xcolor}
\usepackage{colortbl}
\usepackage{placeins}
\usepackage{float}
\usepackage{tikz}
\usetikzlibrary{shapes.geometric,shapes.misc,matrix,positioning,patterns}
\usepackage{pgfplots}
\pgfplotsset{compat=1.18}
\usepackage[hidelinks]{hyperref}
\usepackage[ruled,vlined,linesnumbered]{algorithm2e}

\SetKwInput{KwIn}{Input}
\SetKwInput{KwOut}{Output}
\SetKw{Break}{break}

\allowdisplaybreaks
\DeclareRobustCommand{\TrueParamSym}{%
  \tikz[baseline=-0.8ex]\node[star,star points=5,star point ratio=2.25,
  minimum size=9pt,inner sep=0pt,draw=yellow!80!black,fill=yellow!100!black] {};%
}

\definecolor{fcbmyellow}{HTML}{FFD400}
\DeclareRobustCommand{\FCBMOpenSym}{
  \tikz[baseline=-0.8ex]{\node[star,star points=5,star point ratio=2.25,minimum size=9pt,inner sep=0pt,
  draw=black,line width=1.5pt] {};\node[star,star points=5,star point ratio=2.25,minimum size=9pt,
  inner sep=0pt,draw=fcbmyellow,line width=0.75pt] {};}%
}

\DeclareRobustCommand{\MinEntropySym}{%
  \tikz[baseline=-0.6ex]\node[regular polygon,regular polygon sides=3,
  shape border rotate=180,draw=blue!70!black,fill=blue!70!black,inner sep=2pt] {};%
}

\DeclareRobustCommand{\MaxEntropySym}{%
  \tikz[baseline=-0.6ex]\node[regular polygon,regular polygon sides=3,
  draw=red!70!black,fill=red!70!black,inner sep=2pt] {};%
}

\DeclareRobustCommand{\MeanEntropySym}{%
  \tikz[baseline=-0.6ex]\node[circle,minimum size=7pt,inner sep=0pt,
  draw=orange!90!black,fill=orange!90!black] {};%
}

\DeclareRobustCommand{\MedianEntropySym}{%
  \tikz[baseline=-0.6ex]\node[regular polygon,regular polygon sides=4,
  draw=blue!80!black,fill=blue!80!black,inner sep=3pt] {};%
}

\DeclareRobustCommand{\EAPsym}{%
\tikz[baseline=-0.6ex]\draw[fill=orange,draw=black] (0,0) circle (3pt);}

\DeclareRobustCommand{\ECAsym}{%
\tikz[baseline=-0.6ex]\draw[fill=red,draw=black] (0,0) circle (3pt);}

\DeclareRobustCommand{\LACsym}{%
\tikz[baseline=-0.6ex]\draw[fill=green!60!black,draw=black] (0,0) circle (3pt);}

\DeclareRobustCommand{\MENAsym}{%
\tikz[baseline=-0.6ex]\draw[fill=violet,draw=black] (0,0) circle (3pt);}

\DeclareRobustCommand{\NAsym}{%
\tikz[baseline=-0.6ex]\draw[fill=gray,draw=black] (0,0) circle (3pt);}

\DeclareRobustCommand{\SARsym}{%
\tikz[baseline=-0.6ex]\draw[fill=blue,draw=black] (0,0) circle (3pt);}

\DeclareRobustCommand{\SSAsym}{%
\tikz[baseline=-0.6ex]\draw[fill=yellow,draw=black] (0,0) circle (3pt);}

\begin{document}

\title{Network Reconstruction via Jeffreys Prior under Missing Sufficient Statistics}

\author{Minh Duc Duong}
\affiliation{Computer Science Department, University of Pisa, 56127 Pisa, Italy}
\affiliation{IMT School for Advanced Studies, 55100 Lucca, Italy}

\author{Diego Garlaschelli}
\affiliation{IMT School for Advanced Studies, 55100 Lucca, Italy}
\affiliation{Lorentz Institute for Theoretical Physics, Leiden University, 2333 CC Leiden, the Netherlands}

\date{October 31, 2026}

\begin{abstract}

Methods for the reconstruction of economic networks from aggregate information are important for modeling, analysis and policymaking. The traditional Fitness Model (FM) achieves good performance by using node-specific variables that are easily accessible (e.g., GDP for countries or total assets for banks or firms) and the overall link density as the only sufficient statistics. However, it often ignores additional contextual or mesoscopic features which may be more difficult to observe. 
Here we extend the framework by incorporating block structure as in the Fitness-Corrected Block Model (FCBM), which allows for heterogeneous densities within and across blocks, but in the deliberately more challenging setting where such block-specific densities are not empirically available.
Our method uses a Jeffreys prior to provide an unbiased distribution of all compatible solutions and associated entropies, that are otherwise left unidentified, and selects a single reconstruction at the median of this entropy distribution to compensate for the missing sufficient statistics.  We evaluate the procedure on nine international-trade networks using GDP as node fitness and World Bank regions as an externally specified partition. 
The method systematically outperforms the baseline Block-Agnostic FM (which uses the same observed network input statistic) and sometimes even the FCBM (which always uses more information).
Controlled synthetic networks confirm that the median is the best of four entropy-based representatives for a  range of densities and regional effects similar to the empirical ones, while other representatives can perform better when parameters are pushed far from their realistic range. 
Overall, our  method compensates for missing sufficient statistics via a median-entropy estimate that is empirically very robust within a realistic range of parameters.

\end{abstract}

\maketitle

\section{Introduction}

Reconstructing network structure from incomplete information, generally in the form of aggregate publicly available data, is a fundamental problem across disciplines such as economics, finance, sociology, statistical physics, and computer science~\cite{Squartini2018,Cimini2021}. In practice, network connections between countries, companies, and other entities are often confidential. Analysts are therefore tasked with reconstructing these network structures, typically beginning with a binary formulation in which the presence or absence of a direct link between each pair of nodes is determined. Having information about the connections between countries or units provides significant advantages for policy-making. However, in many real-world situations, such as international trade, financial interbank transactions, or social interaction networks, we often have only a limited amount of data at the macro level, such as national GDP, total assets of companies, population, distance, and geographic area, along with a small amount of related information such as the total number of connections in the region, the number of partners that each node has (degree). Based on this limited information, researchers aim to reconstruct the true network structure as accurately as possible. Apart from a small amount of published information, networks in each domain often do not have a clearly identifiable pattern. At the same time, the network reconstruction model needs to exhibit high generalization capability, satisfy some initial conditions, and ensure the balance of the reconstructed network.

Over the last two decades, the literature has been rich in methodologies for addressing the network reconstruction problem. Approaches can be categorized into two main clusters. The first cluster is based on the type of structural constraints imposed, including node-level information (e.g., \cite{Garlaschelli2004,Garlaschelli2005,Squartini2011}), link-level or density constraints (e.g., \cite{Squartini2017, Squartini2018}), total trade volume (e.g., \cite{divece2022gravity}) and related structural constraints. The second cluster concerns the methodological approaches used to solve the reconstruction problem. Representative approaches include basic economic models \cite{divece2022gravity}, minimum-density approaches \cite{Mistrulli2007}, and maximum-entropy methods \cite{Garlaschelli2004, Garlaschelli2005} -- often with maximum-likelihood frameworks \cite{Squartini2011, Squartini2011a, Squartini2011b}. For comprehensive reviews, see \cite{Squartini2018,Squartini2017,Cimini2021}.

The maximum-entropy principle \cite{Jaynes1982} underpins most of the approaches. However, the first and most traditional method used this principle in a way that focused only on link weights, while keeping the topology fully connected \cite{UPPER2004827}. In contrast, empirical networks often have very different characteristics, which can only be captured via nontrivial and heterogeneous link probabilities across pairs of nodes. Several more sophisticated methods were therefore developed \cite{ANAND2018107}. However, most of these methods have certain limitations in the underlying assumptions, leading to unrealistic outcomes. Significant progress was made with the extension of the maximum-entropy framework to the topology of networks and the related introduction of maximum-likelihood estimation in network models, which maximizes the likelihood of link formation to satisfy the imposed constraints \cite{Squartini2017}. This was followed by a successful method, the fitness model, originally introduced in \cite{Caldarelli2002} and developed into a network reconstruction method in \cite{Cimini2015}. This method assigns a so-called fitness value to each node (usually total assets in the case of a company or GDP in the case of a country), and the probability of connection between two nodes depends on their fitness values. Following that success, a variety of applications have emerged \cite{Squartini2018}.  The combination of maximum likelihood and exponential family has been further developed in \cite{Mastrandrea2014}, \cite{Bargigli2014} and \cite{Parisi2020}.

The application of network reconstruction to international trade began with \cite{Gleditsch2002}, who initiated research on the relationship between trade and GDP, and \cite{Serrano2003}. Subsequently, \cite{Garlaschelli2004, Garlaschelli2005} successfully applied network reconstruction to the World Trade Web, demonstrating the central role of GDP in the network reconstruction process. This approach was further developed by \cite{Barigozzi2010}. \cite{Almog_2015, Almog_2019} refined the model and combined it with empirical trade flow data. Parallel contributions were made by \cite{vanBergeijk2010} and \cite{Anderson_2011} to the development of the gravity model. Furthermore, the integration of network theory into international trade networks was advanced by \cite{Fagiolo2009, Fagiolo2013, Fagiolo2013b}. Performance measures for these models were examined in \cite{Fronczak2012}, who evaluated them using statistical performance indicators. Finally, ensemble modeling approaches for trade networks were developed in \cite{Squartini2015} and \cite{Gabrielli2019}.

The category variables are often treated as blocks and form models that can be applied to network reconstruction. The simplest formulation is the Stochastic Block Model (SBM), but this model is homogeneous within blocks and therefore too simple when assigning a certain connection probability to pairs of nodes. A refinement was introduced by \cite{karrer2011stochastic}, namely the Degree-Corrected SBM. Then, \cite{Mossel_2015} introduced the Planted Partition version of SBM. However, since the empirical knowledge of the degrees is often not guaranteed in privacy-protected networks, the approach has been adapted to the network reconstruction problem in the Fitness-Corrected Block Model (FCBM) \cite{Bernaschi2022fitness}. 
Similarly, \cite{Kojaku2018Structural} integrated categorical structure into the network reconstruction process in the form of inter-group connectivity. Several studies did not employ a traditional block-model, but the resulting methods are closely related to the block modeling framework. For instance, \cite{Ramadiah2020network} studied networks of varying density and different block structures. \cite{Engel2021block} used block models to reconstruct directed and weighted multinational financial networks. \cite{Macchiati2022liquidity} used a block reconstruction method based on the exchange flows between different countries to explain liquidity and systemic contagion. Finally, \cite{Ialongo2022firm} used industry classification, as a form of firm-level interactions, to constrain interactions and treat them as blocks.

These methods generally assume that the link densities within and across blocks are observable and can therefore be used as separate constraints. In privacy-protected or incompletely observed networks, however, those block-specific sufficient statistics may be unavailable even when the categorical membership of each node is known. We study precisely this setting. The model uses the same network-level statistic as the traditional FM---the total number of links---together with node fitnesses and exogenous block labels, but it contains two structural parameters. Consequently, the parameters are not identifiable from the available network statistic alone and form a continuum of compatible solutions.

Rather than claiming to identify a unique parameter pair from information that is insufficient for that purpose, we equip the one-dimensional feasible family with the reparameterization-invariant measure induced by the Fisher information, following the Jeffreys construction \cite{Jeffreys_1946,Fisher_1925}. This measure provides an invariant weighting of compatible solutions. It is important to distinguish this step from parameter or probability-matrix averaging: our final reconstruction is a single representative point selected from the feasible family by the median of the Shannon-entropy distribution induced by the Jeffreys measure. We also explain the rationale for this representative choice, in particular by distinguishing network sparsity from entropy and by comparing the median with the minimum-, maximum-, and mean-entropy points already defined along the feasible curve.

The remainder of the paper is organized as follows. Section~II introduces the reconstruction models and the constrained Jeffreys construction. Section~III describes the data. Section~IV reports the empirical and controlled synthetic reconstruction results. Section~V summarizes the conclusions and limitations.

\section{Network reconstruction methods}
\subsection{Traditional Solution: Fitness Model (Block-Agnostic FM)}

The fitness model is a foundational approach for statistical network reconstruction. The core idea of this method is that each node $i$ is assigned a fitness variable $x_i$, which reflects its power to form links. Examples of fitness values include a country's GDP or a company's total assets. The probability of a link between nodes $i$ and $j$ is given by
\begin{equation}
\label{eq:fm-prob}
p_{ij}(\beta) = \frac{e^{\beta} x_i x_j}{1 + e^{\beta} x_i x_j}.
\end{equation}
with the following definitions:
\begin{itemize}
  \item $x_i$, $x_j$: fitness values of nodes $i$ and $j$;
  \item $\beta$: global parameter controlling overall network density (via $e^{\beta}$).
\end{itemize}
The main objective is to estimate the parameter $\beta$. To do that, links are assumed to be independent Bernoulli random variables. We then aim to maximize the likelihood of the observed network under a Bernoulli model with edge probabilities $p_{ij}$. 
The likelihood for a binary undirected network adjacency matrix $A = \{a_{ij}\}$ (with $a_{ij} \in \{0, 1\}$ and $a_{ij} = a_{ji}$) is given by:
\[
\mathcal{L}(\beta \mid A)
=
\prod_{i<j}
p_{ij}^{a_{ij}}
(1-p_{ij})^{1-a_{ij}}.
\]
Taking the logarithm to form the log-likelihood,
\begin{align}
\ell(\beta)
&=\log \mathcal{L}(\beta\mid A)\notag\\
&=\sum_{i<j}\left[a_{ij}\log p_{ij}+(1-a_{ij})\log(1-p_{ij})\right].
\end{align}
Maximizing $\ell(\beta)$ with respect to $\beta$ yields the first-order condition
\[
\frac{\partial \ell}{\partial \beta}
=
\sum_{i<j}(a_{ij} - p_{ij}) = 0.
\]
Therefore, the maximum-likelihood estimator satisfies
\[
\sum_{i<j} p_{ij} = \sum_{i<j} a_{ij}.
\]
Since $\sum_{i<j} a_{ij}$ equals the observed total number of links $L_{\mathrm{total}}$, the MLE condition is equivalent to imposing the constraint
\begin{equation}
\label{eq:fm-link-constraint}
\sum_{i < j} p_{ij} = L_{\mathrm{total}},
\end{equation}
i.e., the expected total number of links should equal the empirical number of links.
Throughout the paper we work with undirected (symmetrized) binary networks, so $L_{\mathrm{total}}=\sum_{i<j} a_{ij}$ correctly counts each link only once. 
Substituting the model for $p_{ij}$ yields:
\[
L_{\mathrm{total}} = \sum_{i < j} \frac{e^{\beta} x_i x_j}{1 + e^{\beta} x_i x_j}
\]
This yields a nonlinear scalar equation in $\beta$, with a single solution (the right-hand side is a monotonic function of $\beta$) that can be identified numerically. Once $\beta$ is estimated, the entire probability matrix ${p_{ij}}$ is uniquely determined.

Throughout this paper, we assume that the total number of links (or a good proxy of it) is empirically available, and that no other information about network structure is given. 
This is a common assumption in the network reconstruction literature~\cite{Squartini2018}.
We also consider international trade data between world countries as an illustrative example of real-world economic networks to be reconstructed. 
In this case, $x_i$ can be chosen to be the (normalized) GDP of country $i$. The application of the above FM to this specific example has been considered several times in the literature~\cite{Garlaschelli2004,Garlaschelli2005,Almog_2015,Parisi2020} and will represent a convenient benchmark for our analysis in the rest of the paper.

\subsection{Fitness-Corrected Block-Model (FCBM) in its Planted Partition version}
\label{sec:fcbm-full}
Our preliminary modification of the benchmark FM illustrated above consists in a refinement where the network properties are allowed to depend also on the membership of countries in economic regions as defined by the World Bank~\cite{World_bank_data}.  
In other words, we introduce a categorical variable consisting in the economic region of each country. Under region equivalence, countries belonging to the same region are assigned to the same block (and, all else being equal, are expected to have comparatively larger connection probabilities), whereas countries from different regions belong to different blocks (and are expected to have comparatively lower connection probabilities).
This is the so-called Planted Partition variant of the Stochastic Block-Model \cite{Mossel_2015}, with an important `fitness correction' -- which actually makes it a specific `Planted Partition version' of the so-called Fitness-Corrected Block-Model (FCBM)~\cite{Bernaschi2022fitness}.
We model an undirected binary network \( A = \{a_{ij}\} \), where \( a_{ij} \in \{0, 1\} \) indicates the presence of a link between countries \( i \) and \( j \).
We also define the quantities below. Figure~\ref{fig:region_matrix} illustrates the corresponding block indicator for a simple four-block example.
\begin{itemize}
    \item \( x_i \): node-specific economic attribute (here, the rescaled  GDP of country $i$);
    \item \( R_{ij} \in \{0, 1\} \): regional categorical indicator defined as
    \[
    R_{ij} =
    \begin{cases}
        1 & \text{if } i \text{ and } j \text{ are in the same region;} \\
        0 & \text{otherwise.}
    \end{cases}.
    \]
\end{itemize}

\begin{figure}[t]
\centering
\resizebox{0.95\columnwidth}{!}{%
\begin{tikzpicture}
  \matrix (m) [matrix of nodes,
               nodes={minimum size=0.9cm, anchor=center, draw},
               column sep=-\pgflinewidth,
               row sep=-\pgflinewidth,
               nodes in empty cells,
               ampersand replacement=\&] {
    $R_{i,j}$ \& A \& B \& C \& D \\
    A \& \node[fill=black!50] {}; \& \node[pattern=north west lines, pattern color=black] {}; \& \node[pattern=north west lines, pattern color=black] {}; \& \node[pattern=north west lines, pattern color=black] {}; \\
    B \& \node[pattern=north west lines, pattern color=black] {}; \& \node[fill=black!50] {}; \& \node[pattern=north west lines, pattern color=black] {}; \& \node[pattern=north west lines, pattern color=black] {}; \\
    C \& \node[pattern=north west lines, pattern color=black] {}; \& \node[pattern=north west lines, pattern color=black] {}; \& \node[fill=black!50] {}; \& \node[pattern=north west lines, pattern color=black] {}; \\
    D \& \node[pattern=north west lines, pattern color=black] {}; \& \node[pattern=north west lines, pattern color=black] {}; \& \node[pattern=north west lines, pattern color=black] {}; \& \node[fill=black!50] {}; \\
  };
\end{tikzpicture}%
}

\vspace{3pt}
{\footnotesize
\begin{tabular}{@{}ll@{}}
\tikz\draw[pattern=north west lines, pattern color=black] (0,0) rectangle (0.5,0.5); & Different regions ($R_{i,j}=0$) \\
\tikz{\fill[black!50] (0,0) rectangle (0.5,0.5); \draw (0,0) rectangle (0.5,0.5);} & Same region ($R_{i,j}=1$)
\end{tabular}
}
\caption{Schematic illustration (not an empirical network) of the block indicator $R_{ij}$ for four hypothetical regions $A,B,C,D$. Same-region node pairs have $R_{ij}=1$ (plain), and between-region pairs have $R_{ij}=0$ (striped). If the within- and between-region link counts are known separately, both parameters of the fitness-corrected planted-partition model can be estimated. The reconstruction problem studied here assumes that only their sum, the total link count, is available.}
\label{fig:region_matrix}
\end{figure}
We define the link probability \( p_{ij} \) as the following specific variant of the FCBM~\cite{Bernaschi2022fitness}:
\begin{equation}
\label{eq:category}
p_{ij}(\beta,\gamma)
=
\frac{e^{\beta} e^{\gamma R_{ij}}x_i x_j}
     {1+e^{\beta} e^{\gamma R_{ij}}x_i x_j}.
\end{equation}
Equivalently, the model has the log-odds form
$\operatorname{logit}p_{ij}=\beta+\gamma R_{ij}+\log x_i+\log x_j$.
Thus, $\beta$ and $\gamma$ are both parameters on the log-odds scale;
$e^{\beta}$ is the positive baseline odds multiplier, not an additional
model parameter. In plots involving the baseline parameter, we display
$e^{\beta}$ rather than $\beta$: this compresses the long tail at very
negative $\beta$ and makes the geometry of the feasible curve and the
selected points easier to inspect. We retain $\gamma$ rather than
$e^{\gamma}$ because its sign is informative, and $\gamma=0$ denotes
no regional effect. These are plotting conventions only; smooth
reparameterization does not change the induced Jeffreys measure,
selected solutions, or reconstruction results.
For node pairs in the same region ($R_{ij}=1$),
\[
p_{ij}(\beta,\gamma)
=
\frac{e^{\beta} e^{\gamma}x_i x_j}{1+e^{\beta} e^{\gamma}x_i x_j}.
\]
For node pairs in different regions ($R_{ij}=0$),
\[
p_{ij}(\beta,\gamma)
=
\frac{e^{\beta}x_i x_j}{1+e^{\beta}x_i x_j}.
\]
\noindent where
\begin{itemize}
    \item \( \beta \): parameter controlling the overall link density;
    \item \( \gamma \): parameter capturing the same-region effect.
\end{itemize}

\noindent\textbf{Parameter estimation with full knowledge of the constraints.}
Compared to the traditional reconstruction setup of the FM, new information has been added: the total number of links in the same region, which comes with the extra parameter $\gamma$ in the model.
This means that the sufficient statistics needed to identify the two parameters $\beta$ and $\gamma$ are the following quantities:
\begin{itemize}
    \item $L_{R_1}$: number of links between countries within the same region;
    \item $L_{R_0}$: number of links between countries in  different regions.
    \end{itemize}
Clearly, the two numbers add up to the total number of links that is used by the usual FM to estimate its only parameter: $L_{\mathrm{total}} = L_{R_1} + L_{R_0}$.
Assuming that both $L_{R_1}$ and $L_{R_0}$ are separately available ensures that the model is  statistically identifiable. We will make this assumption for the moment, and depart from it later.  

The log-likelihood of observing the network is:
\begin{equation}
\label{eq:loglik_fg}
\begin{aligned}
\ell(\beta,\gamma)
&= \log \mathcal{L}(\beta,\gamma \mid A) \\
&= \sum_{i<j}
\Bigl[
 a_{ij}\log p_{ij}
 +(1-a_{ij})\log(1-p_{ij})
\Bigr].
\end{aligned}
\end{equation}
The first-order conditions for maximum likelihood estimation are obtained by setting the partial derivatives of the log-likelihood to zero:
\[
\frac{\partial \ell}{\partial \beta}
= \sum_{i<j}(a_{ij}-p_{ij}) = 0,
\]
\[
\frac{\partial \ell}{\partial \gamma}
= \sum_{i<j}(a_{ij}-p_{ij})R_{ij} = 0.
\]
These equations imply the moment-matching conditions
\[
\sum_{i<j} p_{ij}
=
\sum_{i<j} a_{ij},
\qquad
\sum_{i<j} p_{ij} R_{ij}
=
\sum_{i<j} a_{ij} R_{ij}.
\]
The first identity matches the expected total number of links to the observed total, while the second matches the expected number of intra-regional links to its observed value. 
Via the explicit definitions
\[
L_{R_1} = \sum_{i<j} a_{ij} R_{ij},
\qquad
L_{R_0} = \sum_{i<j} a_{ij} (1-R_{ij}),
\]
we see that the two moment-matching conditions are equivalent to imposing the two structural constraints
\[
\sum_{R_{ij}=1} p_{ij} = L_{R_1},
\qquad
\sum_{R_{ij}=0} p_{ij} = L_{R_0}.
\]
Inserting the expression for $p_{ij}$, we rewrite the two conditions as
\begin{itemize}
    \item Different Region Constraint:
\[
\sum_{i<j; R_{ij} = 0}   \frac{e^{\beta} x_i x_j}{1 + e^{\beta} x_i x_j} = L_{R_0};
\]  
    \item Same Region Constraint:
\[
\sum_{i<j; R_{ij} = 1}  \frac{ e^{\beta} e^{\gamma} x_i x_j}{1 +  e^{\beta} e^{\gamma} x_i x_j} = L_{R_1}.
\]
\end{itemize}
Clearly, now we have two equations and two unknown parameters to be identified.
However, the first equation can be solved independently, and the second one can be solved subsequently.
Indeed, denoting as \( \beta^\star \) and \( \gamma^\star \)
the parameter values that solve the above equations, we can proceed as follows:
\begin{enumerate}
    \item solve the different-region constraint to obtain $\beta^\star$;
    \item inserting $\beta^\star$ obtained from the previous step, solve the same-region constraint to obtain $\gamma^\star$.
\end{enumerate}
Similar to the previous section, once $\beta^\star$ and $\gamma^\star$ are determined, the full probability matrix $p_{ij}$ is uniquely specified. From this point onward, we refer to $(\beta^\star,\gamma^\star)$ as the \textbf{full-information reference point}. For empirical networks this is the MLE obtained when the additional within-/between-block link counts are available; the term ``reference point'' is used to distinguish this empirical estimate from a known data-generating parameter.

\subsection{Fitness-Corrected Planted Partition Model with a Jeffreys-Induced Measure}
\label{sec:jeffreys-method}
\subsubsection{Feasible curve and induced Fisher measure}
We now consider the same FCBM as in Eq.~\eqref{eq:category}, but assume that $L_{R_0}$ and $L_{R_1}$ are not separately available. Only their sum,
$L_{\mathrm{total}}=L_{R_0}+L_{R_1}$, is observed. The block labels $R_{ij}$ and node fitnesses $x_i$ remain known exogenous covariates. Thus, relative to the block-agnostic FM, the proposed model uses the same \emph{network-level sufficient statistic}, but it additionally exploits known block memberships.

With independent Bernoulli edges, the log-likelihood is
\begin{equation}
\label{eq:loglik-fcbm}
\ell(\beta,\gamma)=\sum_{i<j}\left[a_{ij}\log p_{ij}+(1-a_{ij})\log(1-p_{ij})\right],
\end{equation}
with score components
\begin{equation}
\label{eq:scores}
\frac{\partial\ell}{\partial\beta}=\sum_{i<j}(a_{ij}-p_{ij}),\qquad
\frac{\partial\ell}{\partial\gamma}=\sum_{i<j}(a_{ij}-p_{ij})R_{ij}.
\end{equation}
Because only the total link count is observed, two parameters must satisfy one observable moment condition. The compatible parameter pairs therefore form the feasible set
\begin{equation}
\label{eq:feasible-curve}
C(\beta,\gamma)
\equiv
\sum_{i<j}p_{ij}(\beta,\gamma)-L_{\mathrm{total}}=0.
\end{equation}
Whenever $\partial C/\partial\gamma\neq0$, Eq.~\eqref{eq:feasible-curve} locally defines a one-dimensional curve $\gamma(\beta)$, whose slope is
\begin{equation}
\label{eq:curve-slope-main}
\frac{d\gamma}{d\beta}
=-\frac{Q_0(\beta,\gamma)}{Q_1(\beta,\gamma)}.
\end{equation}
Here
\begin{align}
Q_0(\beta,\gamma)&=\sum_{i<j}p_{ij}(1-p_{ij}),\label{eq:q0}\\
Q_1(\beta,\gamma)&=\sum_{i<j}p_{ij}(1-p_{ij})R_{ij},\label{eq:q1}\\
Q_2(\beta,\gamma)&=\sum_{i<j}p_{ij}(1-p_{ij})R_{ij}^2.\label{eq:q2}
\end{align}
The Fisher information matrix of the unconstrained two-parameter family is
\begin{equation}
\label{eq:fisher-main}
I(\beta,\gamma)=
\begin{pmatrix}
Q_0 & Q_1\\
Q_1 & Q_2
\end{pmatrix}.
\end{equation}
Since $R_{ij}\in\{0,1\}$, $Q_2=Q_1$.

The available statistic does not identify a point in the two-dimensional family. We therefore use the Fisher information of the unconstrained family as a metric from which to construct the Jeffreys measure \emph{induced on the constrained one-dimensional submodel}; we do not interpret the corresponding two-dimensional Jeffreys density as a posterior distribution over $\beta$ and $\gamma$. Parameterizing the curve by $\gamma$, let $\theta(\gamma)=(\beta(\gamma),\gamma)$. The Fisher information along the tangent direction is
\begin{equation}
\label{eq:induced-info}
I_{\mathcal C}(\gamma)
=\theta'(\gamma)^{\mathsf T}I(\theta(\gamma))\theta'(\gamma)
=Q_2-\frac{Q_1^2}{Q_0}
=Q_1-\frac{Q_1^2}{Q_0}.
\end{equation}
The corresponding Jeffreys line element is
\begin{align}
\label{eq:jeffreys-line-element}
ds&=J_\gamma(\gamma)\,|d\gamma|=J_\beta(\beta)\,|d\beta|,\\
J_\gamma&=\sqrt{I_{\mathcal C}},\qquad
J_\beta=\left|\frac{d\gamma}{d\beta}\right|\sqrt{I_{\mathcal C}}.\notag
\end{align}
Equation~\eqref{eq:jeffreys-line-element} makes the construction invariant to whether the same feasible curve is parameterized by $\beta$, $\gamma$, or any other smooth one-to-one coordinate. The normalized measure on a feasible segment $\mathcal C$ is therefore
\begin{equation}
\label{eq:normalized-jeffreys-measure}
d\mu_J=\frac{ds}{S_{\mathrm{tot}}},
\qquad
S_{\mathrm{tot}}=\int_{\mathcal C}ds.
\end{equation}
The empirical numerical procedure evaluates a finite segment of the feasible curve, with admissible values of $\beta$ determined from the total-link constraint. Appendix~\ref{app:math} gives the limiting conditions as $\gamma\to\pm\infty$. This numerical segment should not be interpreted as a proof that the full mathematical feasible family is bounded.

The measure $\mu_J$ assigns an invariant weighting to the compatible solutions on $\mathcal C$. The algorithm below does \emph{not} compute a Jeffreys-weighted average of $(\beta,\gamma)$ or of the probability matrices $\{p_{ij}\}$. Instead, $\mu_J$ induces a distribution of entropy values on the feasible family, from which we select one representative solution.

\subsubsection{Jeffreys-uniform discretization and entropy distribution}
\label{sec:jeffreys-discretization}
For the empirical reconstructions we discretize the feasible curve on a numerical grid in $\beta$. At each grid point, the total-link constraint is solved for $\gamma$ by bisection. We accumulate the induced Jeffreys length in the $\beta$ coordinate using the trapezoidal rule and locate equally spaced arclength targets by linear interpolation of the cumulative length. Let $M$ denote the number of samples. We define
\begin{equation}
\label{eq:s-grid}
s_m=\frac{m}{M}S_{\mathrm{tot}},\qquad m=1,\ldots,M,
\end{equation}
and choose $(\beta_m,\gamma_m)$ so that
\begin{equation}
\label{eq:s-inversion}
s_m=\int_{\beta_{\min}}^{\beta_m}J_\beta(\beta)\,|d\beta|.
\end{equation}
Both coordinates of each sampled pair $(\beta_m,\gamma_m)$ are obtained by linear interpolation between the adjacent feasible grid points. Consequently, the discrete samples approximate both Jeffreys-uniform spacing and the total-link constraint up to interpolation and quadrature accuracy; no additional root solve is performed at the interpolated points. At each point we compute
\begin{equation}
\label{eq:entropy-main}
H_m=-\sum_{i<j}\left[p_{ij}^{(m)}\log p_{ij}^{(m)}+(1-p_{ij}^{(m)})\log(1-p_{ij}^{(m)})\right].
\end{equation}
Thus $\{H_m\}$ is a numerical sample from the entropy distribution induced by the normalized Jeffreys measure.

For comparison, we consider four entropy-based representatives: minimum, maximum, mean, and median. The extreme points are
\begin{equation}
\label{eq:entropy-extrema}
\theta^{\min H}=\operatorname*{arg\,min}_{m}H_m,
\qquad
\theta^{\max H}=\operatorname*{arg\,max}_{m}H_m,
\end{equation}
where $\theta_m=(\beta_m,\gamma_m)$. Let $\mathcal I$ denote the connected arc of the feasible curve between these two extrema. Expressing this restriction intrinsically as an arc, rather than as a particular coordinate interval, avoids dependence on the choice of $\beta$ or $\gamma$ as plotting coordinates. The mean-entropy representative is the point on $\mathcal I$ whose entropy is closest to
\begin{equation}
\label{eq:mean-entropy}
\bar H=\frac{1}{M}\sum_{m=1}^{M}H_m,
\end{equation}
and the median-entropy representative is the point on $\mathcal I$ whose entropy is closest to
\begin{equation}
\label{eq:median-entropy}
H^{\mathrm{med}}=\operatorname{median}\{H_m:m=1,\ldots,M\}.
\end{equation}
The arc restriction provides a stable tie-breaking convention when the same entropy level is attained on multiple branches. The numerical procedure is summarized in Appendix~\ref{app:algorithms}.

\subsubsection{Rationale for the Median-Entropy Point}
\label{sec:median-rationale}
The median-entropy point is introduced as a criterion for selecting a representative solution along the feasible curve, rather than as an additional identification condition. This distinction is particularly relevant when comparing it with the minimum-entropy solution. All points on the feasible curve satisfy Eq.~\eqref{eq:feasible-curve}, and therefore have the same expected total number of links. Consequently, selecting the minimum-entropy point does not yield a sparser reconstruction. Instead, it produces a more concentrated allocation of the same expected link probability, with individual probabilities driven closer to zero or one. In the present planted-partition formulation, the minimum-entropy point is associated with a strong concentration of connectivity within regions and a corresponding suppression of between-region probabilities. Such an extreme allocation is not implied by the observed total link count alone.

At the opposite extreme, the maximum-entropy point is characterized by a weak regional effect and approaches a more block-agnostic allocation of link probabilities. The mean-entropy representative provides an intermediate entropy value, but it can be sensitive to asymmetry in the entropy distribution along the feasible curve. The median is therefore used as a robust measure of central tendency of the entropy values sampled uniformly with respect to the Jeffreys measure. This choice avoids imposing either limiting connectivity pattern while retaining the regional information encoded by the model. We do not interpret the median-entropy point as a theoretically identified estimate of the missing sufficient statistic; its proximity to the full-information reference point and its reconstruction performance are evaluated empirically in Sec.~IV.

\subsection{Evaluation Metrics}
We assess reconstruction quality using ROC AUC and precision--recall AUC (PR AUC), computed from the model probabilities $p_{ij}$. ROC AUC measures ranking performance across positive and negative pairs, whereas PR AUC focuses on the precision--recall trade-off and is particularly informative for sparse networks, where non-links dominate. Higher values indicate better discrimination; a random ranking has ROC AUC $0.5$, while the PR-AUC baseline equals the empirical link prevalence.

We also report likelihood-based penalized scores in the algebraic forms of the Akaike~\cite{akaike1974} and Bayesian (Schwarz)~\cite{BIC} information criteria,
\begin{equation}
\label{eq:aic-bic}
\mathrm{AIC}=-2\ell(\hat\theta)+2k,
\qquad
\mathrm{BIC}=-2\ell(\hat\theta)+k\ln\!\left[\frac{n(n-1)}{2}\right].
\end{equation}
For the block-agnostic FM and the full-information FCBM, $\hat\theta$ is a maximum-likelihood estimate, so these are the standard AIC and BIC. The median-entropy representative is not obtained by maximizing likelihood along the feasible curve. Consequently, for that method the same formulas are reported only as descriptive \emph{AIC/BIC-form penalized log-likelihood scores}, with $k=1$ denoting the one-dimensional constrained family. We do not use those two values as formal evidence that the median-entropy rule is statistically preferred to the full-information MLE.

\section{Dataset Summary}
\subsection{ELEnet16}
The \textbf{ELEnet16} dataset is part of the \texttt{ITNr} R package, which is commonly used for international trade network analysis. The dataset contains the international trade network of automotive electrical goods for the year 2016. The classification of electrical automotive goods is defined according to \cite{Amighini2014}. The data include countries involved in the trade of electrical automotive goods, with edges directed and weighted, representing trade flows between countries. A threshold is applied by the authors, whereby only edges representing at least 0.01\% of global trade are included. 

\subsection{UN Comtrade}
The \textbf{UN Comtrade} database is the official repository of international trade statistics collected and reported by national customs and statistical agencies. The database provides raw trade data, and there may be asymmetries in imports and exports between countries. The dataset provides export data across multiple classification systems (HS, SITC) and is updated regularly. The data can be downloaded at  \cite{UN_COMTRADE_data}

\subsection{BACI (CEPII)}
The \textbf{BACI} dataset, developed by the CEPII institute, is widely used in academic and policy research. BACI is a harmonized bilateral trade dataset constructed by refining, normalizing, and reconciling raw UN Comtrade records. It addresses national reporting asymmetries through mirror-flow reconciliation and provides harmonized bilateral trade flows at the 6-digit HS (Harmonized System) product level. The product--year combinations used in this study are listed in Fig.~\ref{fig:trade_networks} and Table~\ref{tab:beta_gamma_summary}.

\subsection{Selected sub-dataset}

We filter different products and years from the BACI and UN Comtrade datasets for application to the proposed algorithm. Details about HS codes and official descriptions by World Customs Organization (WCO) for all items are in Table \ref{tab:HS_code}. The data are selected across multiple years and product categories to show the stability of the algorithm. Specifically:
\begin{itemize}
    \item \textbf{Fresh products:} Milk and plums. These are essential goods and are relatively costly to transport over long distances.
    \item \textbf{Geographically specific products:} Cocoa and oil. These products are produced in geographically concentrated locations, where specific countries hold clear production advantages, while demand remains global.
    \item \textbf{Technology-intensive manufactured products:} Refrigerators and automotive (ElEnet16). Trade in these goods is strongly influenced by G7 economies. 
    \item \textbf{Common products:} Steel, fabric, wood. These products are widely produced across countries, and no country holds a strongly dominant production advantage.
\end{itemize}
\begin{table}[t]
\centering
\scriptsize
\caption{HS product codes and descriptions for the empirical international-trade datasets used in this study.}
\label{tab:HS_code}
\renewcommand{\arraystretch}{1.18}
\setlength{\tabcolsep}{4pt}

\begin{tabular}{|c|l|l|}
\hline
\textbf{HS code} & \textbf{In Short} & \textbf{Official description by WCO} \\
\hline

040110 & Milk &
\parbox[t]{0.55\columnwidth}{\raggedright
\strut\hspace{2pt}Dairy produce: milk and cream, not concentrated, not containing added sugar or other sweetening matter, of a fat content, by weight, not exceeding 1\%.\hspace{2pt}\strut} \\
\hline

721810 & Steel &
\parbox[t]{0.55\columnwidth}{\raggedright
\strut\hspace{2pt}Steel, stainless: ingots and other primary forms.\hspace{2pt}\strut} \\
\hline

841810 & Refrigerators &
\parbox[t]{0.55\columnwidth}{\raggedright
\strut\hspace{2pt}Refrigerators and freezers: combined refrigerator-freezers, fitted with separate external doors, electric or other.\hspace{2pt}\strut} \\
\hline

080940 & Plums &
\parbox[t]{0.55\columnwidth}{\raggedright
\strut\hspace{2pt}Fruit, edible: plums and sloes, fresh.\hspace{2pt}\strut} \\
\hline

2707 & Oils &
\parbox[t]{0.55\columnwidth}{\raggedright
\strut\hspace{2pt}Oils and products of the distillation of high temperature coal tar.\hspace{2pt}\strut} \\
\hline

551211 & Fabrics &
\parbox[t]{0.55\columnwidth}{\raggedright
\strut\hspace{2pt}Fabrics, woven: of synthetic staple fibres, containing 85\% or more by weight of polyester staple fibres, unbleached or bleached.\hspace{2pt}\strut} \\
\hline

441300 & Wood &
\parbox[t]{0.55\columnwidth}{\raggedright
\strut\hspace{2pt}Wood: densified wood, in blocks, plates, strips or profile shapes.\hspace{2pt}\strut} \\
\hline

180500 & Cocoa &
\parbox[t]{0.55\columnwidth}{\raggedright
\strut\hspace{2pt}Cocoa: powder, not containing added sugar or other sweetening matter.\hspace{2pt}\strut} \\
\hline

\end{tabular}
\end{table}

Additionally, the \textbf{Gravity} dataset (CEPII) was also used to obtain annual GDP data for each country. Countries without information on GDP or economic region were excluded from the analysis.

\begin{figure*}[tbp]
  \centering
  \includegraphics[width=0.96\textwidth]{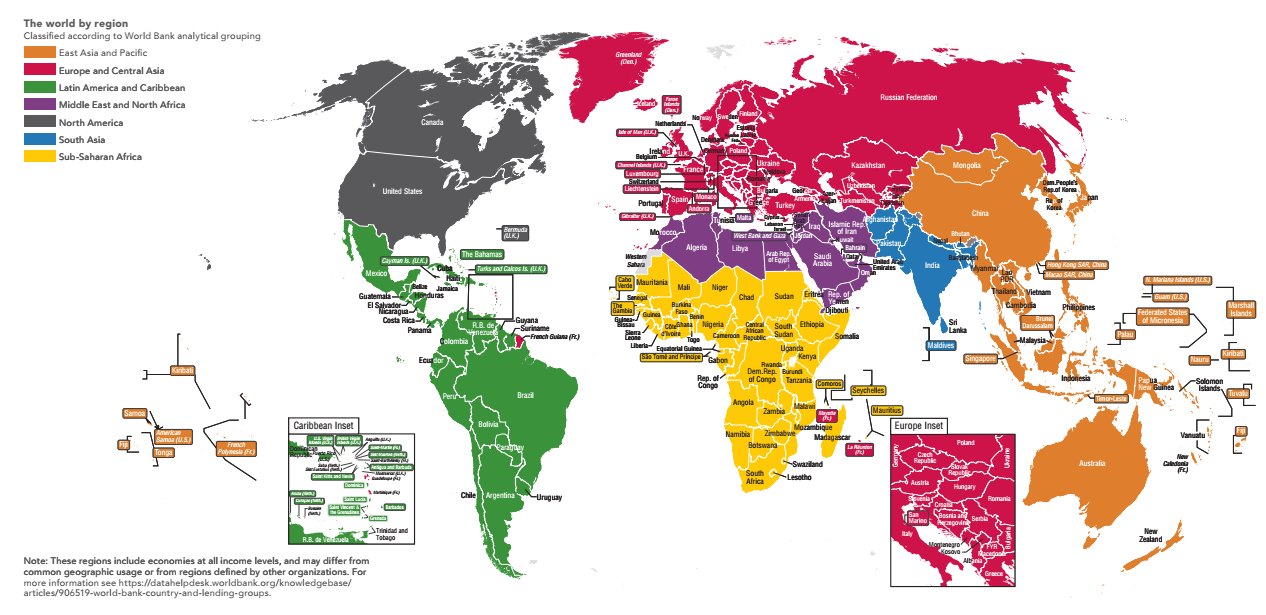}
  \caption{Reference map of the World Bank economic regions used to assign the block labels to countries in the empirical trade networks and synthetic experiments. Source: World Bank Data Topics (\url{datatopics.worldbank.org/world-development-indicators/images/figures-png/world-by-region-map.pdf}). Used under Creative Commons Attribution 4.0 International (CC BY 4.0).}
  \label{fig:economic_regions}
\end{figure*}

\subsection{Economic regions}
The World Bank defines seven economic regions (Figure~\ref{fig:economic_regions}). Each region groups countries based on geographical proximity, economic similarities, and trade integration. The definitions of these economic regions are not related to the definitions of political or cultural blocs. Detailed country classifications are available at \cite{World_bank_data}.


The seven economic regions include:

\begin{itemize}
    \item \textbf{East Asia and Pacific (EAP)}: Includes countries in East Asia and the Pacific, as well as Australia. Notable economies in this region include China, Japan, South Korea, and Australia. This region also includes populous countries with dynamically growing economies such as Indonesia and Vietnam.

    \item \textbf{Europe and Central Asia (ECA)}: Includes all EU countries, along with Russia, West Asian countries, and Turkey. This group is characterized by technologically advanced economies and has developed industrialized and service economies. Most prominent are the G7 countries: Germany, France, Italy and the United Kingdom.

    \item \textbf{Latin America and the Caribbean (LAC)}: Countries in the Americas from Mexico southward. Includes South American, Central American, and Caribbean countries. This region includes resource-rich exporting countries (e.g., Brazil, Chile) and service-oriented and maritime trade-dependent economies.

    \item \textbf{Middle East and North Africa (MENA)}: All of the Middle East (except Turkey) and North African countries, notably Egypt and Morocco. This region is known for its complex political structure. The region's economy is dependent on oil and gas exports.

    \item \textbf{North America (NA)}: There are only three countries, the United States, Canada, and Bermuda. This region is known for its highly developed economy, with advanced infrastructure and high income levels. The United States and Canada are members of the G7.

    \item \textbf{South Asia (SAR)}: Includes India, Pakistan, Bangladesh, and surrounding countries. It is a region with a strong human resource base due to its large population, growing service and industrial sectors. However, the region remains economically less developed relative to other regions.
    
    \item \textbf{Sub-Saharan Africa (SSA)}: Includes both Eastern \& Southern Africa (AFE) and Western \& Central Africa (AFW). This region is very diverse in terms of economy. It consists primarily of low- and middle-income countries. One advantage of this region is its rich natural resources.
    
\end{itemize}

\section{Reconstruction results}
\subsection{Visualization of real-world networks}
\begin{figure*}[p]
\centering

\newcommand{\panelw}{0.327\textwidth}
\newcommand{\panelgap}{0.006\textwidth}

\begin{minipage}[t]{\panelw}
    \centering
    \includegraphics[width=\linewidth]{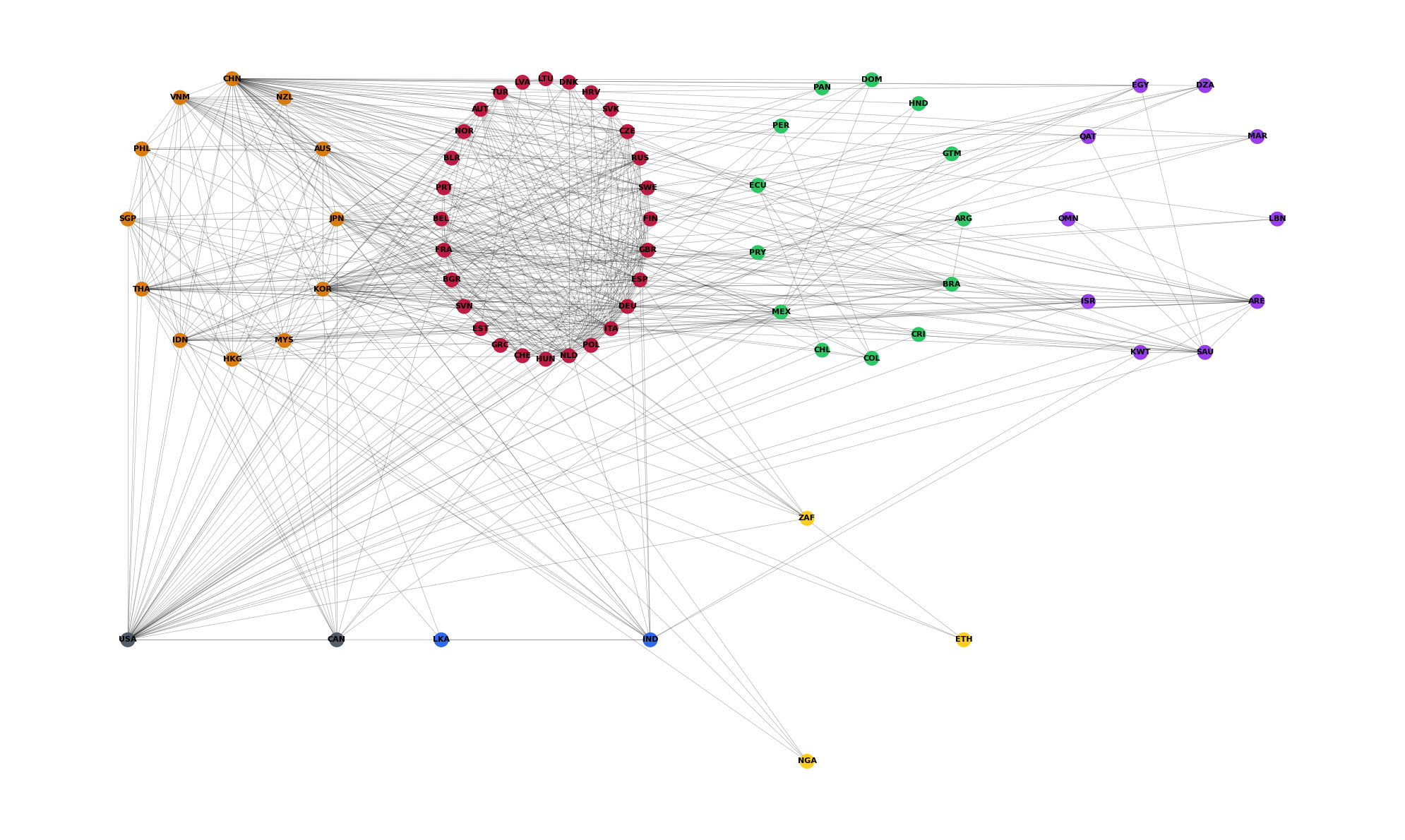}
    \vspace{0.12em}
    \subcaptext{(a) ELEnet 2016 Automotive}
\end{minipage}%
\hspace{\panelgap}%
\begin{minipage}[t]{\panelw}
    \centering
    \includegraphics[width=\linewidth]{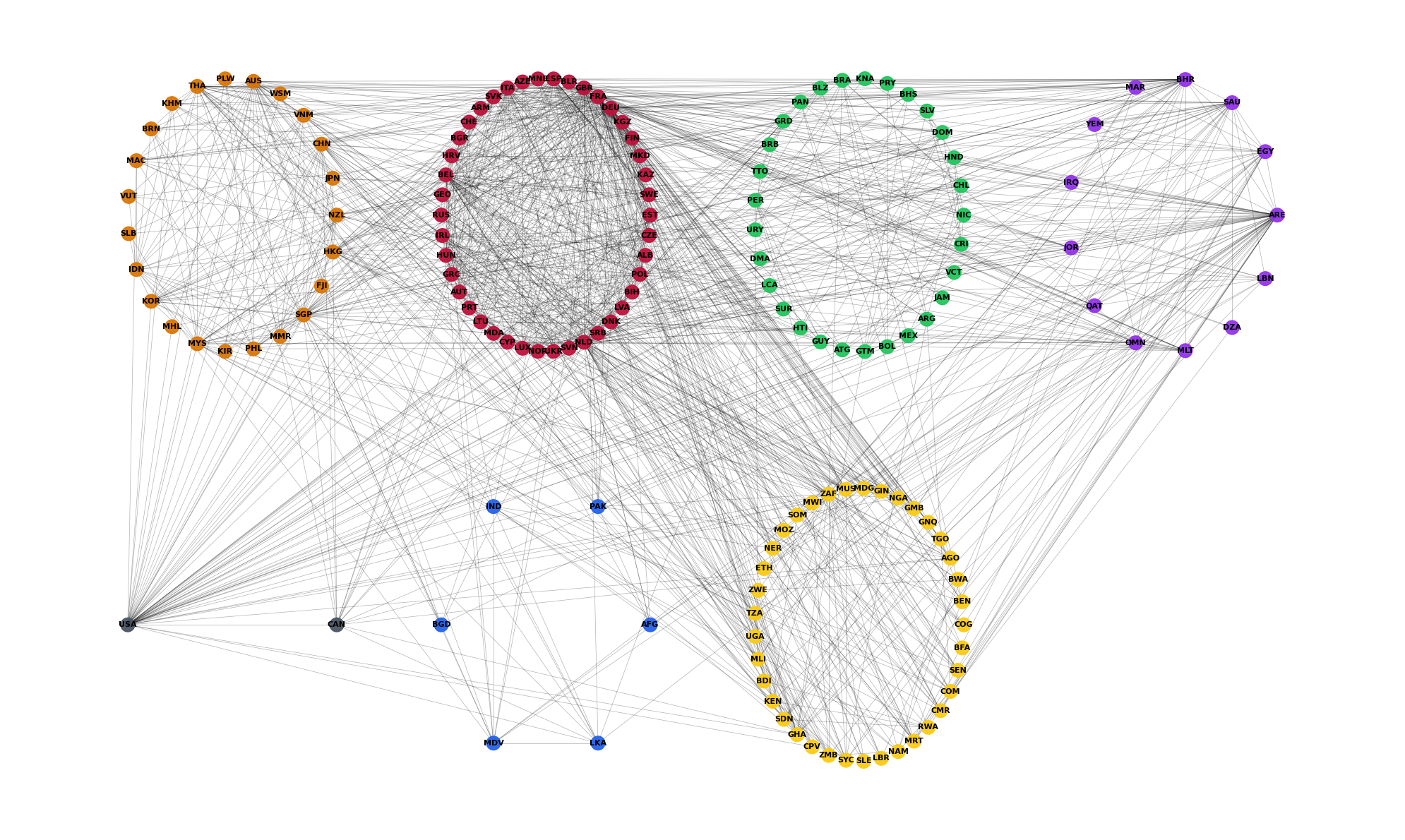}
    \vspace{0.12em}
    \subcaptext{(b) BACI 2016 Milk}
\end{minipage}%
\hspace{\panelgap}%
\begin{minipage}[t]{\panelw}
    \centering
    \includegraphics[width=\linewidth]{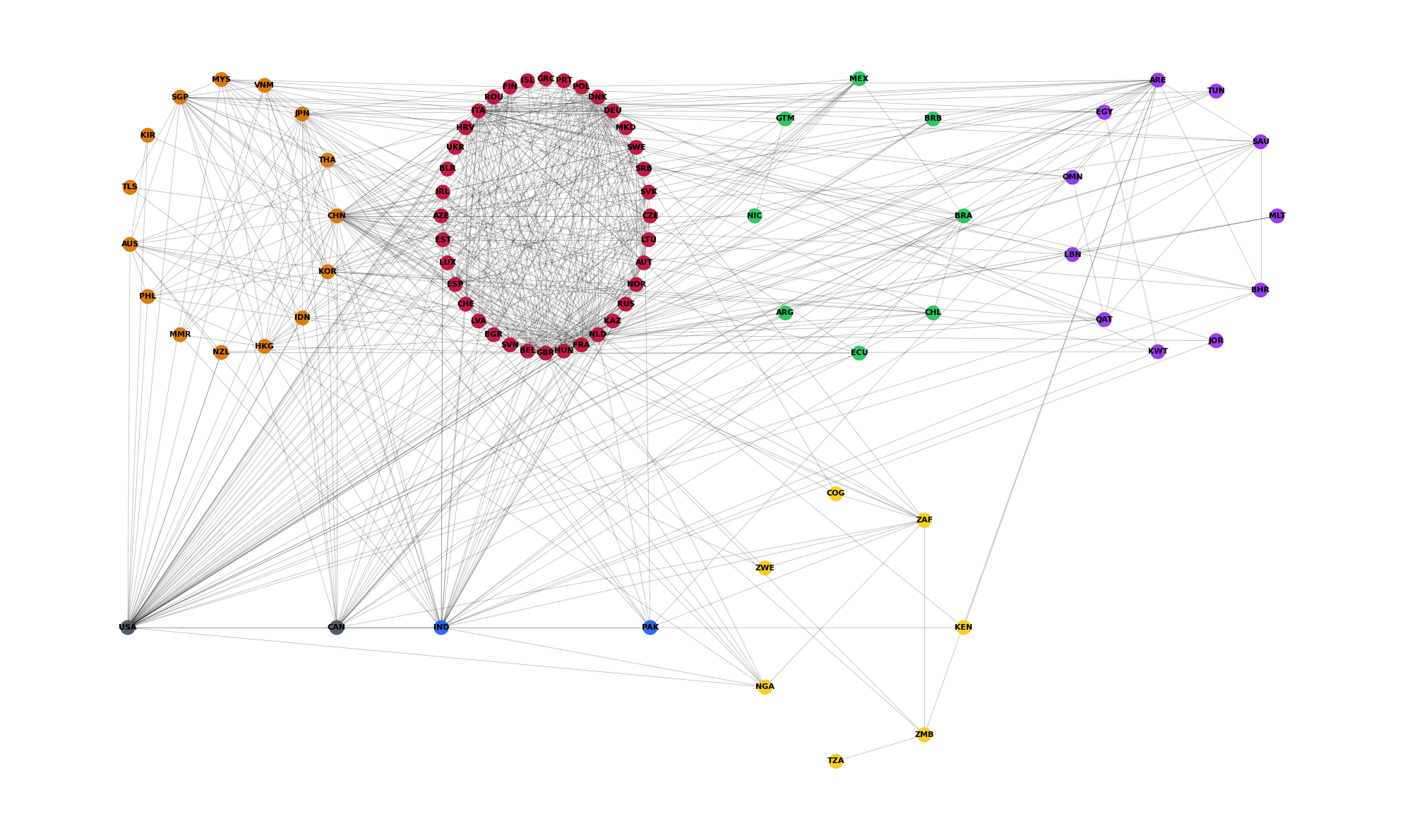}
    \vspace{0.12em}
    \subcaptext{(c) UN Comtrade 2017 Steel}
\end{minipage}

\vspace{0.35em}

\begin{minipage}[t]{\panelw}
    \centering
    \includegraphics[width=\linewidth]{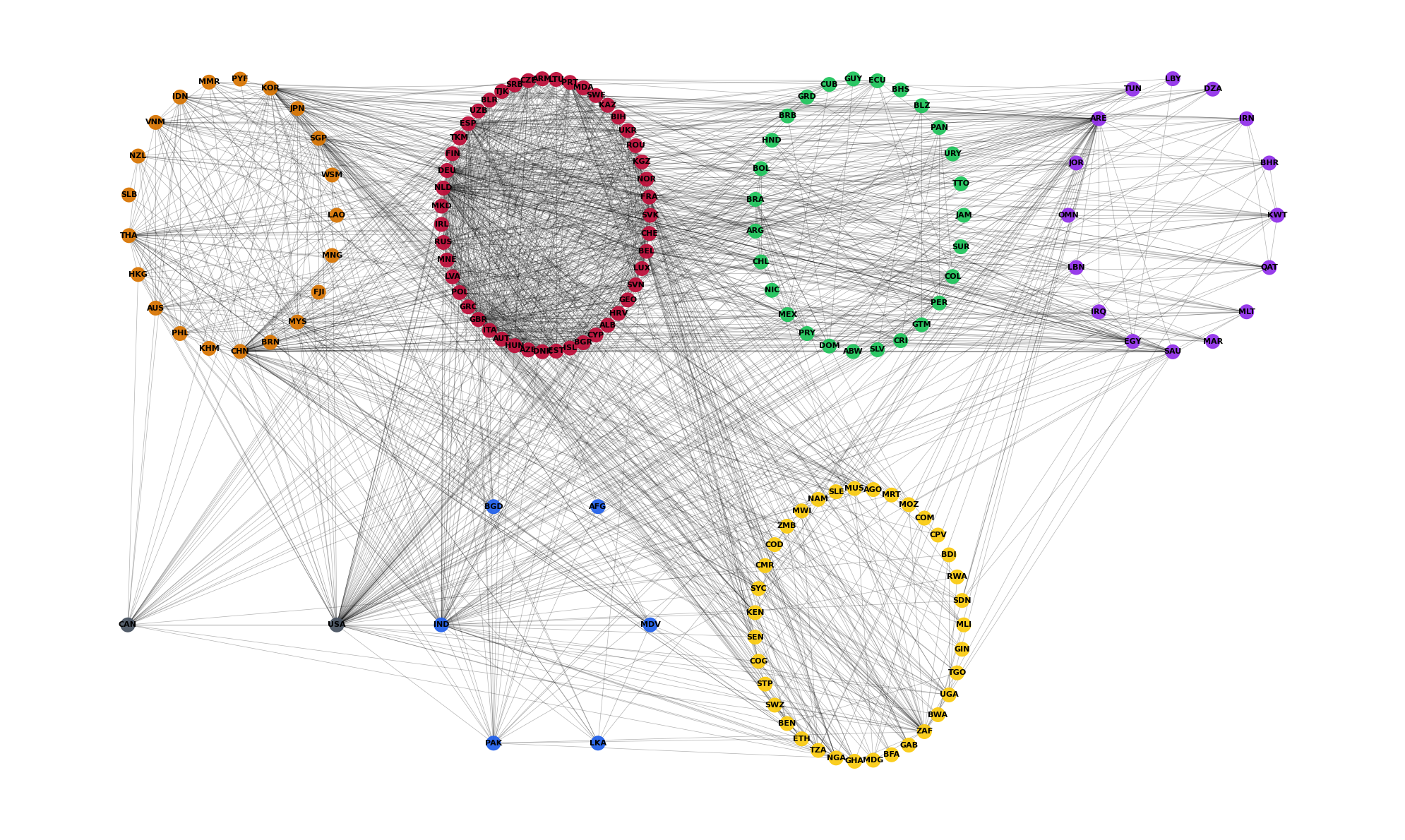}
    \vspace{0.12em}
    \subcaptext{(d) UN Comtrade 2018 Oils}
\end{minipage}%
\hspace{\panelgap}%
\begin{minipage}[t]{\panelw}
    \centering
    \includegraphics[width=\linewidth]{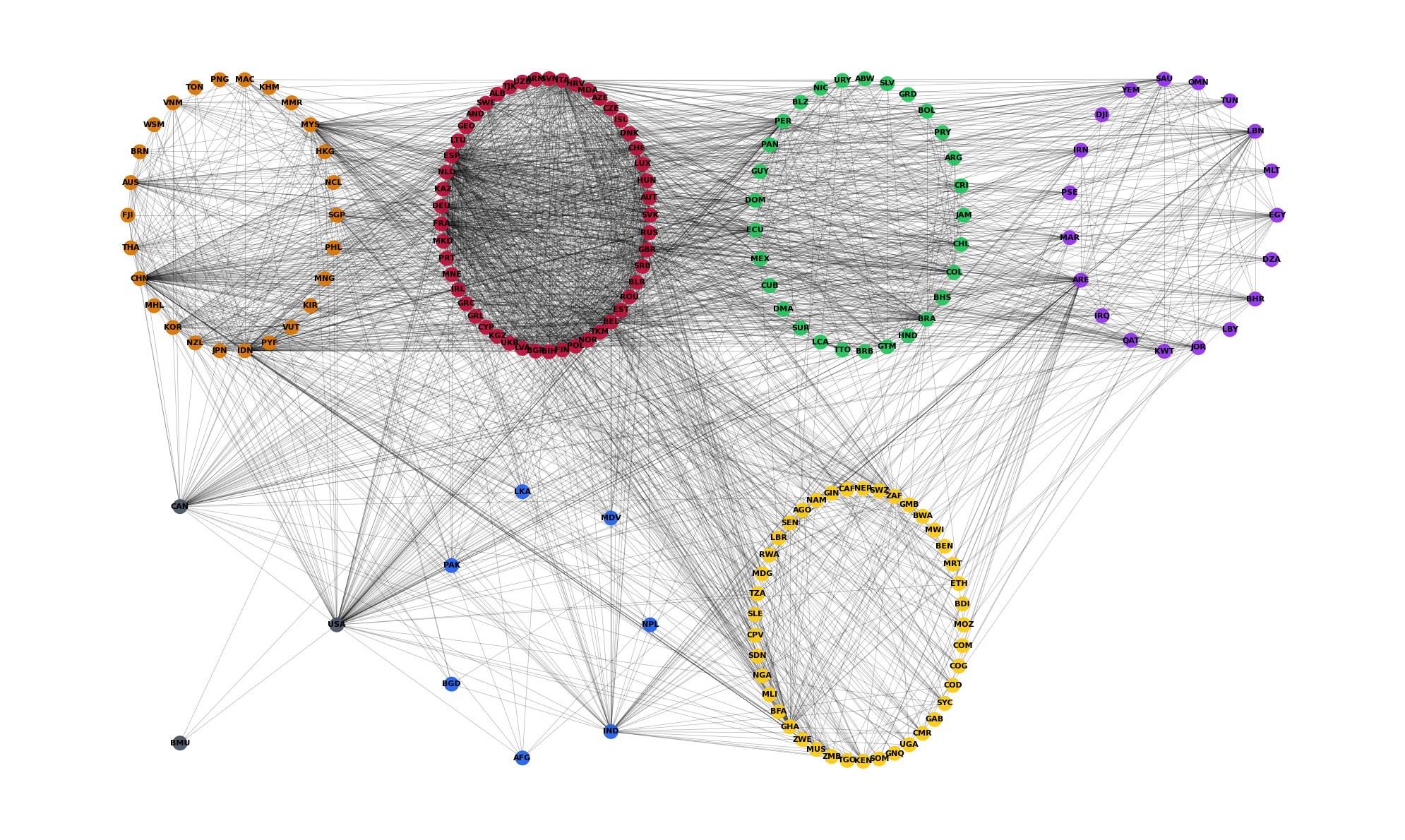}
    \vspace{0.12em}
    \subcaptext{(e) BACI 2019 Cocoa}
\end{minipage}%
\hspace{\panelgap}%
\begin{minipage}[t]{\panelw}
    \centering
    \includegraphics[width=\linewidth]{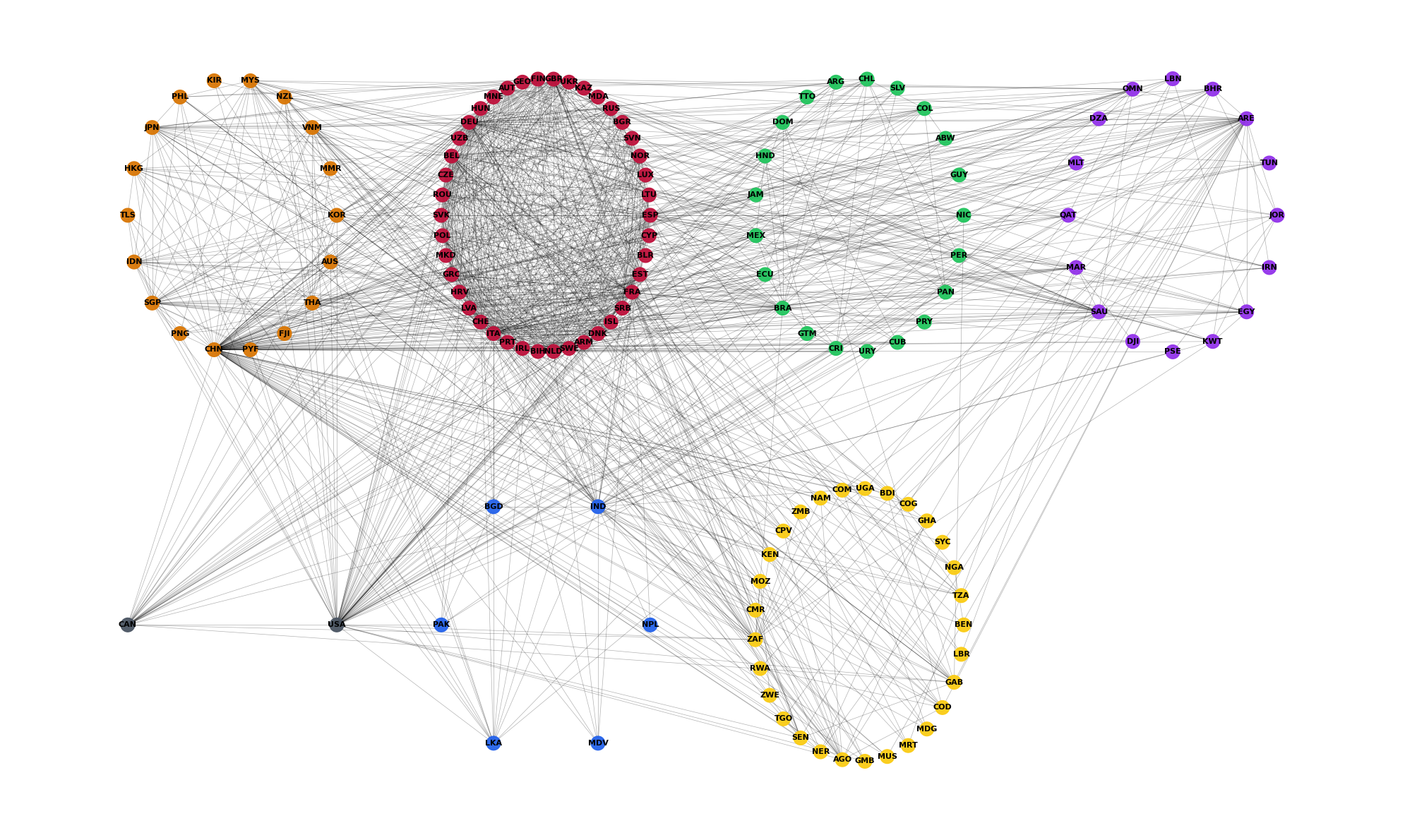}
    \vspace{0.12em}
    \subcaptext{(f) BACI 2020 Wood}
\end{minipage}

\vspace{0.35em}

\begin{minipage}[t]{\panelw}
    \centering
    \includegraphics[width=\linewidth]{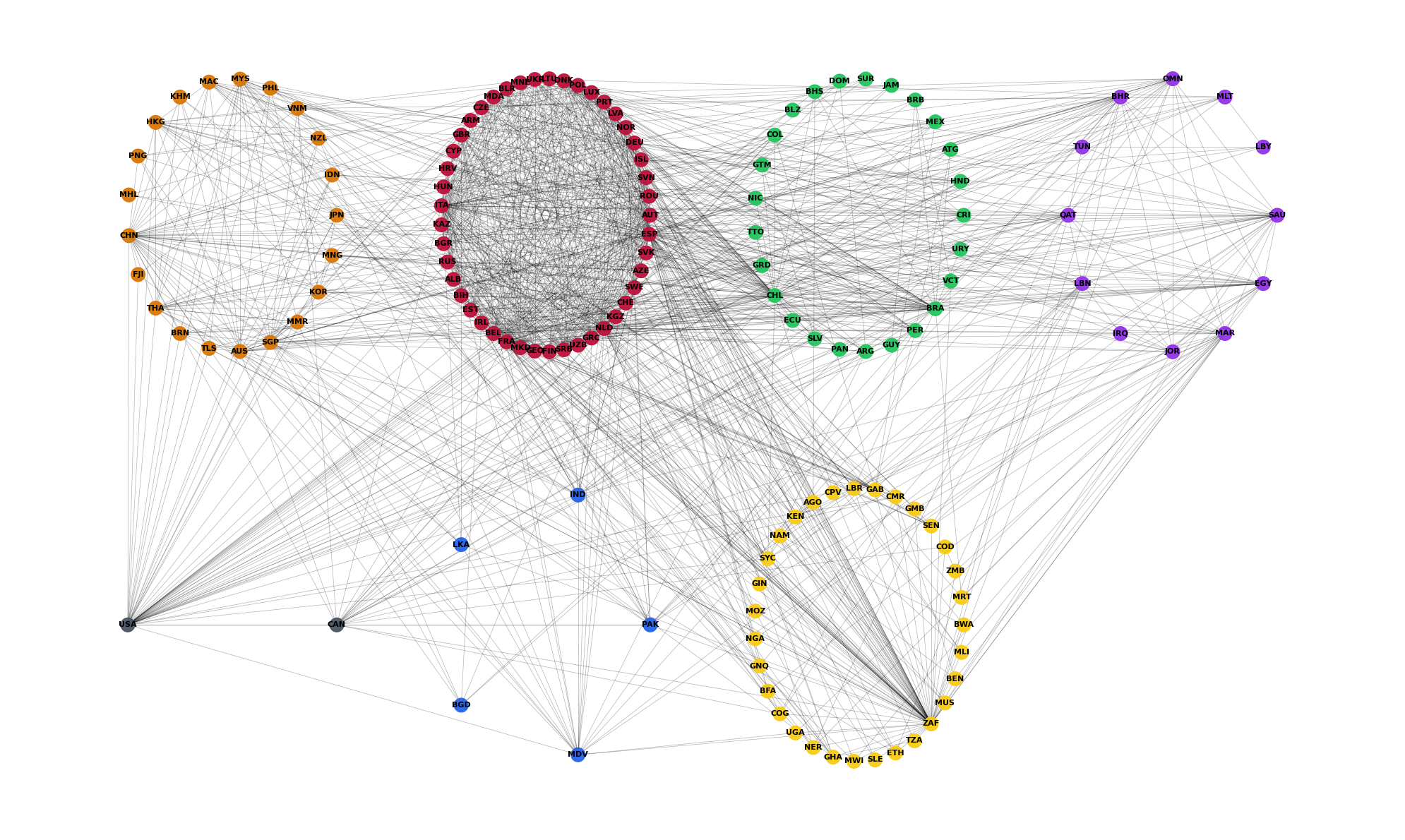}
    \vspace{0.12em}
    \subcaptext{(g) UN Comtrade 2021 Plums}
\end{minipage}%
\hspace{\panelgap}%
\begin{minipage}[t]{\panelw}
    \centering
    \includegraphics[width=\linewidth]{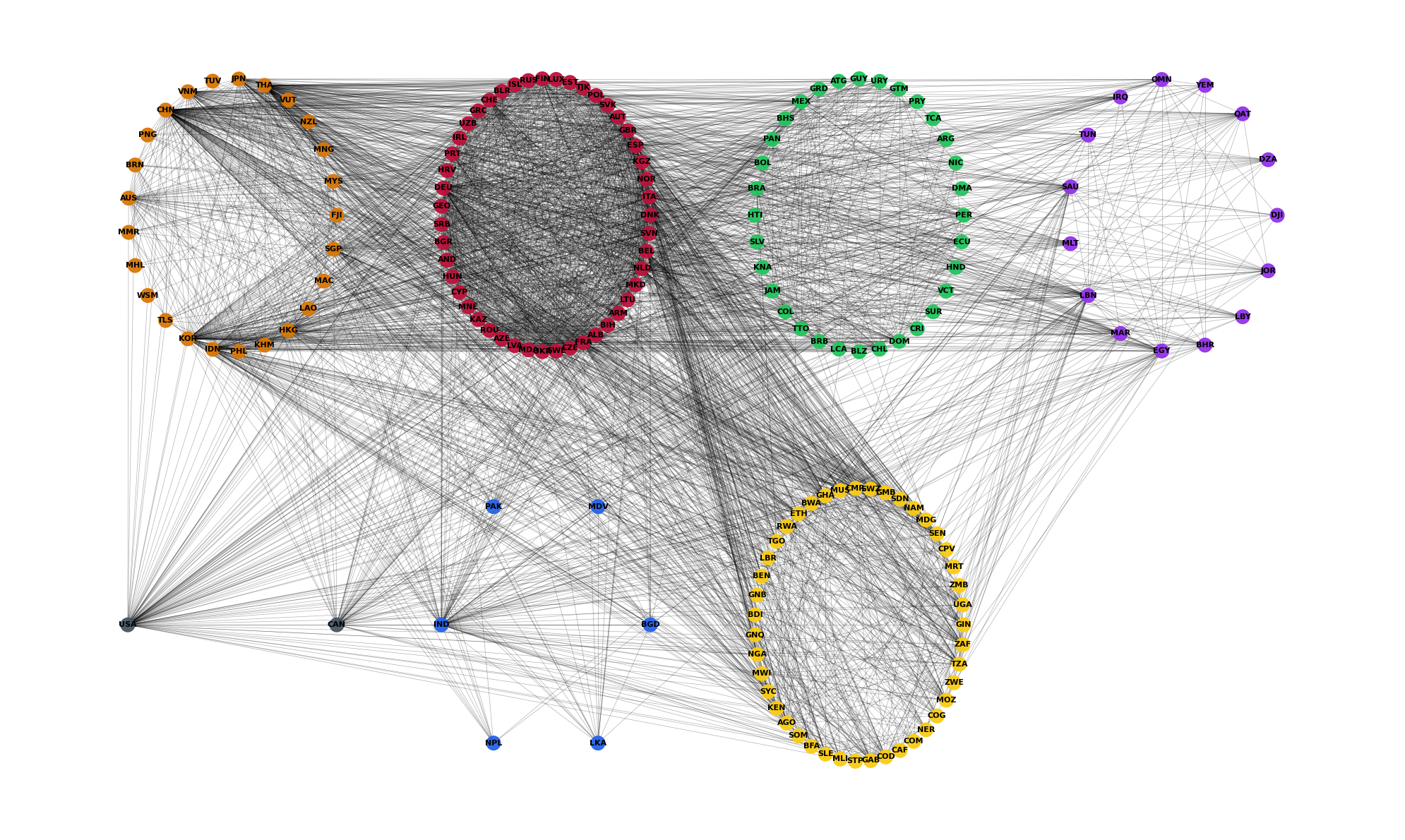}
    \vspace{0.12em}
    \subcaptext{(h) BACI 2022 Refrigerators}
\end{minipage}%
\hspace{\panelgap}%
\begin{minipage}[t]{\panelw}
    \centering
    \includegraphics[width=\linewidth]{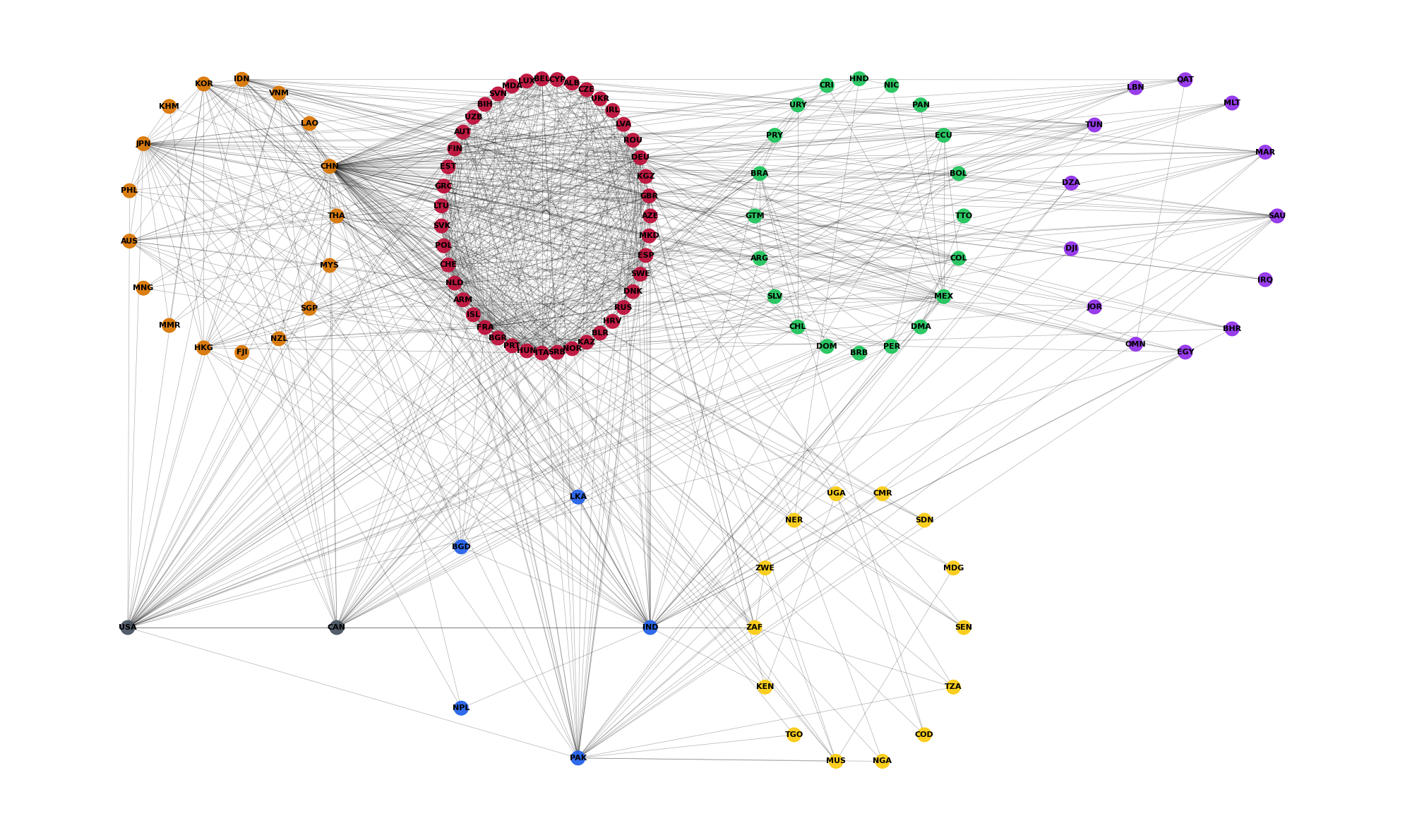}
    \vspace{0.12em}
    \subcaptext{(i) UN Comtrade 2023 Fabric}
\end{minipage}

\caption{Observed empirical international-trade networks for nine product--year combinations (not synthetic samples). For readability, only nodes with degree $\geq 3$ are displayed; this threshold is for visualization only. Node colors indicate World Bank economic regions: \EAPsym\ East Asia \& Pacific (EAP); \ECAsym\ Europe \& Central Asia (ECA); \LACsym\ Latin America \& Caribbean (LAC); \MENAsym\ Middle East \& North Africa (MENA); \NAsym\ North America (NA); \SARsym\ South Asia (SAR); \SSAsym\ Sub-Saharan Africa (SSA).}

\label{fig:trade_networks}
\end{figure*}

Figure~\ref{fig:trade_networks} provides a visualization of the empirical networks used in this paper. For visual clarity, only nodes with degree at least 3 are displayed in this figure; the degree threshold is a plotting choice and is not an additional constraint of the reconstruction model. We use colors for the nodes that correspond to the World Bank economic regions. A detailed list of datasets is provided in Table~\ref{tab:HS_code}. The full-information reference point corresponds to the full-information FCBM fit of Sec.~\ref{sec:fcbm-full}, which uses the observed within- and between-region link counts.
\subsection{Reconstruction results}

\begin{figure*}[p]
\centering

\newcommand{\panelw}{0.327\textwidth}
\newcommand{\panelgap}{0.006\textwidth}

\begin{minipage}[t]{\panelw}
    \centering
    \includegraphics[width=\linewidth]{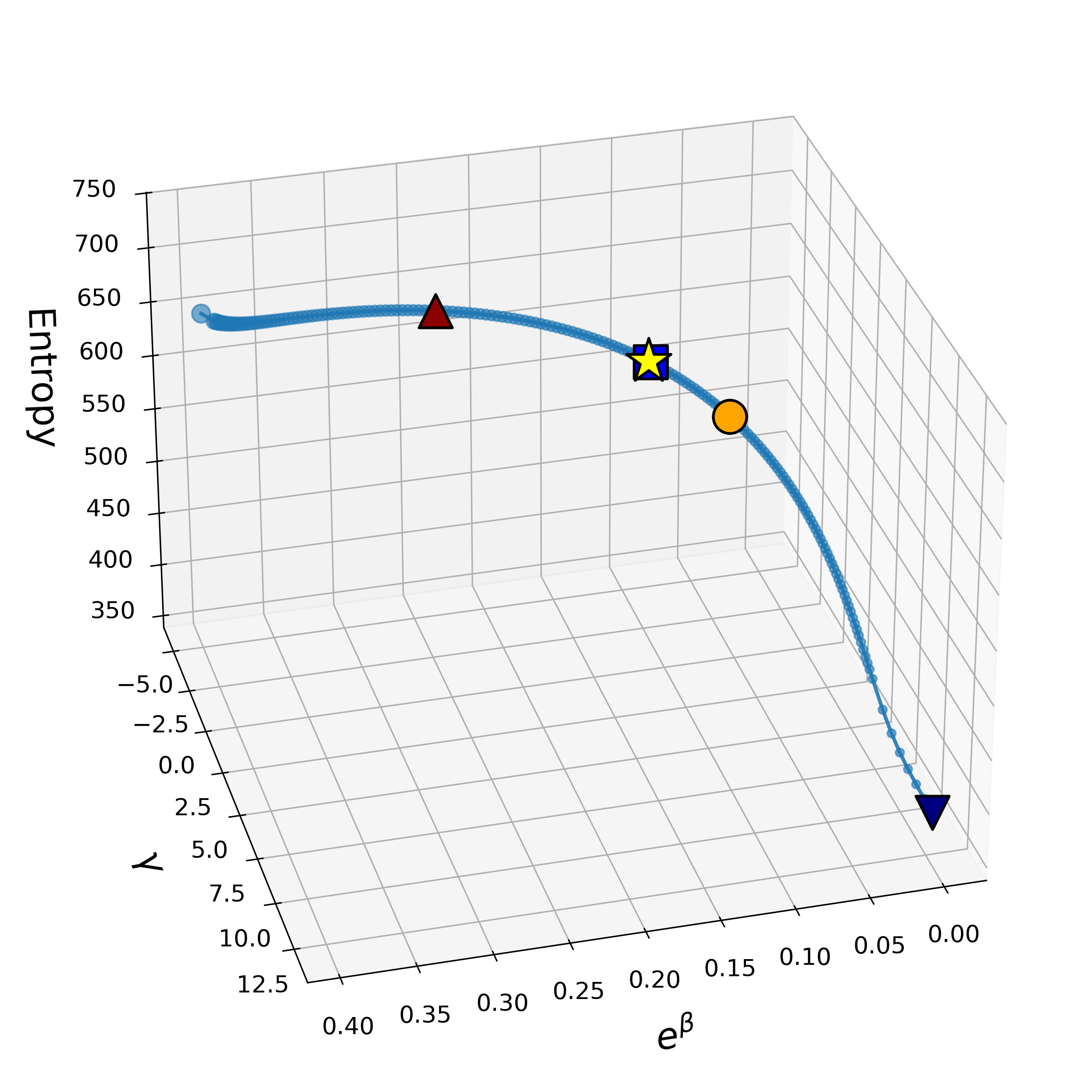}
    \vspace{0.12em}
    {\footnotesize (a) Automotive 2016}
\end{minipage}%
\hspace{\panelgap}%
\begin{minipage}[t]{\panelw}
    \centering
    \includegraphics[width=\linewidth]{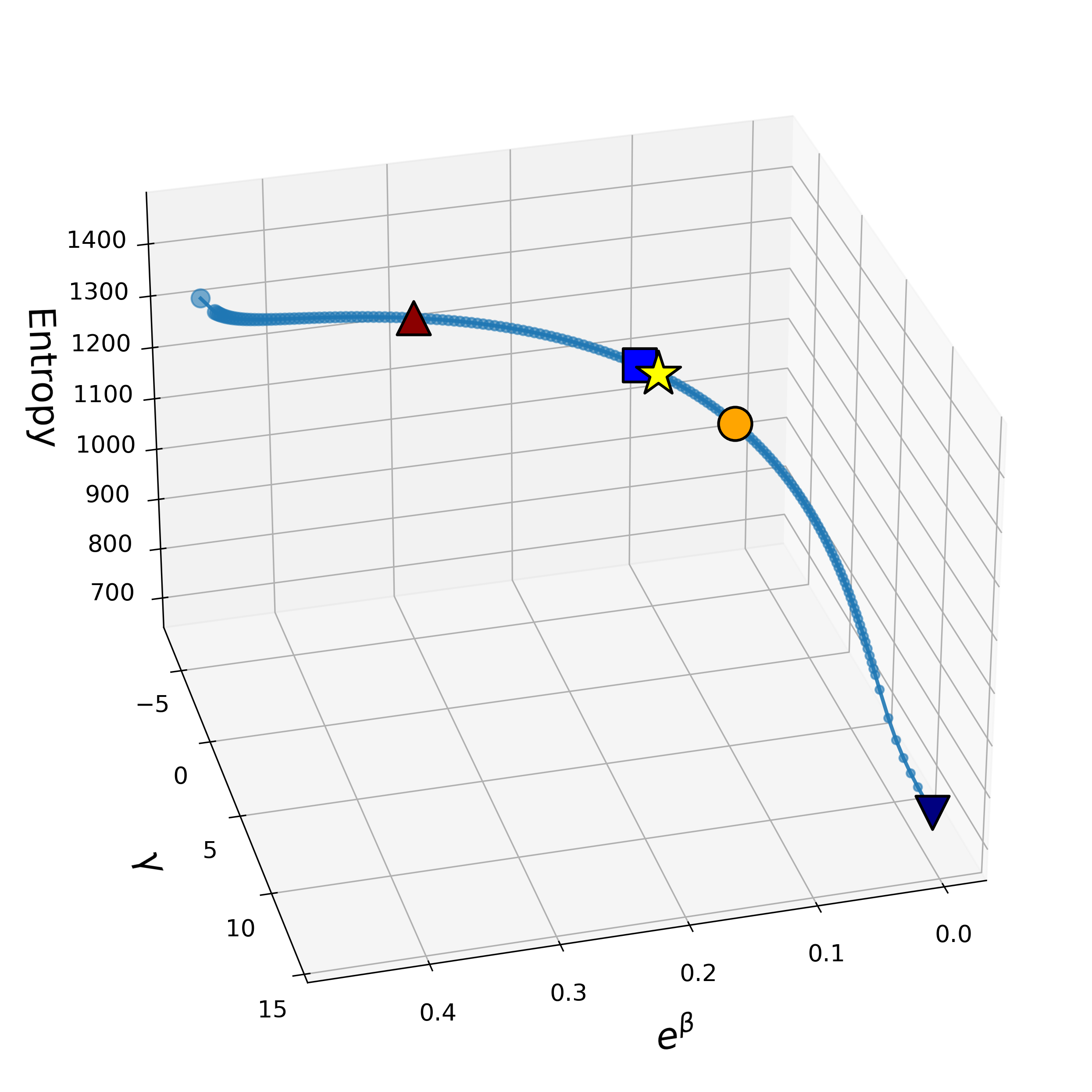}
    \vspace{0.12em}
    {\footnotesize (b) Milk 2016}
\end{minipage}%
\hspace{\panelgap}%
\begin{minipage}[t]{\panelw}
    \centering
    \includegraphics[width=\linewidth]{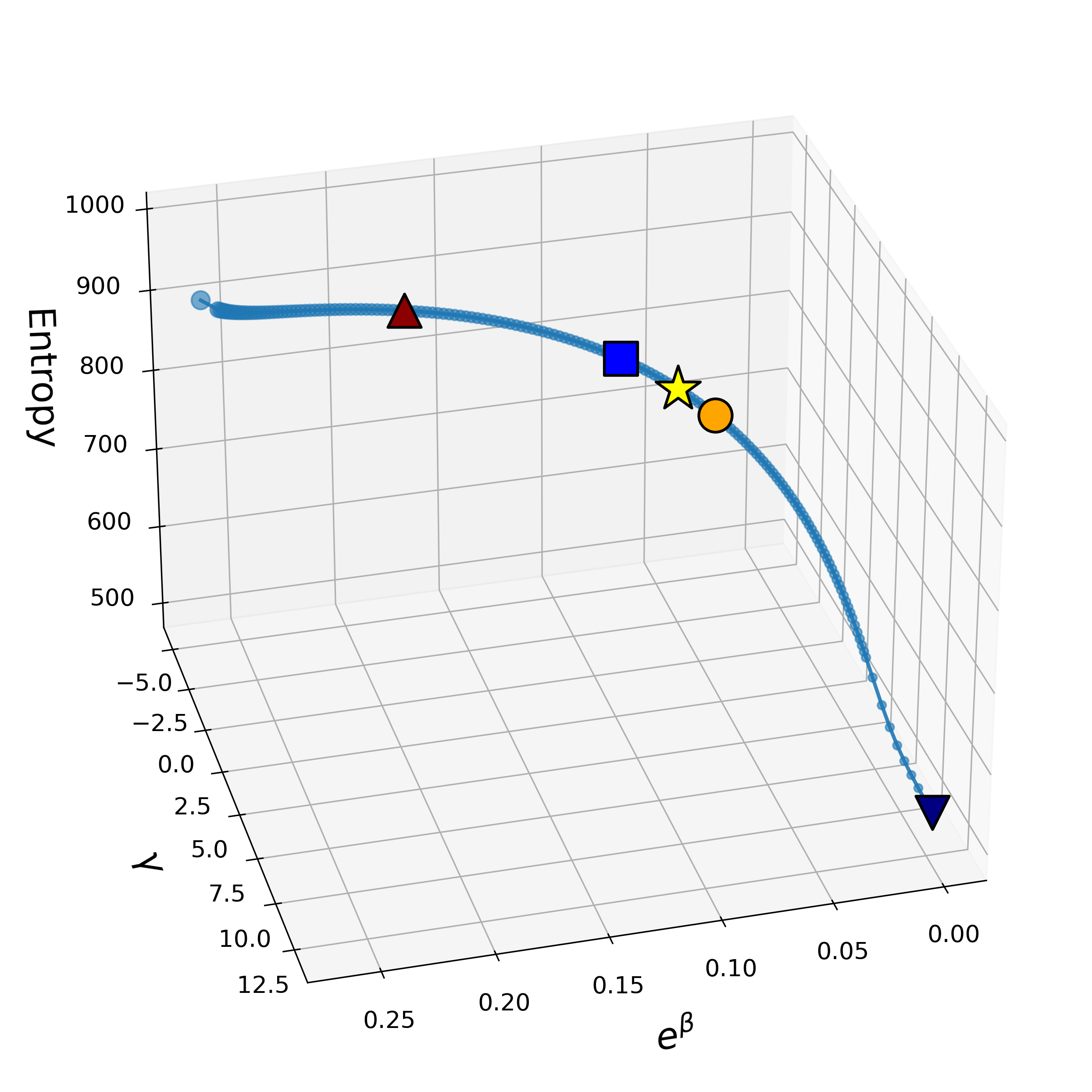}
    \vspace{0.12em}
    {\footnotesize (c) Steel 2017}
\end{minipage}

\vspace{0.35em}

\begin{minipage}[t]{\panelw}
    \centering
    \includegraphics[width=\linewidth]{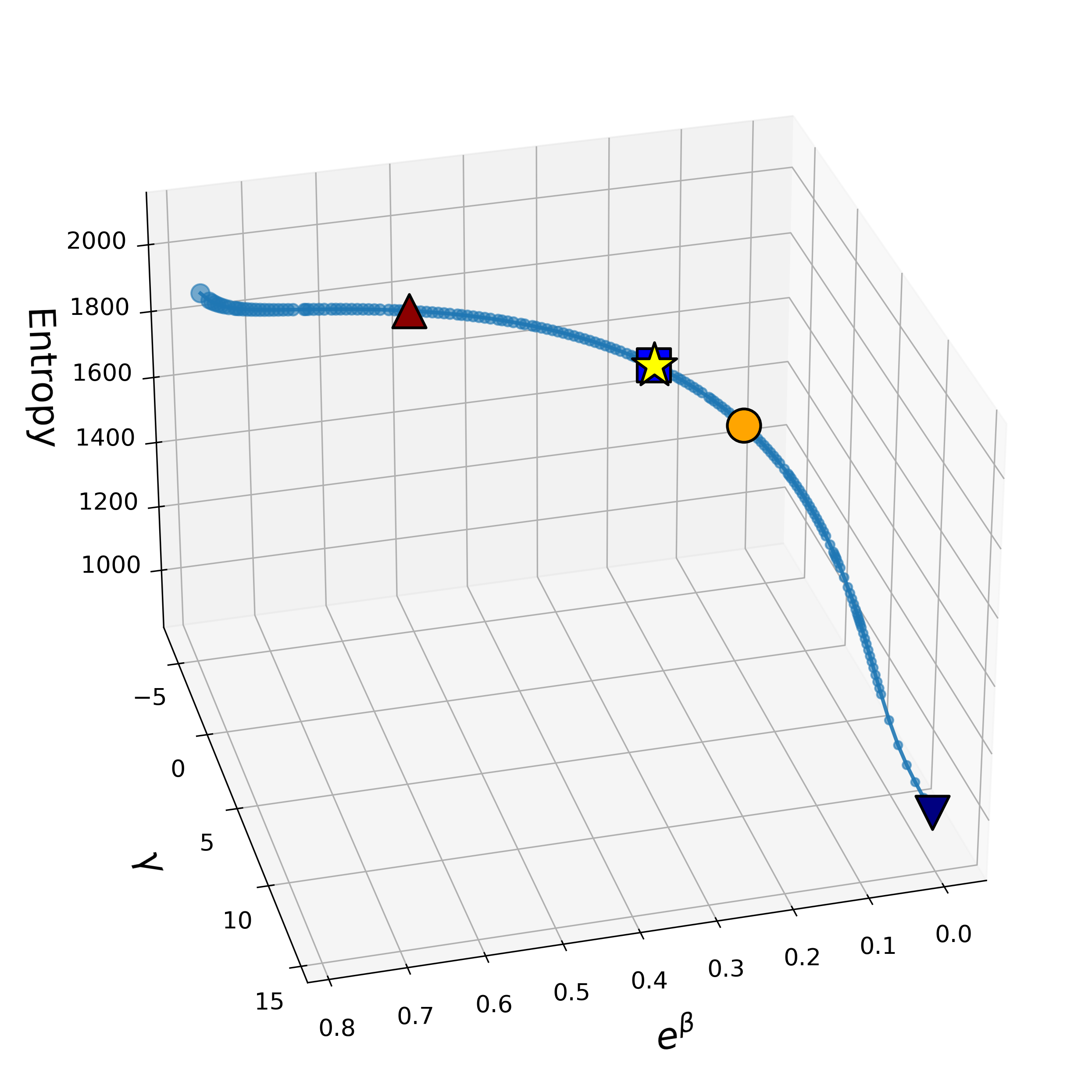}
    \vspace{0.12em}
    {\footnotesize (d) Oils 2018}
\end{minipage}%
\hspace{\panelgap}%
\begin{minipage}[t]{\panelw}
    \centering
    \includegraphics[width=\linewidth]{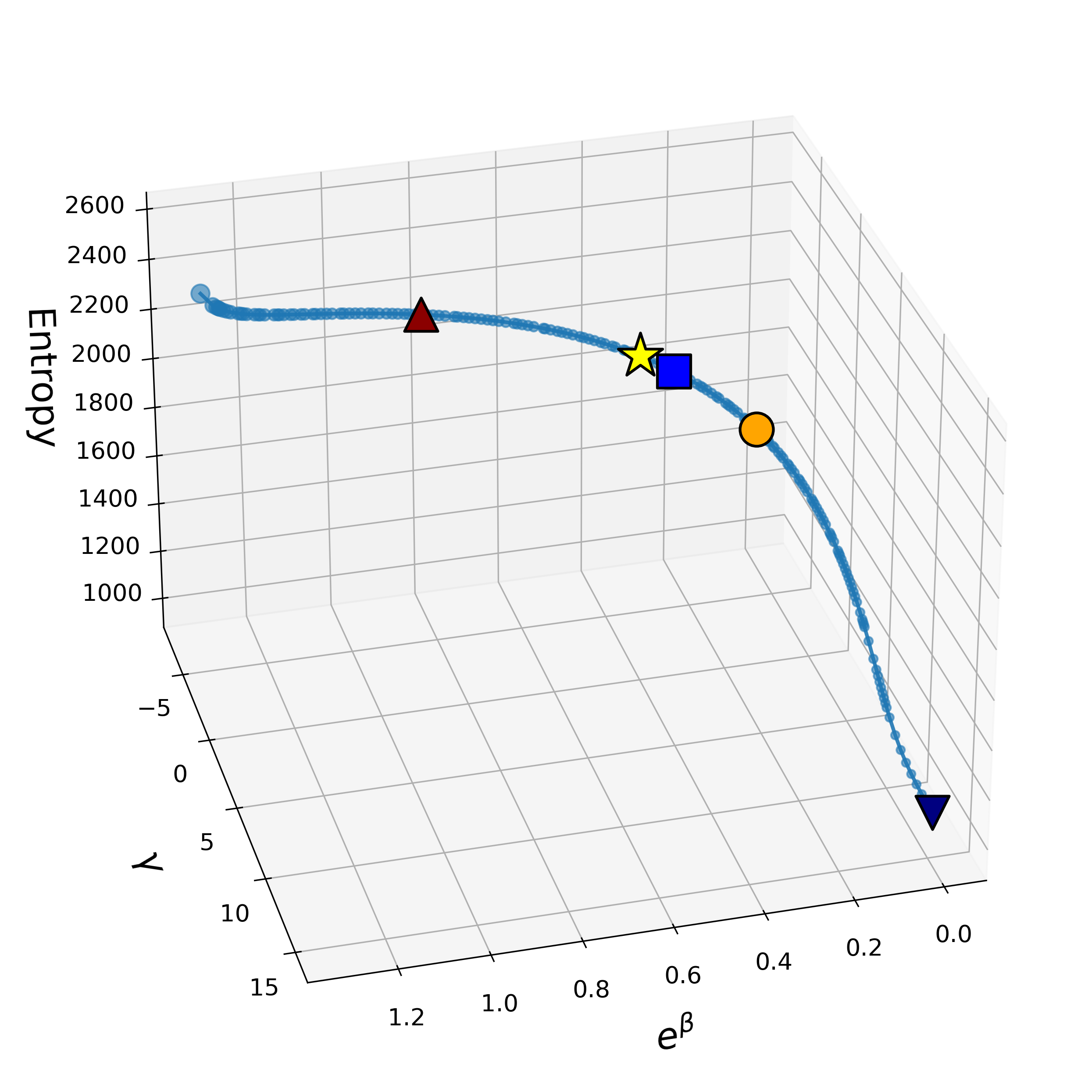}
    \vspace{0.12em}
    {\footnotesize (e) Cocoa 2019}
\end{minipage}%
\hspace{\panelgap}%
\begin{minipage}[t]{\panelw}
    \centering
    \includegraphics[width=\linewidth]{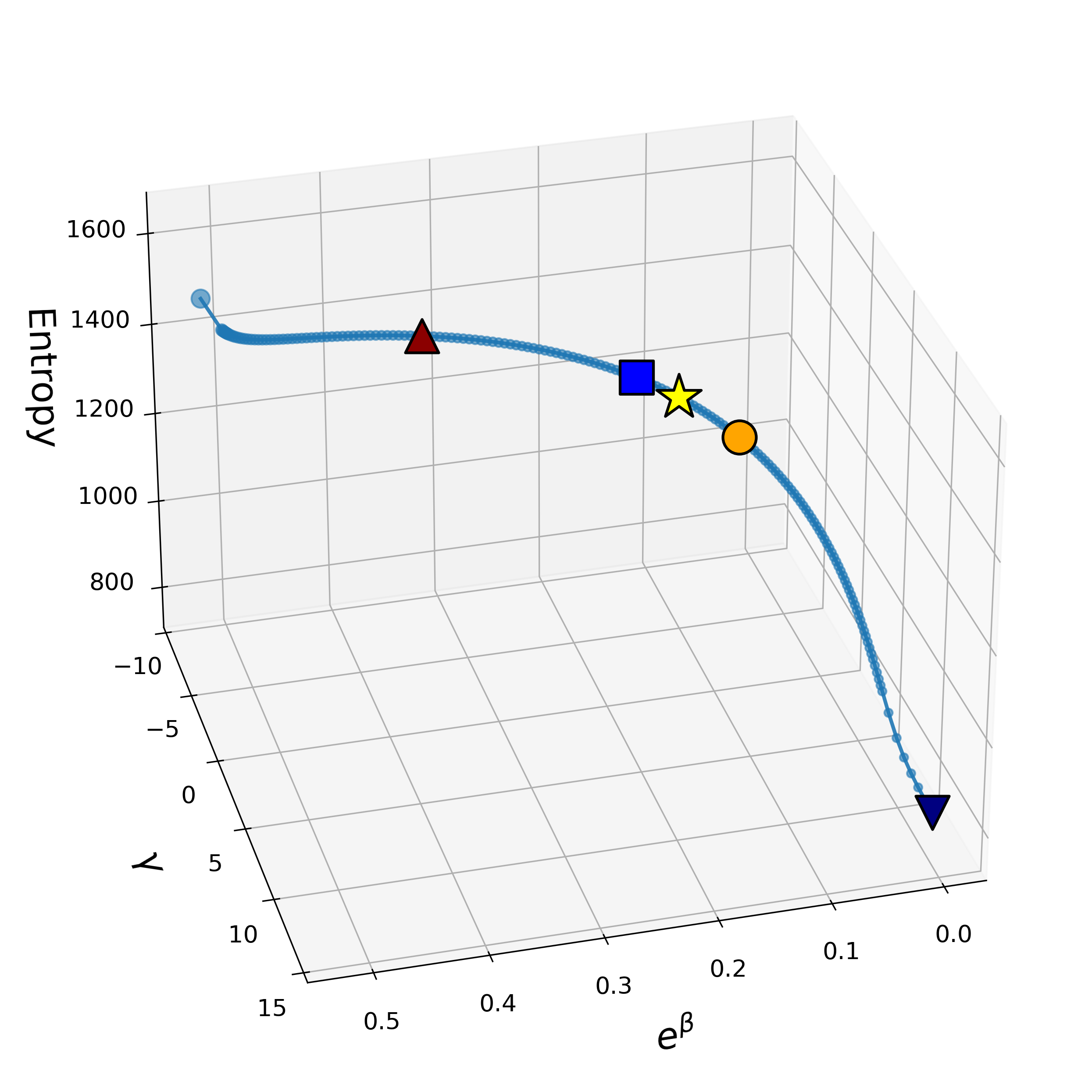}
    \vspace{0.12em}
    {\footnotesize (f) Wood 2020}
\end{minipage}

\vspace{0.35em}

\begin{minipage}[t]{\panelw}
    \centering
    \includegraphics[width=\linewidth]{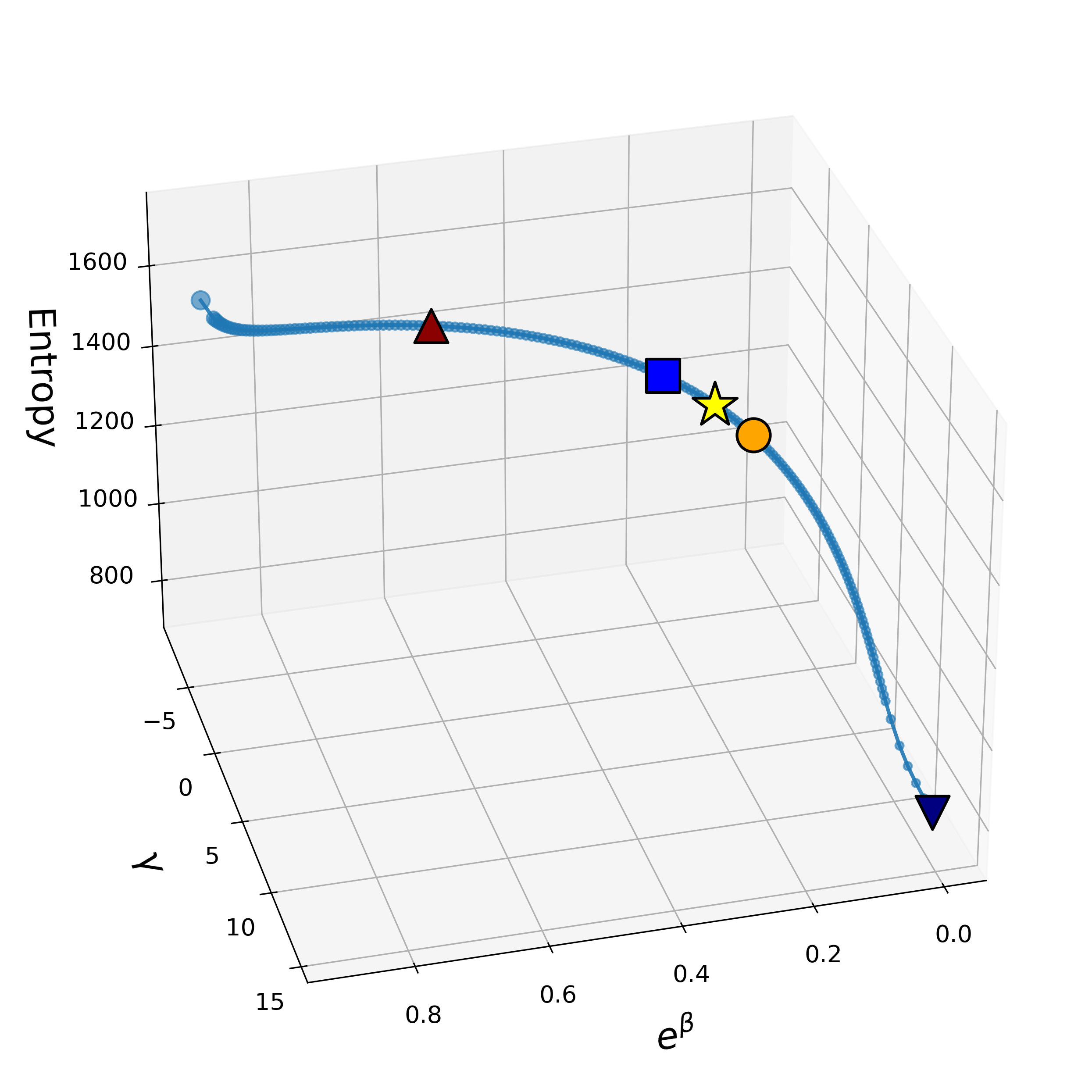}
    \vspace{0.12em}
    {\footnotesize (g) Plums 2021}
\end{minipage}%
\hspace{\panelgap}%
\begin{minipage}[t]{\panelw}
    \centering
    \includegraphics[width=\linewidth]{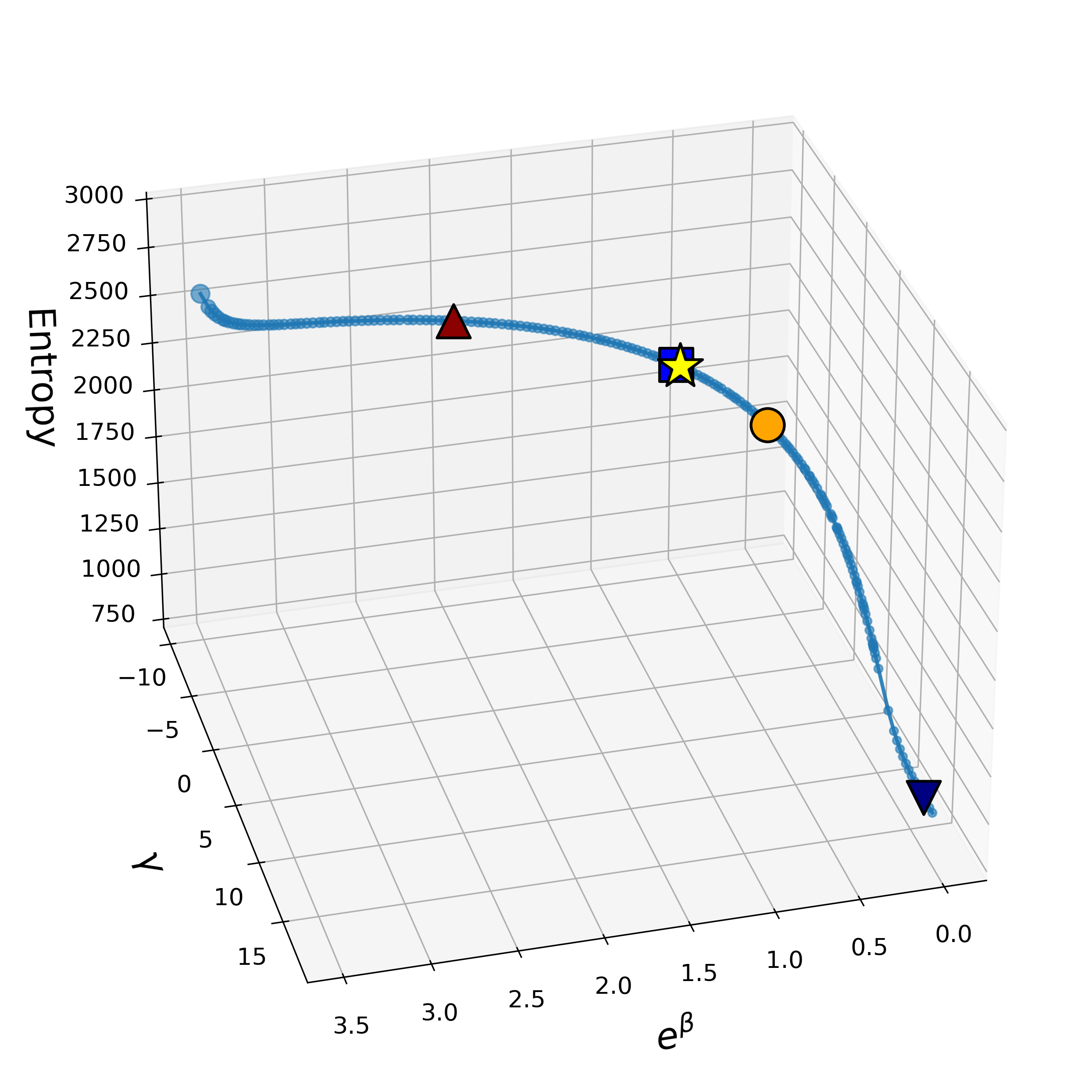}
    \vspace{0.12em}
    {\footnotesize (h) Refrigerators 2022}
\end{minipage}%
\hspace{\panelgap}%
\begin{minipage}[t]{\panelw}
    \centering
    \includegraphics[width=\linewidth]{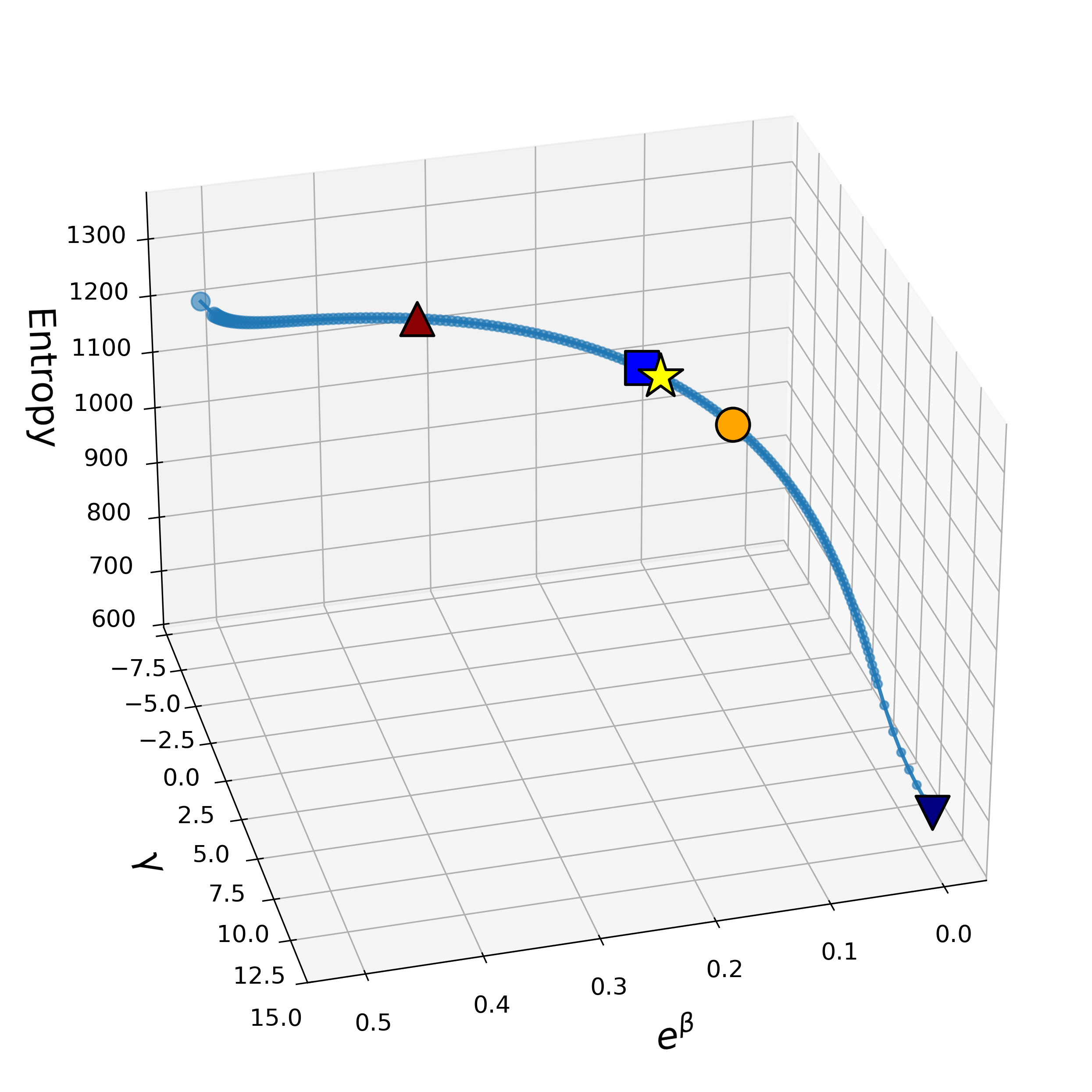}
    \vspace{0.12em}
    {\footnotesize (i) Fabric 2023}
\end{minipage}


\caption{Empirical trade networks: Shannon entropy along the total-link feasible curve for each of the nine observed networks, shown in three dimensions with coordinates $e^{\beta}$, $\gamma$, and entropy. The full-information FCBM point (\TrueParamSym) uses the observed within- and between-region link counts; the minimum (\MinEntropySym), maximum (\MaxEntropySym), mean (\MeanEntropySym), and median (\MedianEntropySym) entropy representatives use only the observed total link count. Panels (a)--(i) follow the empirical datasets in Fig.~\ref{fig:trade_networks}.}

\label{fig:Jeffreys}

\end{figure*}

\begin{figure*}[p]
\centering

\newcommand{\panelw}{0.327\textwidth}
\newcommand{\panelgap}{0.006\textwidth}

\begin{minipage}[t]{\panelw}
    \centering
    \includegraphics[width=\linewidth]{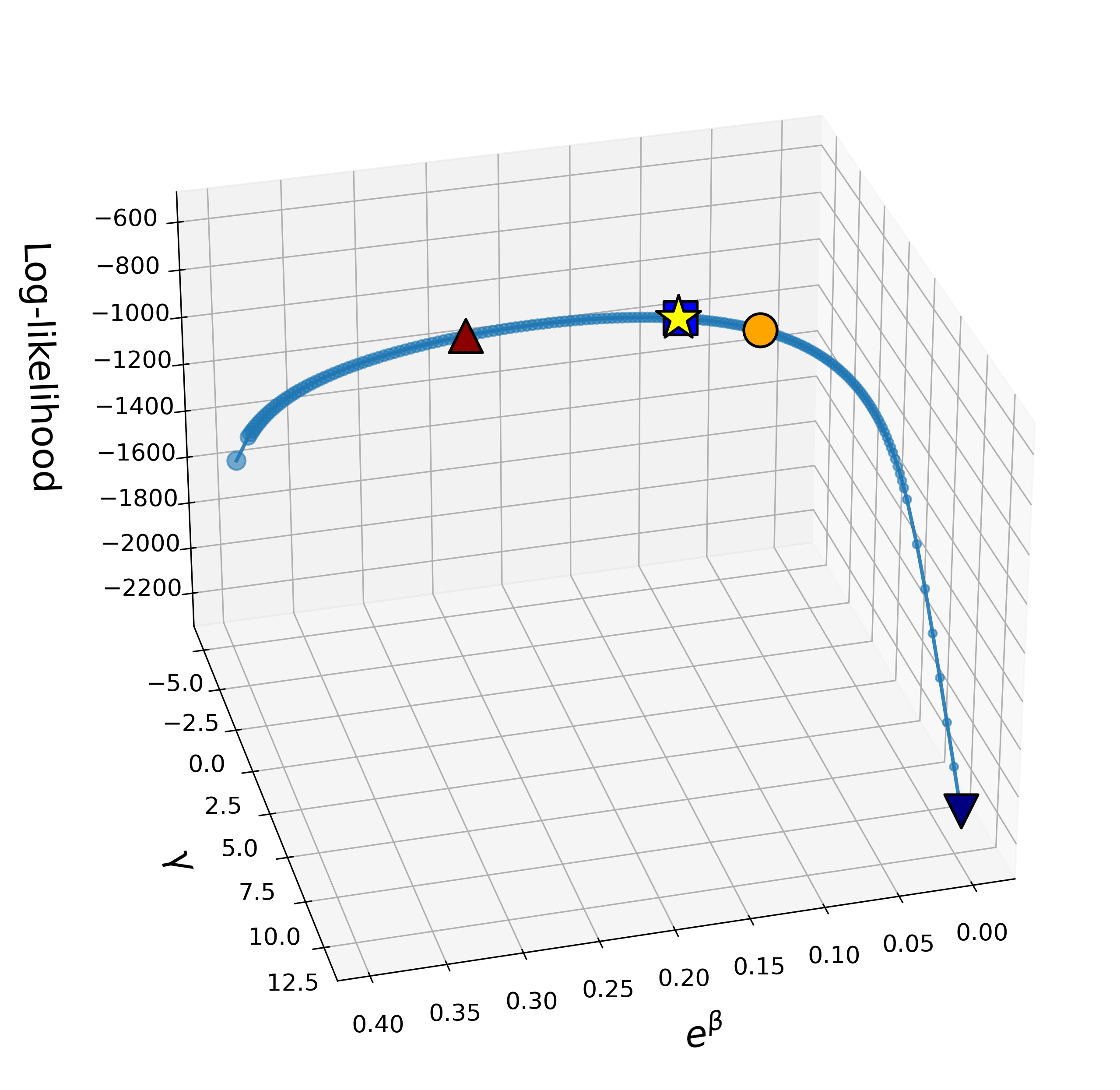}
    \vspace{0.12em}
    {\footnotesize (a) Automotive 2016}
\end{minipage}%
\hspace{\panelgap}%
\begin{minipage}[t]{\panelw}
    \centering
    \includegraphics[width=\linewidth]{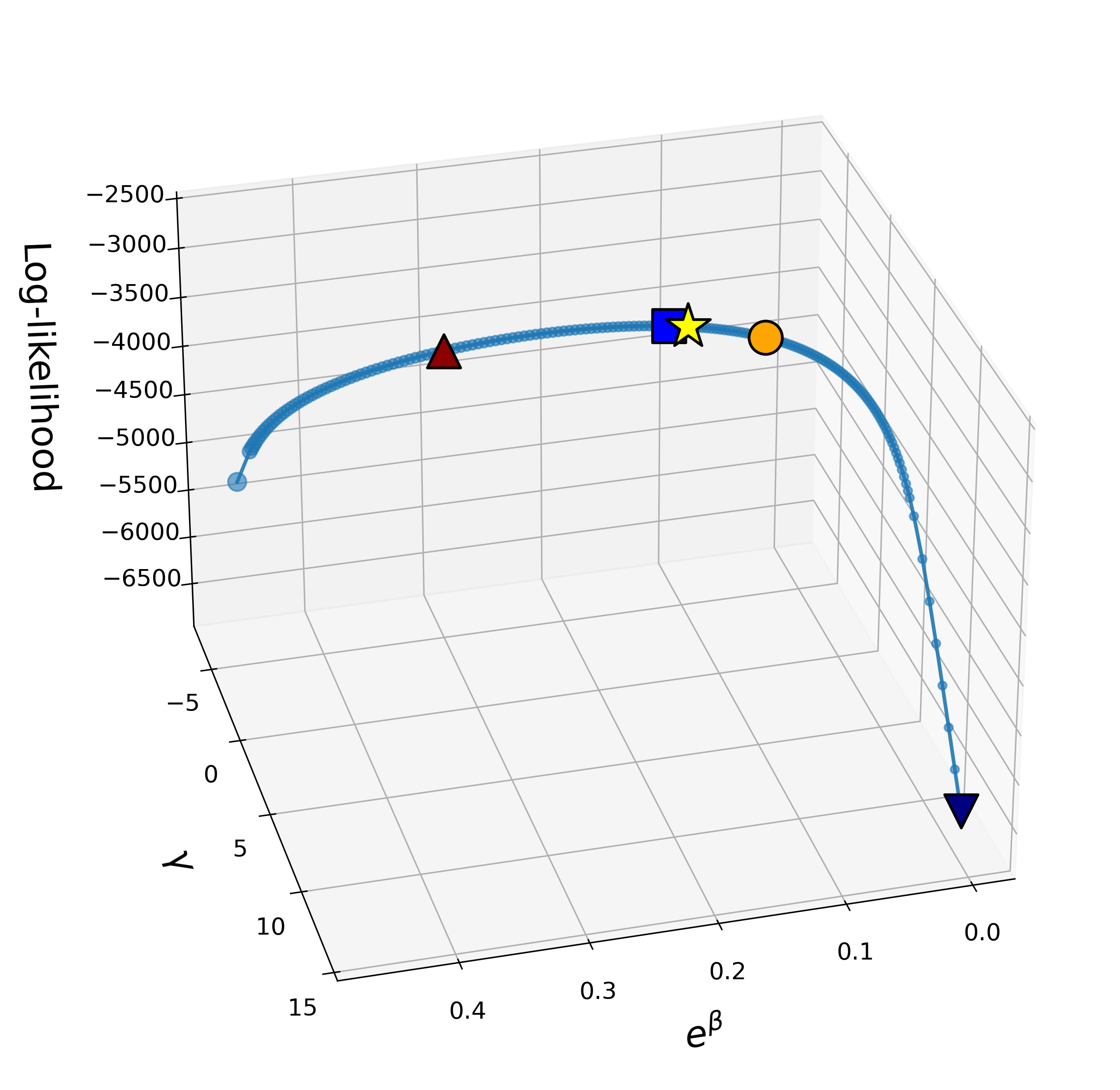}
    \vspace{0.12em}
    {\footnotesize (b) Milk 2016}
\end{minipage}%
\hspace{\panelgap}%
\begin{minipage}[t]{\panelw}
    \centering
    \includegraphics[width=\linewidth]{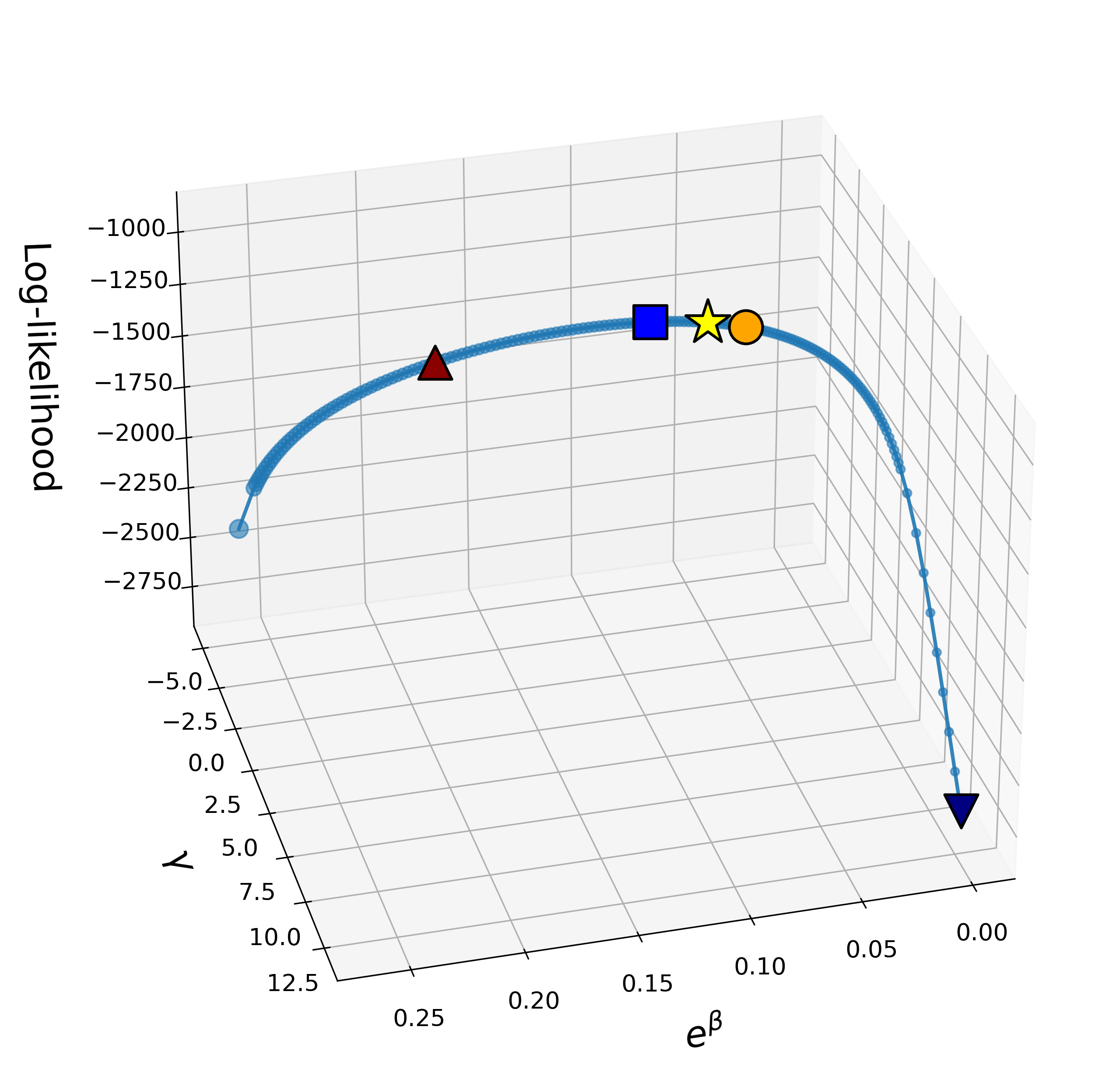}
    \vspace{0.12em}
    {\footnotesize (c) Steel 2017}
\end{minipage}

\vspace{0.35em}

\begin{minipage}[t]{\panelw}
    \centering
    \includegraphics[width=\linewidth]{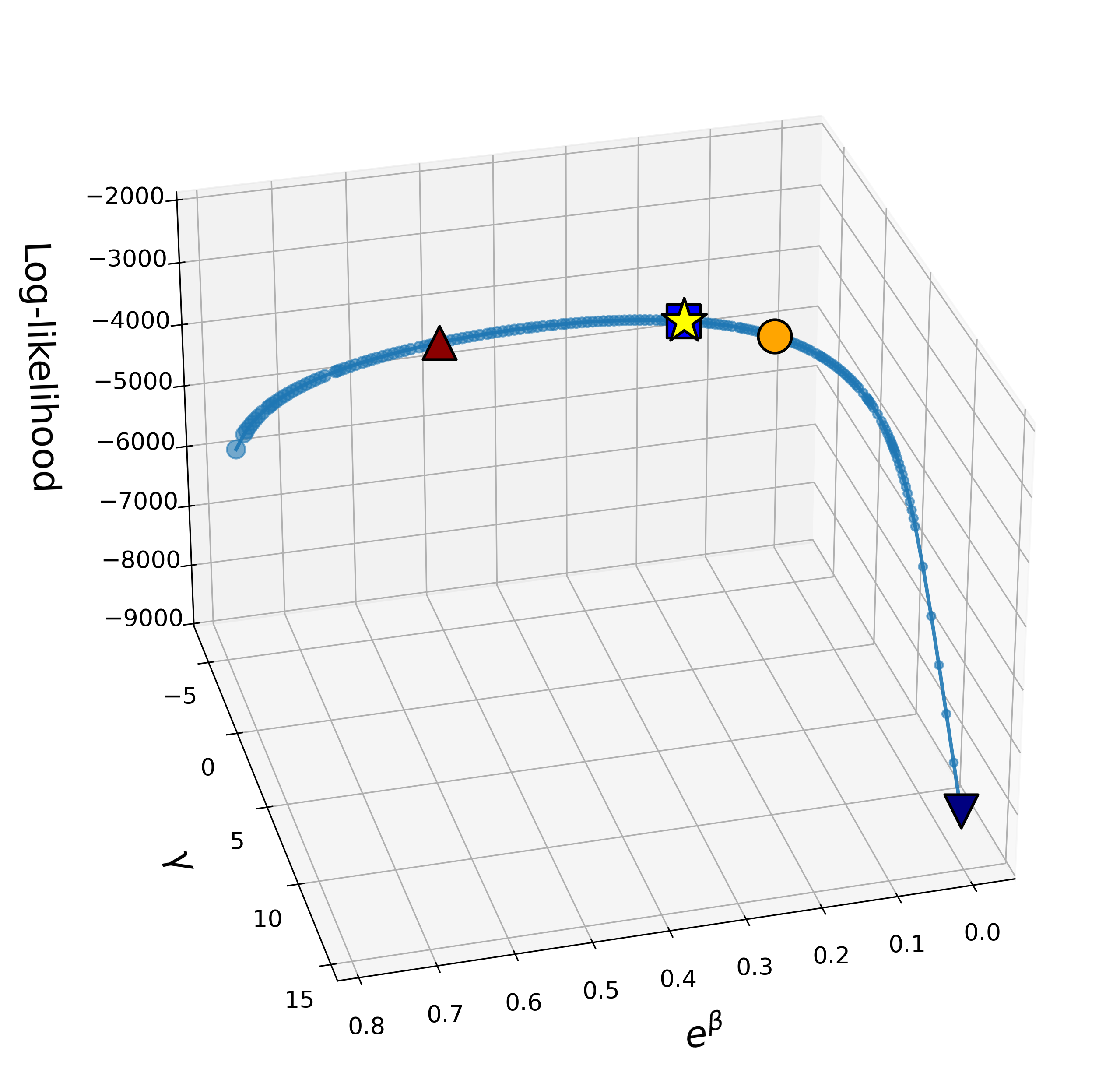}
    \vspace{0.12em}
    {\footnotesize (d) Oils 2018}
\end{minipage}%
\hspace{\panelgap}%
\begin{minipage}[t]{\panelw}
    \centering
    \includegraphics[width=\linewidth]{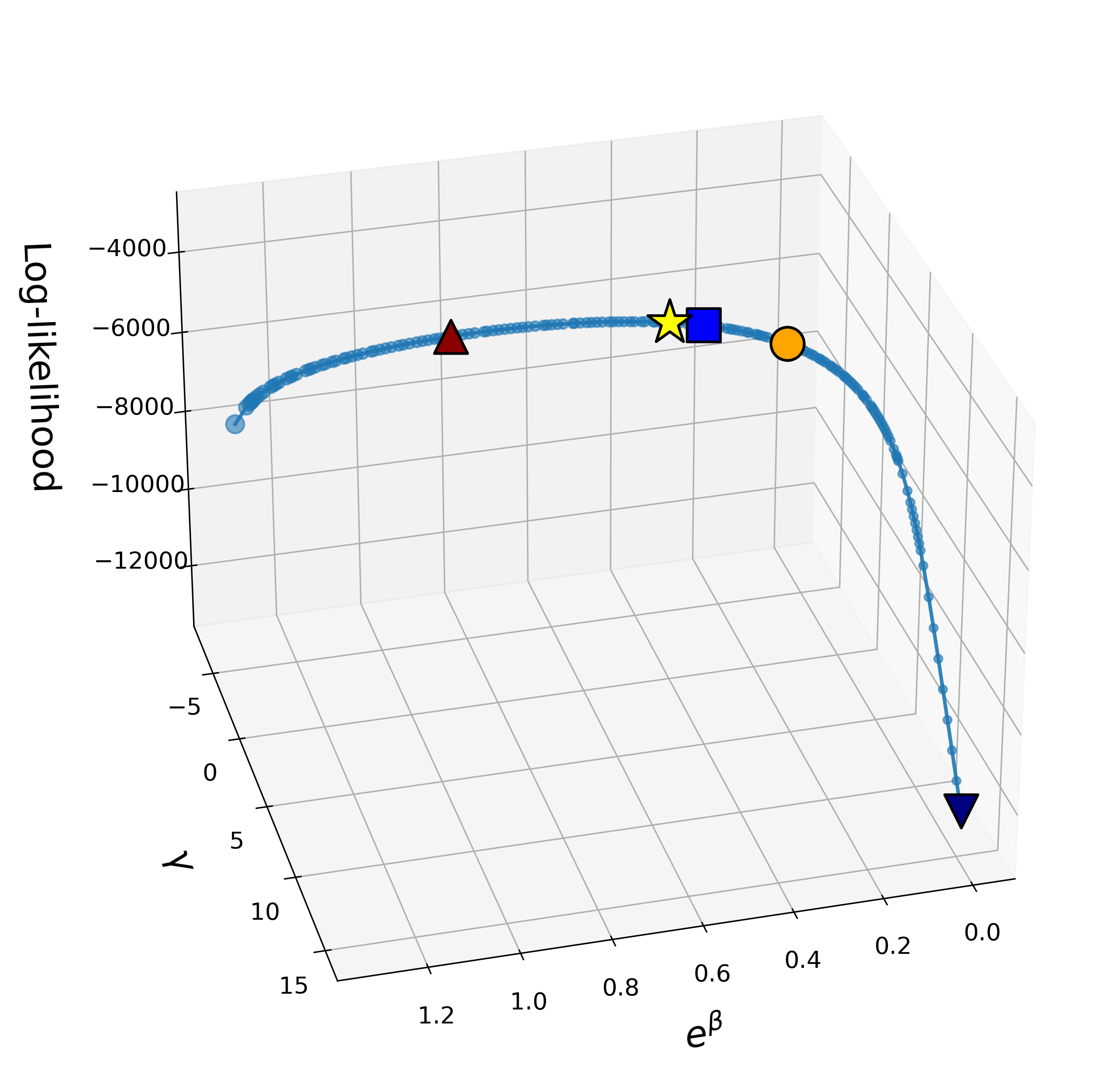}
    \vspace{0.12em}
    {\footnotesize (e) Cocoa 2019}
\end{minipage}%
\hspace{\panelgap}%
\begin{minipage}[t]{\panelw}
    \centering
    \includegraphics[width=\linewidth]{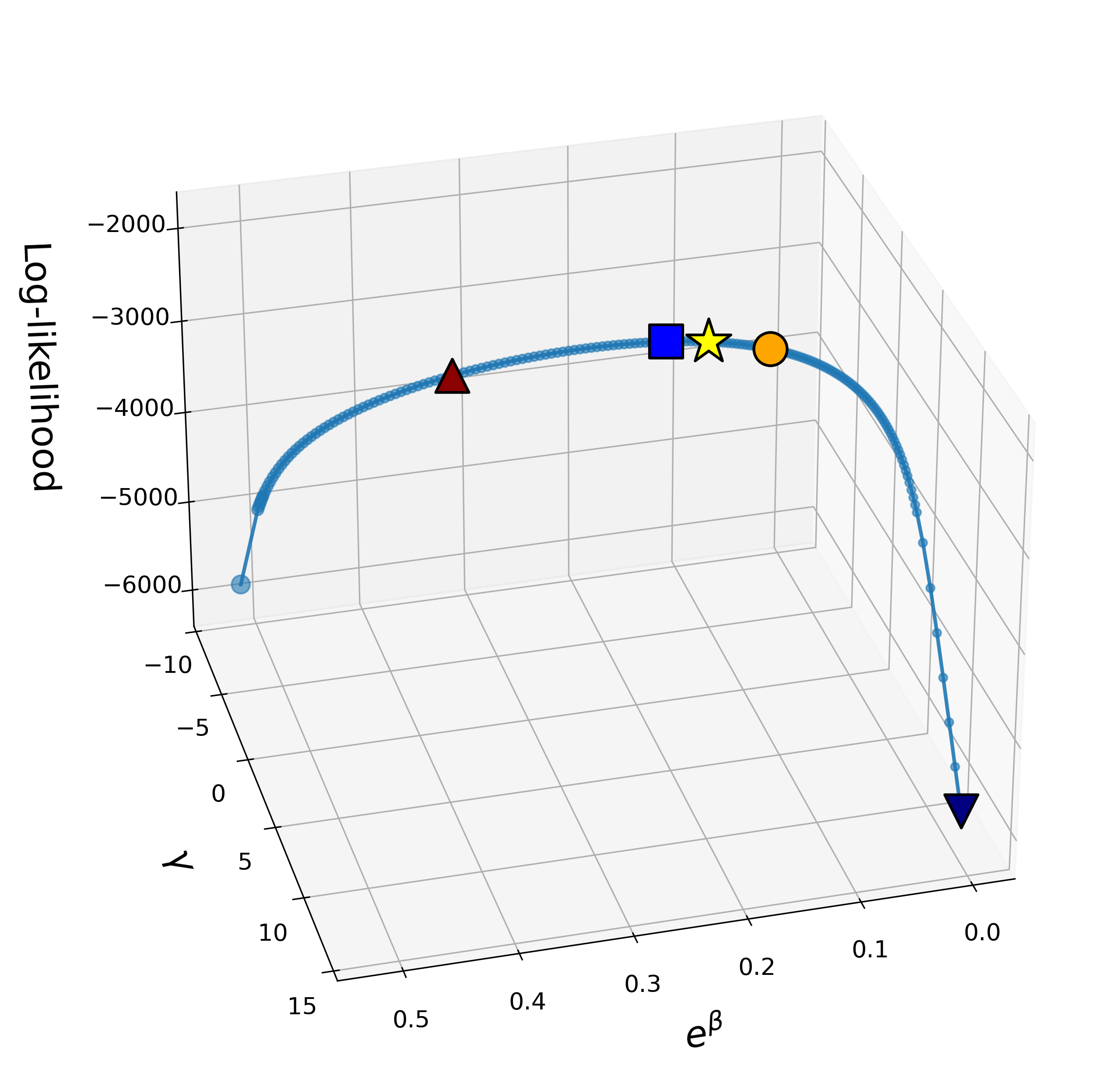}
    \vspace{0.12em}
    {\footnotesize (f) Wood 2020}
\end{minipage}

\vspace{0.35em}

\begin{minipage}[t]{\panelw}
    \centering
    \includegraphics[width=\linewidth]{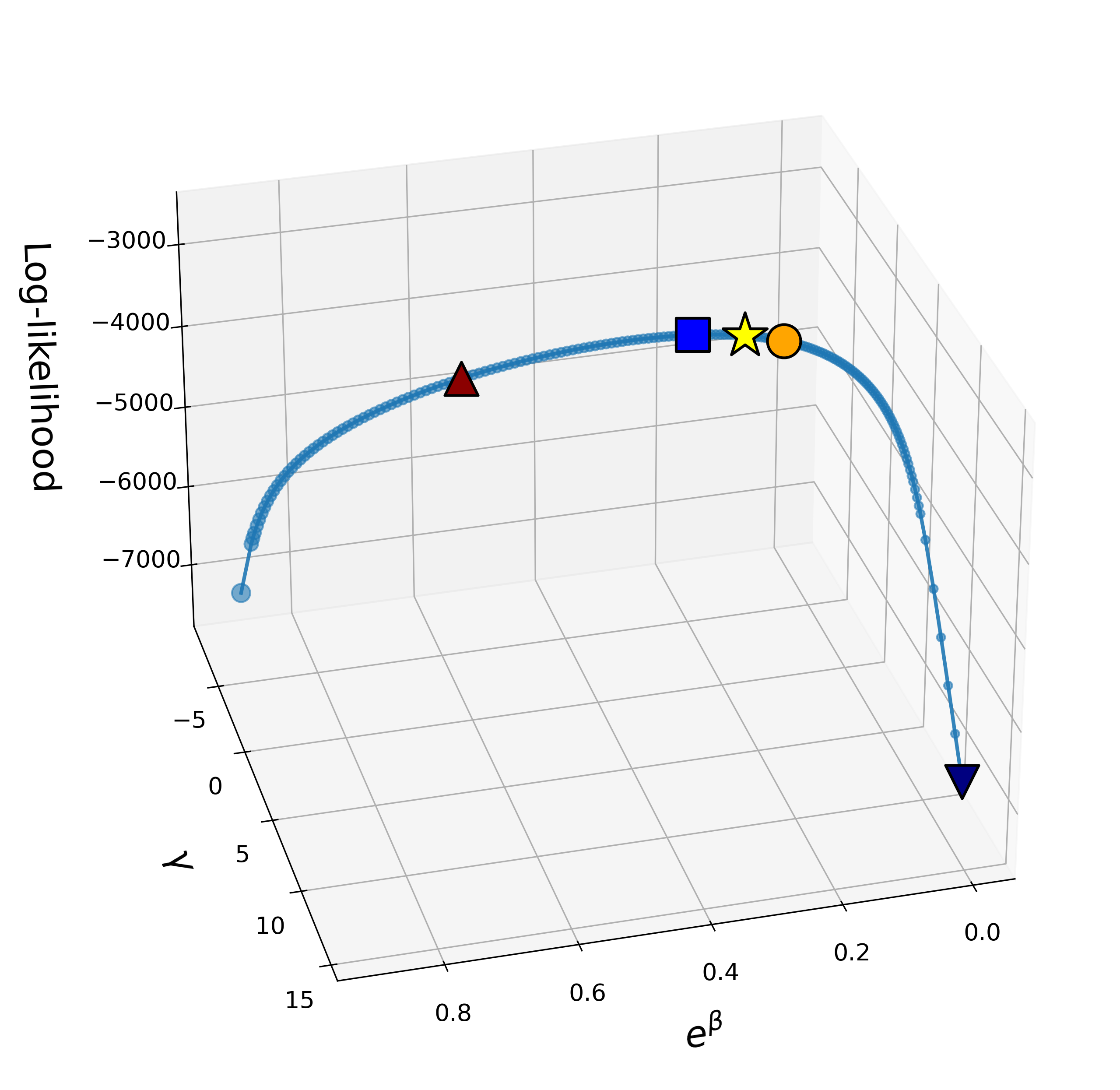}
    \vspace{0.12em}
    {\footnotesize (g) Plums 2021}
\end{minipage}%
\hspace{\panelgap}%
\begin{minipage}[t]{\panelw}
    \centering
    \includegraphics[width=\linewidth]{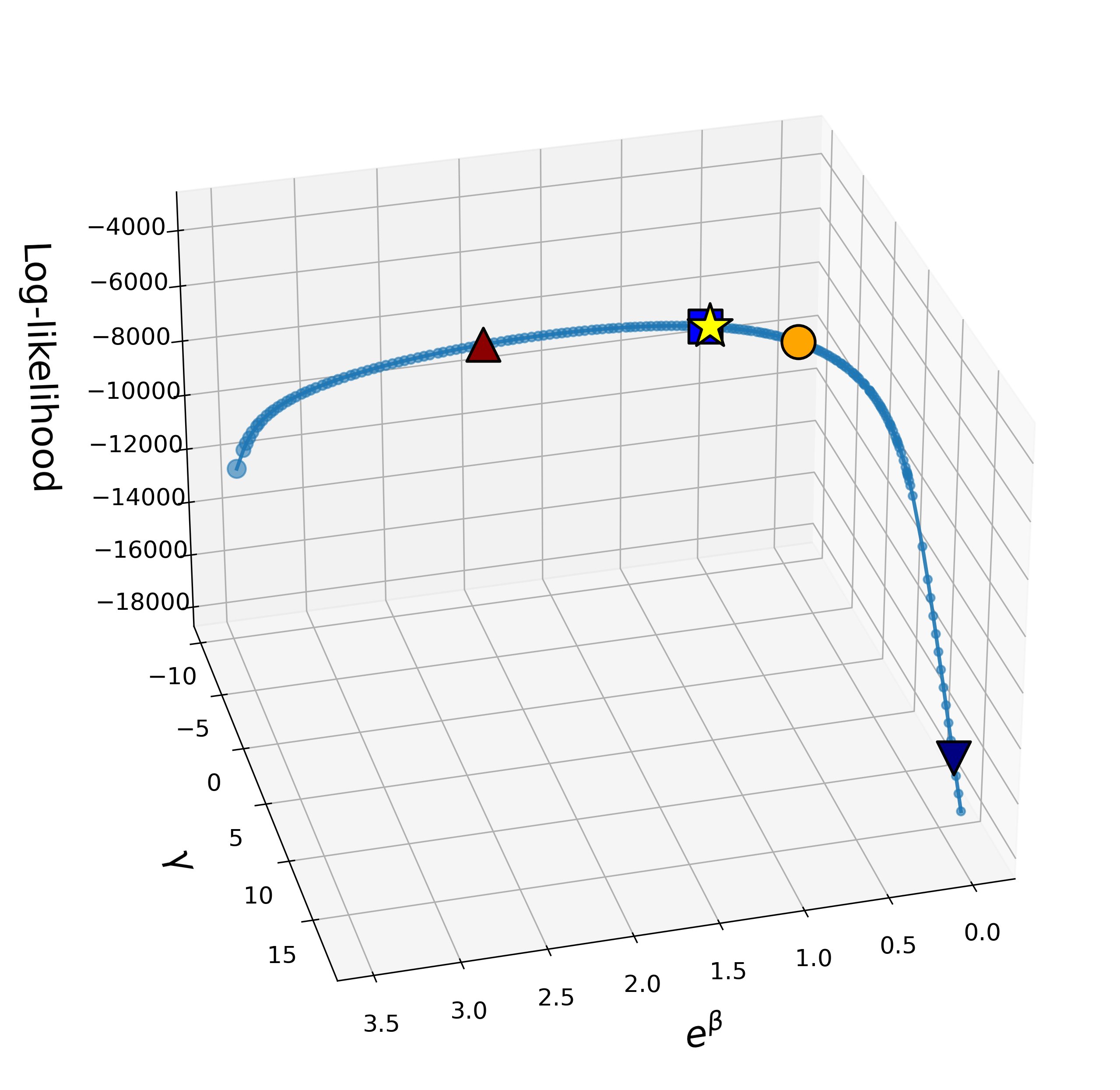}
    \vspace{0.12em}
    {\footnotesize (h) Refrigerators 2022}
\end{minipage}%
\hspace{\panelgap}%
\begin{minipage}[t]{\panelw}
    \centering
    \includegraphics[width=\linewidth]{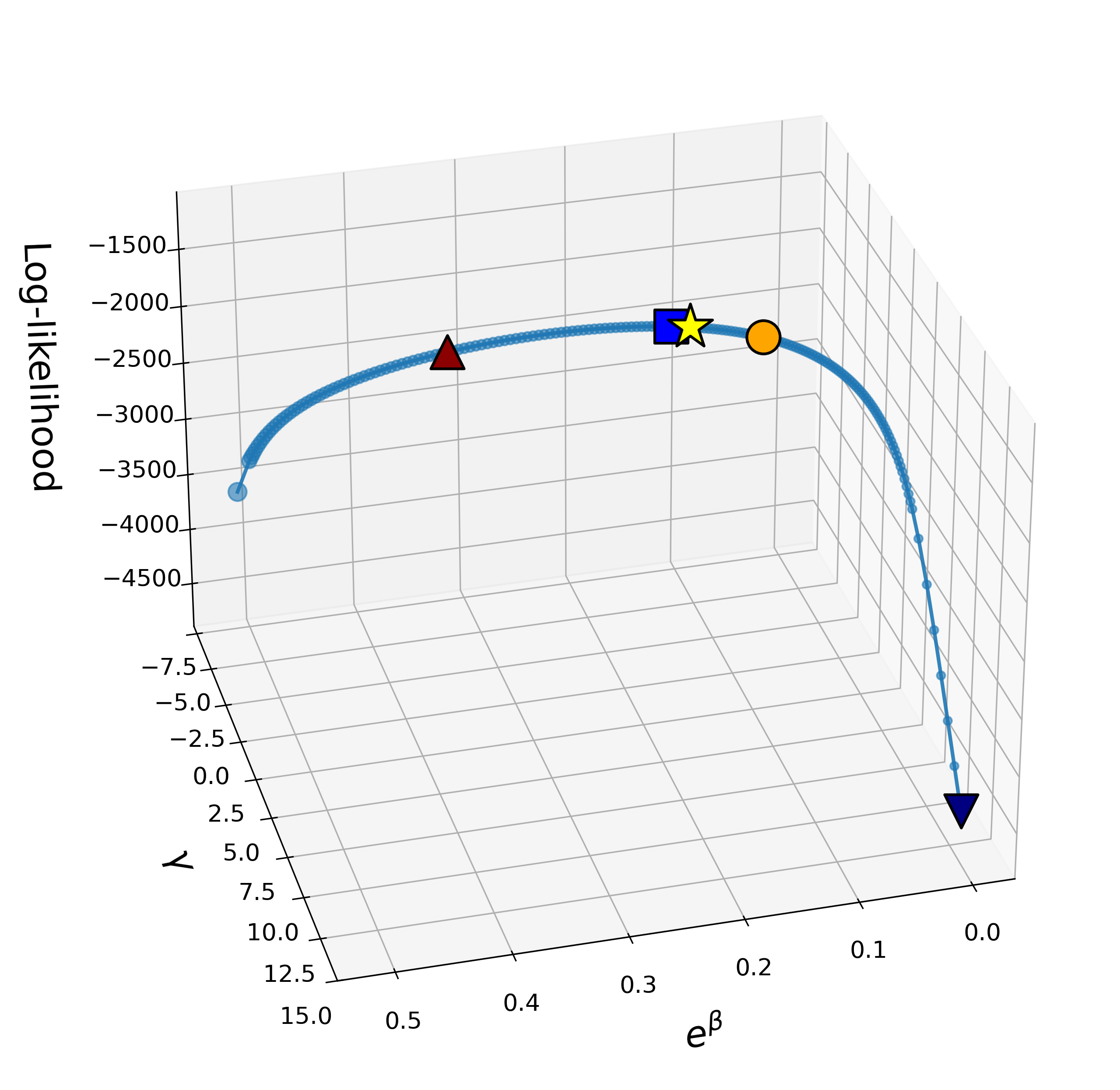}
    \vspace{0.12em}
    {\footnotesize (i) Fabric 2023}
\end{minipage}


\caption{Empirical trade networks: observed-network log-likelihood along the total-link feasible curve for each of the nine observed networks, shown in three dimensions with coordinates $e^{\beta}$, $\gamma$, and log-likelihood. The full-information FCBM point (\TrueParamSym), the likelihood maximum, uses the observed within- and between-region link counts; the minimum (\MinEntropySym), maximum (\MaxEntropySym), mean (\MeanEntropySym), and median (\MedianEntropySym) entropy representatives use only the observed total link count. Panels (a)--(i) follow the empirical datasets in Fig.~\ref{fig:trade_networks}. The likelihood gaps between these points are shown in Fig.~\ref{fig:Jeffreys_loglikelihood}.}

\label{fig:Jeffreys_loglikelihood_3d}

\end{figure*}

\begin{figure*}[tp]
\centering
\includegraphics[width=0.98\textwidth]{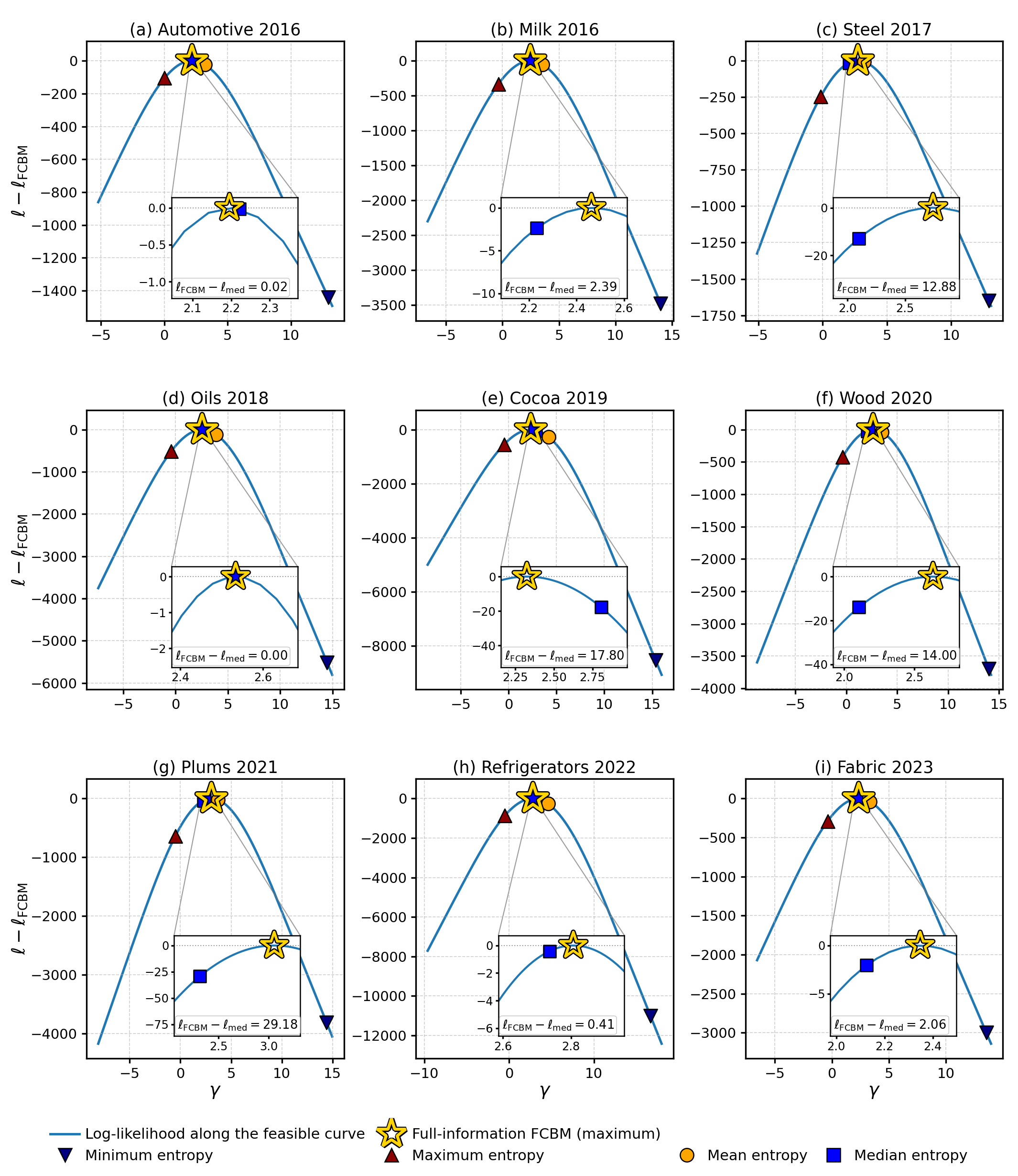}
\caption{Empirical trade networks (observed, not synthetically generated): full-information log-likelihood profiles for the nine datasets of Fig.~\ref{fig:trade_networks}, evaluated along each network's total-link feasible curve as a function of $\gamma$. This is a two-dimensional view of the curves in Fig.~\ref{fig:Jeffreys_loglikelihood_3d}.
The vertical axis shows $\ell(\beta(\gamma),\gamma)-\ell(\hat\beta_{\mathrm{FCBM}},\hat\gamma_{\mathrm{FCBM}})$, so the full-information FCBM maximum is zero; the difference below zero at the median point is its log-likelihood gap. Each inset enlarges the region around the maximum and gives this gap, $\ell_{\mathrm{FCBM}}-\ell_{\mathrm{med}}$.
The full-information FCBM maximum (\FCBMOpenSym) is fitted using both within- and between-region counts. The minimum (\MinEntropySym), maximum (\MaxEntropySym), mean (\MeanEntropySym), and median (\MedianEntropySym) entropy representatives use only the total link count. The proximity of the median point to the likelihood maximum varies by dataset.}
\label{fig:Jeffreys_loglikelihood}
\end{figure*}

Figure~\ref{fig:Jeffreys} shows the entropy along the feasible family for the
nine empirical networks, together with the full-information reference point
and the four entropy-based representatives. Figure~\ref{fig:Jeffreys_loglikelihood_3d}
shows the observed empirical network log-likelihood along the same total-link
feasible family, in the same three-dimensional coordinates.
Because likelihood differences are difficult to read in three dimensions,
Fig.~\ref{fig:Jeffreys_loglikelihood} shows this log-likelihood profile as a
function of $\gamma$. There, the plotted quantity is the log-likelihood relative to the
full-information FCBM maximum, so the vertical difference from zero
indicates the likelihood gap.
The full-information FCBM estimate is the likelihood maximum,
whereas the median-entropy representative is selected without using
the intra-regional link count.
This comparison makes the location of the reduced-information selection
relative to the likelihood maximum directly visible. The remaining
entropy and log-likelihood projections are provided in
Appendix~\ref{app:2d-projections}; they are alternative coordinate views
of the same feasible solutions, not independent reconstruction tests.

\begin{figure*}[t]
\centering
\includegraphics[width=0.92\textwidth]{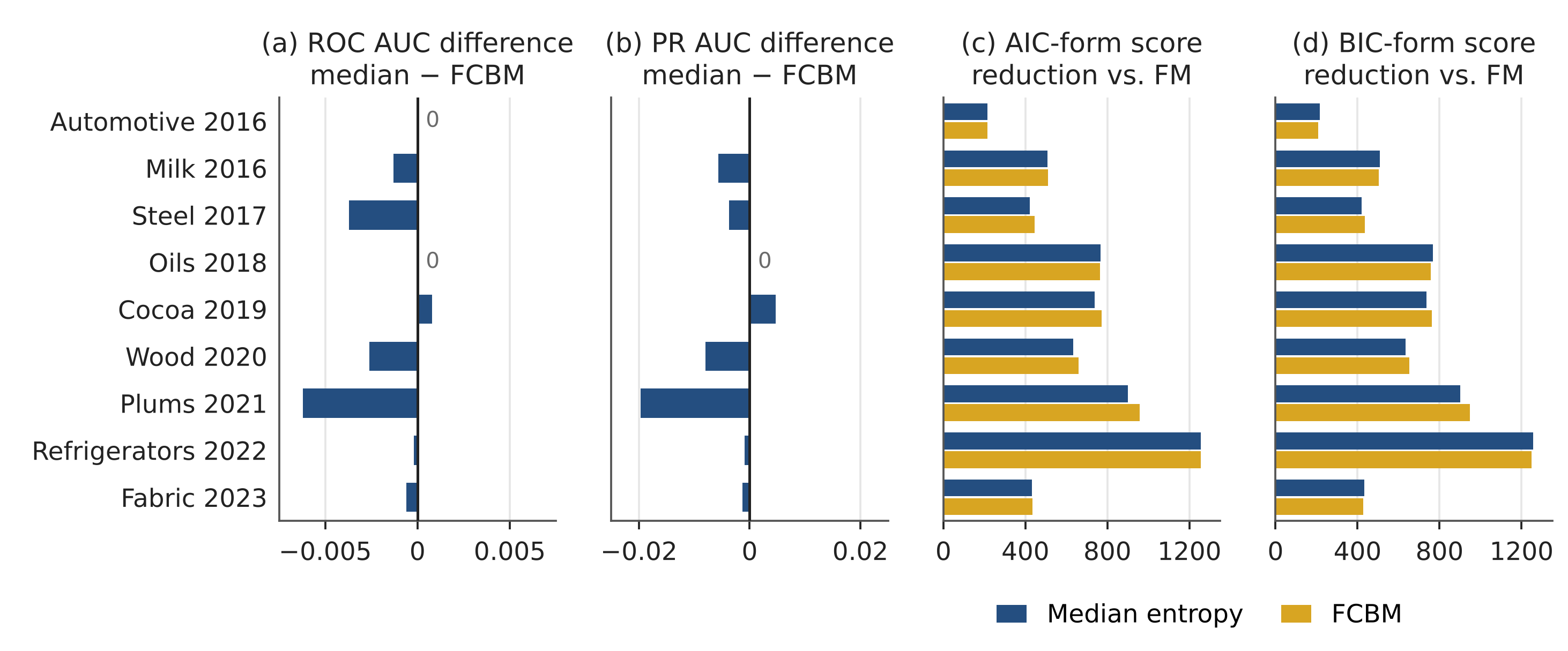}
\caption{Empirical trade networks: comparisons that respect the evaluation domain and descriptive likelihood-score summaries. Panels (a)--(b) show differences between median entropy and the full-information FCBM in ROC AUC and PR AUC (median minus FCBM), both evaluated on unordered pairs $i<j$. Panels (c)--(d) show reductions in AIC- and BIC-form scores relative to the block-agnostic FM, computed from undirected pair log-likelihoods. The historical FM ROC/PR AUC values in Appendix Table~\ref{tab:model_perf} were originally evaluated on all $n^2$ adjacency entries and therefore are \emph{not} subtracted from block-aware AUCs in this figure. Positive values in (c)--(d) indicate a smaller penalized score. Median-entropy AIC/BIC-form scores are descriptive rather than conventional information criteria, as explained in Sec.~II.D. FCBM also uses the observed within-region link count.}
\label{fig:model_perf}
\end{figure*}

The empirical curves exhibit the qualitative extremes discussed in
Sec.~\ref{sec:median-rationale}. The minimum-entropy point concentrates
link probabilities and, in these data, corresponds to a strong
within-region preference. The maximum-entropy point lies near
$\gamma=0$ and largely suppresses regional differentiation. The
median-entropy point lies between these extremes. Table~\ref{tab:beta_gamma_summary}
reports the regional-effect parameters for the reduced-information median
and full-information FCBM. Their agreement is particularly close for
Oils 2018 and Automotive 2016, whereas larger differences occur for several other networks.
In Figs.~\ref{fig:Jeffreys_loglikelihood_3d} and~\ref{fig:Jeffreys_loglikelihood}, the median is near the
full-information peak in several cases; nevertheless, the likelihood
gap varies and cannot be inferred reliably from marker separation alone.
The full-information FCBM maximizes the realized-network likelihood,
while the median-entropy rule does not claim to maximize it. Their
link-ranking performance is assessed separately in Fig.~\ref{fig:model_perf}.

\begin{table}[t]
\centering
\caption{Empirical trade networks: regional-effect parameter $\gamma$ from the full-information FCBM fit using separate observed within-/between-region link counts, and from the reduced-information median-entropy selection using only the observed total link count. Values are rounded to four decimals; the median was selected from a numerical Jeffreys-arclength discretization of the feasible curve.}
\label{tab:beta_gamma_summary}
\renewcommand{\arraystretch}{1.12}
\begin{tabular}{lcc}
\toprule
Network & $\gamma_{\mathrm{FCBM}}$ & $\gamma_{\mathrm{median}}$ \\
\midrule
Automotive 2016 & 2.1956 & 2.2220 \\
Milk 2016 & 2.4624 & 2.2316 \\
Steel 2017 & 2.7433 & 2.0980 \\
Oils 2018 & 2.5339 & 2.5283 \\
Cocoa 2019 & 2.3247 & 2.8057 \\
Wood 2020 & 2.6329 & 2.1054 \\
Plums 2021 & 3.0556 & 2.3143 \\
Refrigerators 2022 & 2.8069 & 2.7369 \\
Fabric 2023 & 2.3477 & 2.1254 \\
\bottomrule
\end{tabular}
\end{table}

Figure~\ref{fig:model_perf} compares the ROC AUC and PR AUC of the two block-aware methods directly, using the same unordered country pairs $i<j$, and also displays the undirected penalized-likelihood differences relative to FM. The exact values for all three models are listed in Appendix Table~\ref{tab:model_perf}. The historical FM AUCs in that table were evaluated on all $n^2$ matrix entries, including the diagonal, and are retained to preserve the originally submitted values. Consequently we do not interpret their numerical differences from the two block-aware AUCs as a controlled, same-sample performance gain. The median-entropy reconstruction has smaller AIC/BIC-form penalized log-likelihood scores than FM in all nine networks; this is a descriptive in-sample comparison and does not imply a matched-pair ROC/PR gain. 

The comparison between the median-entropy rule and the full-information FCBM requires more care. The latter uses the additional empirical within-region link count, whereas the proposed reduced-information method does not. Their ROC/PR values are nevertheless close in many datasets, and the median-entropy rule is marginally higher on Automotive 2016 (PR AUC) and Cocoa 2019 (ROC and PR AUC). These differences are descriptive in-sample reconstruction results and, by themselves, do not establish overfitting by the full-information model. Likewise, the AIC/BIC-form values for the median-entropy rule are not standard information criteria because that point is not a maximum-likelihood estimate; they are retained only as penalized-likelihood summaries, as explained after Eq.~\eqref{eq:aic-bic}.

Taken together, the empirical results support a deliberately limited interpretation of the median-entropy rule. The Jeffreys-induced measure gives a coordinate-invariant way to organize the unidentified feasible family, while the median entropy provides a central representative that performs well for the trade networks considered here. When additional block-level sufficient statistics are available, the full-information FCBM remains the statistically identified benchmark.

\subsection{Controlled Synthetic Validation}
\label{sec:synthetic-validation}
\begin{figure*}[p]
\centering
\includegraphics[width=0.96\textwidth]{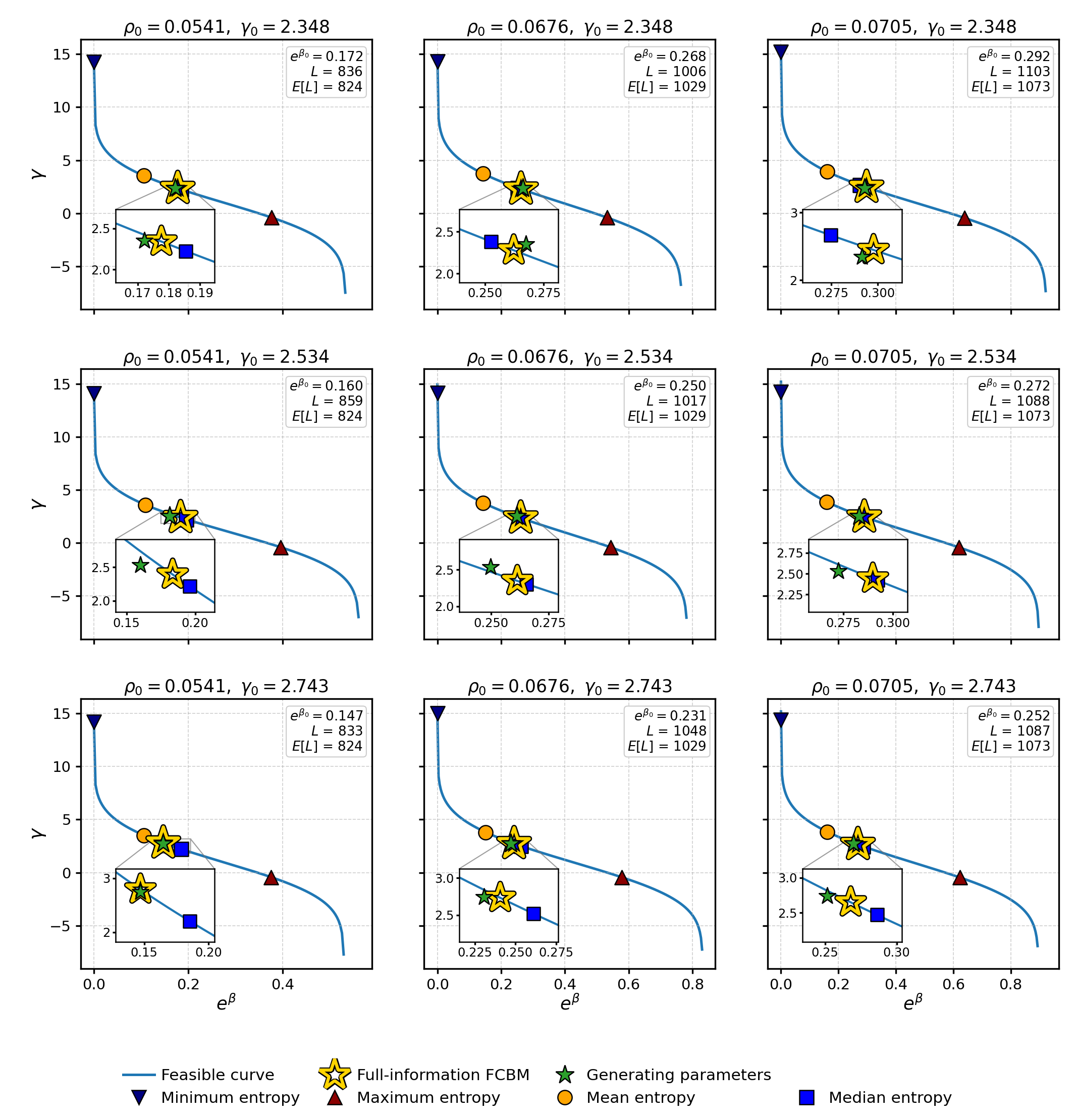}
\caption{Synthetic FCBM networks calibrated to empirical trade-network characteristics (not observed trade networks): one independently sampled network in each of nine generating regimes. Each panel shows the total-link feasible curve determined by that synthetic network's sampled link count $L$, plotted in $(e^{\beta},\gamma)$ coordinates. Rows correspond to generating regional effects $\gamma_0$ at the empirical 25th, 50th, and 75th percentiles, and columns to target densities $\rho_0$ at the corresponding empirical percentiles. The panel annotations give the generating baseline multiplier $e^{\beta_0}$, sampled $L$, and expected $\mathbb E[L]$. Green star: known generating parameters. Open yellow star: full-information FCBM MLE fitted separately to the sampled synthetic network. Downward triangle, upward triangle, circle, and square: minimum-, maximum-, mean-, and median-entropy representatives selected using only that sampled total link count. Insets: enlarged view of the region around the generating point, the FCBM fit and the median-entropy point. The generating point may lie off the sampled network's feasible curve because $L$ fluctuates around $\mathbb E[L]$. These same nine samples are evaluated in the synthetic likelihood profiles of Fig.~\ref{fig:synthetic-loglik}.}
\label{fig:synthetic-validation}
\end{figure*}

To examine reconstruction under known generating parameters, we sampled binary networks from the same fitness-corrected planted-partition model, using the GDP fitnesses and World Bank regional assignments of the BACI 2020 Wood dataset ($n=175$ countries) as fixed covariates; its empirical adjacency matrix is not used. We selected a $3\times3$ grid from the empirical 25th, 50th, and 75th percentiles of the full-information regional-effect parameter $\gamma$ ($\gamma_0=2.348$, $2.534$, $2.743$) and of the observed network density ($\rho_0=0.0541$, $0.0676$, $0.0705$) across the nine trade networks. For each target density and $\gamma_0$, the generating $\beta_0$ was fixed by the expected total-link constraint, $\binom{n}{2}^{-1}\sum_{i<j}p_{ij}(\beta_0,\gamma_0)=\rho_0$. In each cell we sampled one network for Fig.~\ref{fig:synthetic-validation} and 100 independent networks for the numerical summary. The full-information FCBM was refitted from the realized within- and between-region link counts of each network, whereas only the realized total number of links, together with node fitnesses and regional memberships, was supplied to the reduced-information Jeffreys procedure. Accuracy is measured against the generating link probabilities,
\begin{equation}
\label{eq:rmse-p}
\mathrm{RMSE}_P(\hat\theta)=\Biggl[\binom{n}{2}^{-1}\sum_{i<j}\bigl(p_{ij}(\hat\theta)-p_{ij}(\theta_0)\bigr)^{2}\Biggr]^{1/2}.
\end{equation}

Figure~\ref{fig:synthetic-validation} compares the known generating
parameters (green stars), the full-information FCBM fits of the realized
networks (open yellow stars), and the four entropy representatives on the
nine feasible curves. Clearly, the full-information estimate is not identical to the
generating point, because the realized adjacency matrix fluctuates around
its expectation. In particular, the green generating point
need not belong to the feasible curve conditioned on the realized total
link count. The reduced-information rules use this realized count but
not the realized within-region count.

\begin{figure*}[p]
\centering
\includegraphics[width=0.97\textwidth]{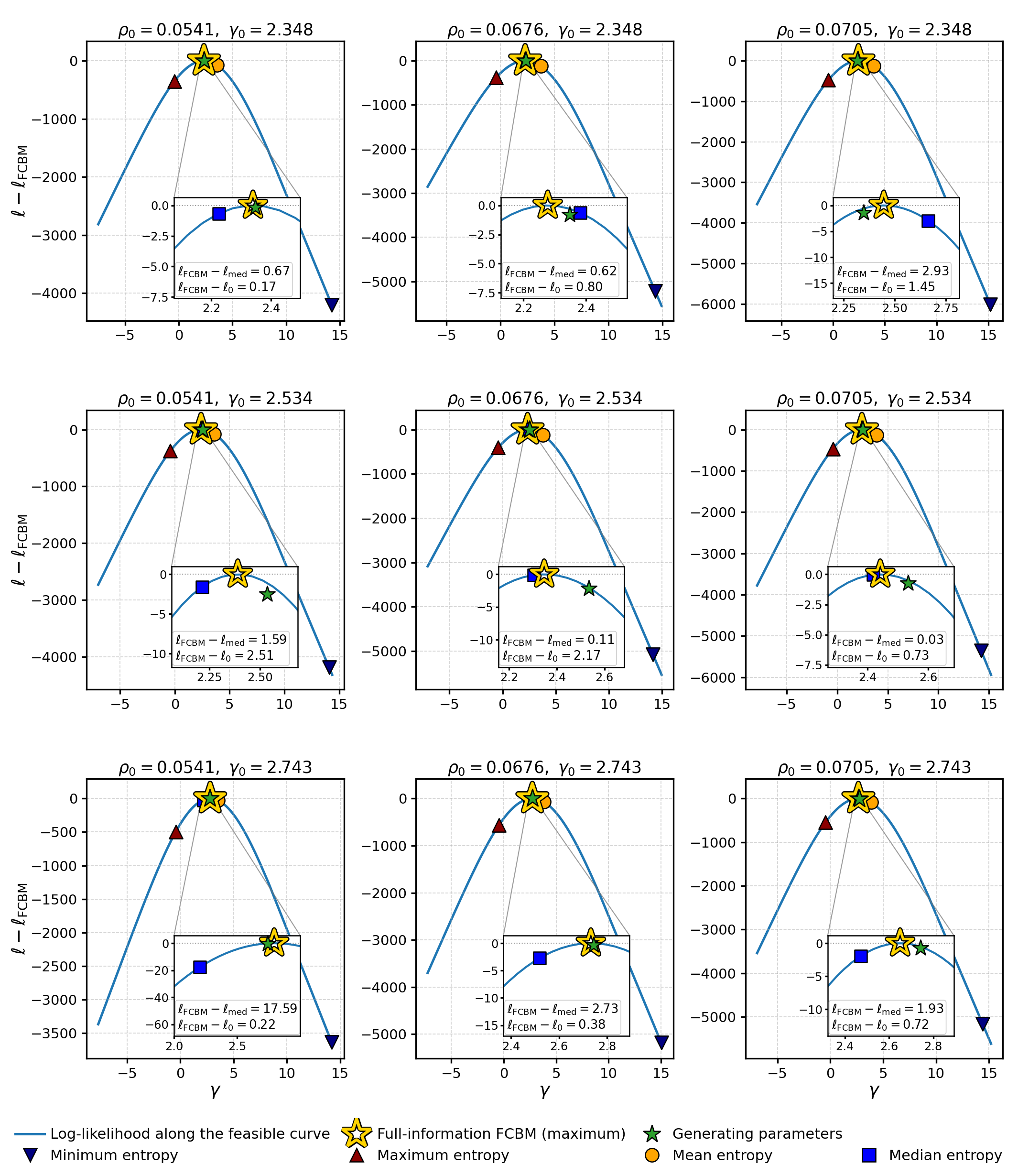}
\caption{Synthetic FCBM networks (not empirical observations): log-likelihood profiles for exactly the same nine sampled networks and generating regimes shown in Fig.~\ref{fig:synthetic-validation}. Each curve is evaluated using its sampled synthetic adjacency matrix along the feasible curve fixed by that sample's total link count, plotted against $\gamma$. The ordinate is the log-likelihood relative to the full-information FCBM maximum, which is zero. Open yellow star: full-information MLE fitted to the sampled network. Green star: log-likelihood of the known generating parameters $(\beta_0,\gamma_0)$, evaluated at that point even if it lies off the sampled feasible curve. Insets: enlarged view of the maximum, giving $\ell_{\mathrm{FCBM}}-\ell_{\mathrm{med}}$ and $\ell_{\mathrm{FCBM}}-\ell_{0}$. The minimum (\MinEntropySym), maximum (\MaxEntropySym), mean (\MeanEntropySym), and median (\MedianEntropySym) entropy representatives are selected using only the sampled total link count. The likelihood profiles are computed \emph{after} reconstruction for validation; within-region link counts are not supplied to the reduced-information selection.}
\label{fig:synthetic-loglik}
\end{figure*}

Figure~\ref{fig:synthetic-loglik} gives the corresponding likelihood
profiles for the \emph{same} nine realizations, making the
full-information optimum and the reduced-information median directly
comparable in likelihood space. The generating point is evaluated at
its actual $(\beta_0,\gamma_0)$, not projected onto the constrained
curve. The realized network and its within-region count enter this
likelihood visualization only after the representatives have been
selected; they are not additional reconstruction inputs. In particular,
proximity to the likelihood peak is a descriptive observation, not
an identification result or a claim that median entropy maximizes the
sampled synthetic-network likelihood.

The nine empirical trade networks considered here have comparatively
similar fitted regional effects and observed densities. Accordingly,
the synthetic grid is deliberately calibrated to their empirical
interquartile ranges rather than being presented as a general survey
of the FCBM parameter space. Across 100 independent realizations at
each of these nine generating settings, the median-entropy point
attained the smallest probability RMSE among the \emph{four
entropy-based representatives} in 893 of the 900 cases. Its per-cell median
$\mathrm{RMSE}_P$ ranged from $0.0039$ to $0.0171$, versus
$0.0032$--$0.0041$ for the full-information FCBM. The median
reconstruction had a smaller probability RMSE than the full-information
FCBM in 171 of the 900 individual realizations, a distinct comparison.
This does not contradict the FCBM maximum-likelihood property: its
objective is the likelihood of the sampled network, not the oracle
probability-matrix RMSE. The repetition establishes robustness to
Bernoulli sampling within the calibrated generating range, not
universal identification of the missing block-level statistic.

\FloatBarrier
There is no reason to expect the same entropy representative to be best
for all generating parameters. Indeed, the Appendix stress test
(Table~\ref{tab:synthetic-stress}) extends beyond the empirical range
by varying the regional effect and density more widely into less realistic values. Maximum
entropy is preferred for weak regional effects ($\gamma_0=0.5$).
For intermediate effects ($\gamma_0=2$), median entropy is preferred except at the
highest density, where maximum entropy wins most realizations, and for strong effects ($\gamma_0=3.5$) the
preferred representative also depends on density.

These results can be read as an empirical fact:
for the real-world trade networks investigated here and for synthetic networks with
comparable parameter ranges, median entropy is a stable and effective
choice; for networks generated by the same model outside those ranges,
other representatives can be preferable. We therefore regard the
median as a practically motivated default in the present
missing-sufficient-statistic setting, not as an optimal estimator for
arbitrary networks or parameter regimes.

\FloatBarrier
\section{Conclusions}
We studied binary network reconstruction when node fitnesses and categorical block memberships are known but one block-level sufficient statistic is missing. In the full-information FCBM, the within- and between-block link counts identify two structural parameters. When only the total link count is available, the same model is underidentified and the compatible parameters form a one-dimensional feasible curve.

Our contribution is to equip this constrained submodel with the Jeffreys measure induced by the Fisher metric and to use that invariant measure to characterize the distribution of Shannon entropy along the feasible family. The Jeffreys construction is a weighting of compatible solutions, not an averaging of parameter vectors or probability matrices. To obtain a single reconstruction, we use the median of the induced entropy distribution as a robust central representative. The minimum- and maximum-entropy points correspond to limiting allocations of edge probabilities, whereas the median avoids either extreme. In particular, minimum entropy does not follow from network sparsity because every feasible point already matches the same expected total number of links.

Across nine international-trade networks, the median-entropy reconstruction has ROC and PR AUC values close to those of the full-information FCBM, including isolated cases where it is higher. The historical FM AUCs were obtained with a different evaluation mask and therefore cannot establish a like-for-like ranking advantage without re-evaluation on common pairs. The proposed method uses the same network-level sufficient statistic as the FM---the total number of links---while additionally exploiting exogenous regional labels. The full-information FCBM uses a further empirical block-level count and therefore remains the appropriate identified benchmark when that information is available. Occasional differences in penalized or ranking scores should not be interpreted as evidence of overfitting without an out-of-sample experiment.

The controlled synthetic exercise clarifies that the choice of median
entropy is supported by these empirical applications. The networks
studied here have comparatively similar densities and fitted regional
effects. In a generative regime calibrated to those empirical ranges,
median entropy was the most stable choice among the four tested
entropy-based representatives. A wider synthetic stress test shows
that the preference changes when the regional effect and density are
forced away from this range. This is an empirical, regime-dependent
finding, not a general optimality theorem: in the absence of the
missing block count, the induced Jeffreys measure supplies an invariant
family of compatible solutions, while median entropy supplies a
well-performing default for the trade networks examined here and for
networks with comparable properties. The distinction between a known
generating point, a realized-network full-information MLE, and a
reduced-information representative remains essential for interpreting
both likelihood and probability-reconstruction performance.

Future work should study model averaging over the Jeffreys measure as an alternative to single-point selection, extend the construction to weighted and directed networks, and develop out-of-sample and missing-data validation protocols. Applications to financial, supply-chain, biological, and social networks are natural when meaningful categorical memberships are known but block-specific connectivity statistics are unavailable.

\section*{Acknowledgements}
\sloppy
This publication is part of the projects ``Network renormalization: from theoretical physics to the resilience of societies'' with file number NWA.1418.24.029 of the research programme NWA L3 - Innovative projects within routes 2024, which is (partly) financed by the Dutch Research Council (NWO) under the grant \url{https://doi.org/10.61686/AOIJP05368}, and ``Redefining renormalization for complex networks'' with file number OCENW.M.24.039 of the research programme Open Competition Domain Science Package 24-1, which is (partly) financed by the Dutch Research Council (NWO) under the grant \url{https://doi.org/10.61686/PBSEC42210}.

\bibliography{Ref}

\clearpage
\onecolumngrid
\appendix
\section{Additional Methodological Details and Results}
\subsection{Pseudocode: Scanning by Jeffreys Arclength}
\label{app:algorithms}
\begin{algorithm}[H]
\caption{Construct feasible curve and Jeffreys arclength}
\label{alg:build-curve}
\KwIn{Increasing grid $\{\beta_k\}_{k=0}^{K}$; total links $L_{\mathrm{total}}$; fitness $\{x_i\}$; blocks $\{R_{ij}\}$}
\KwOut{Feasible sequence $\{(\beta_k,\gamma_k,J_\beta(\beta_k),s_k)\}$}
\BlankLine

Initialize $\mathcal{K}\leftarrow\varnothing$.\;

\For{$k=0$ {\bf to} $K$}{
  Define 
  \[
  F(\gamma;\beta_k)
  =
  \sum_{i<j}
  \frac{e^{\beta_k} e^{\gamma R_{ij}} x_i x_j}
       {1+e^{\beta_k} e^{\gamma R_{ij}} x_i x_j}.
  \]

    Compute feasibility bounds
  \[
    L_{\min}(\beta_k)
    =
    \lim_{\gamma \to -\infty}
    \sum_{i<j}
    p_{ij}(\beta_k,\gamma),
  \quad
    L_{\max}(\beta_k)
    =
    \lim_{\gamma \to +\infty}
    \sum_{i<j}
    p_{ij}(\beta_k,\gamma)
    \]

  \If{$L_{\min}(\beta_k)<L_{\mathrm{total}}<L_{\max}(\beta_k)$}{
      Solve $F(\gamma;\beta_k)=L_{\mathrm{total}}$ for $\gamma_k$ by bracketed bisection.\;
      Add $k$ to $\mathcal{K}$.\;

           Compute
      \[
      Q_0(\beta_k,\gamma_k)
      =
      \sum_{i<j} p_{ij}(\beta_k,\gamma_k)\bigl[1-p_{ij}(\beta_k,\gamma_k)\bigr],
      \]
      \[
      Q_1(\beta_k,\gamma_k)
      =
      \sum_{i<j} p_{ij}(\beta_k,\gamma_k)\bigl[1-p_{ij}(\beta_k,\gamma_k)\bigr]R_{ij},
      \]
      \[
      Q_2(\beta_k,\gamma_k)
      =
      \sum_{i<j} p_{ij}(\beta_k,\gamma_k)\bigl[1-p_{ij}(\beta_k,\gamma_k)\bigr]R_{ij}^2.
      \]

      Define the induced Fisher information (Eq.~\eqref{eq:induced-info})
      \[
      I_{\mathcal C}
      =
      Q_2(\beta_k,\gamma_k)
      -
      \frac{Q_1(\beta_k,\gamma_k)^2}{Q_0(\beta_k,\gamma_k)},
      \]

      and Jeffreys density
      \[
      J_\beta(\beta_k)
      =
      \frac{Q_0(\beta_k,\gamma_k)}{|Q_1(\beta_k,\gamma_k)|}
      \sqrt{I_{\mathcal C}}.
      \]
  }
}

Retain $k\in\mathcal{K}$ (iterate over consecutive feasible indices) and compute
\[
\Delta s_{k:k+1}
=
\frac{1}{2}
\big(J_\beta(\beta_k)+J_\beta(\beta_{k+1})\big)
(\beta_{k+1}-\beta_k),
\quad
s_k=\sum_{j<k}\Delta s_{j:j+1}.
\]

\Return{$\{(\beta_k,\gamma_k,J_\beta(\beta_k),s_k)\}_{k\in\mathcal{K}}$}
\end{algorithm}
\begin{algorithm}[H]
\caption{Uniform sampling in Jeffreys arclength}
\label{alg:scan-by-s}
\KwIn{Feasible sequence $\{(\beta_k,\gamma_k,s_k)\}$; number of target samples $M$}
\KwOut{Uniform-in-$s$ samples $\{(\tilde\beta_m,\tilde\gamma_m)\}$}
\BlankLine

Let $S_{\mathrm{tot}}=s_{k_{\max}}$.\;

Define uniform partition
\[
s_m=\frac{m}{M}S_{\mathrm{tot}},
\qquad
m=1,\ldots,M.
\]

\For{$m=1$ {\bf to} $M$}{
    Find $k$ such that
    \[
    s_k \le s_m \le s_{k+1}.
    \]

    Define interpolation weight
    \[
    \lambda
    =
    \frac{ s_m-s_k}{s_{k+1}-s_k}
    \in[0,1].
    \]

    Compute
    \[
    \tilde\beta_m
    =
    (1-\lambda)\beta_k+\lambda\beta_{k+1},
    \]
    \[
    \tilde\gamma_m
    =
    (1-\lambda)\gamma_k+\lambda\gamma_{k+1}.
    \]
}

\Return{$\{(\tilde\beta_m,\tilde\gamma_m)\}$}
\end{algorithm}
\begin{algorithm}[H]
\caption{Evaluate entropy on the Jeffreys-uniform grid}
\label{alg:evaluate-on-s}
\KwIn{$\{(\tilde\beta_m,\tilde\gamma_m)\}_{m=1}^{M}$}
\KwOut{$\{H_m\}_{m=1}^{M}$}
\BlankLine

\For{$m=1$ {\bf to} $M$}{
    Compute
    \[
    p_{ij}^{(m)}
    =
    \frac{e^{\tilde\beta_m} e^{\tilde\gamma_m R_{ij}} x_i x_j}
         {1+e^{\tilde\beta_m} e^{\tilde\gamma_m R_{ij}} x_i x_j}.
    \]

    Compute Shannon entropy
    \[
    H_m
    =
    -\sum_{i<j}
    \left[
    p_{ij}^{(m)}\log p_{ij}^{(m)}
    +
    \big(1-p_{ij}^{(m)}\big)
    \log\big(1-p_{ij}^{(m)}\big)
    \right].
    \]
}

\Return{$\{H_m\}$}
\end{algorithm}

\subsection{Detailed mathematical transformations for the constrained Jeffreys construction}
\label{app:math}

For each unordered pair $(i,j)$ with $i<j$, the probability that an edge exists is
\begin{equation}
\label{eq:pij}
p_{ij}(\beta,\gamma)
=
\frac{e^{\beta} e^{\gamma R_{ij}} x_i x_j}
{1+e^{\beta} e^{\gamma R_{ij}} x_i x_j},
\qquad
R_{ij}\in\{0,1\}
\end{equation}
This implies the odds identity
\begin{equation}
\frac{p_{ij}(\beta,\gamma)}{1-p_{ij}(\beta,\gamma)}
=
e^{\beta} e^{\gamma R_{ij}} x_i x_j
=
\exp(\beta+\gamma R_{ij})\,x_i\,x_j
\end{equation}
Start from the definition in \eqref{eq:pij}:
\[
p_{ij}(\beta,\gamma)
=
\frac{e^{\beta} e^{\gamma R_{ij}} x_i x_j}
{1+e^{\beta} e^{\gamma R_{ij}} x_i x_j}.
\]
Apply the quotient rule in its original form:
\[
\frac{\partial}{\partial \beta}
\left(
  \frac{e^{\beta} e^{\gamma R_{ij}} x_i x_j}
       {1 + e^{\beta} e^{\gamma R_{ij}} x_i x_j}
\right)
=
\frac{
  \displaystyle
  \frac{\partial}{\partial \beta}
  \big(e^{\beta} e^{\gamma R_{ij}} x_i x_j\big)
  \cdot
  \big(1 + e^{\beta} e^{\gamma R_{ij}} x_i x_j\big)
  -
  \big(e^{\beta} e^{\gamma R_{ij}} x_i x_j\big)
  \cdot
  \frac{\partial}{\partial \beta}
  \big(1 + e^{\beta} e^{\gamma R_{ij}} x_i x_j\big)
}{
  \big(1 + e^{\beta} e^{\gamma R_{ij}} x_i x_j\big)^{2}
}
\]
Compute the two partial derivatives that appear:
\[
\frac{\partial}{\partial\beta}\big[e^{\beta} e^{\gamma R_{ij}} x_i x_j\big]
=
e^{\beta} e^{\gamma R_{ij}} x_i x_j,
\]
\[
\frac{\partial}{\partial\beta}\big[1+e^{\beta} e^{\gamma R_{ij}} x_i x_j\big]
=
e^{\beta} e^{\gamma R_{ij}} x_i x_j.
\]
Therefore,
\begin{align*}
\frac{\partial p_{ij}}{\partial \beta}
&=
\frac{
\big(e^{\beta} e^{\gamma R_{ij}} x_i x_j\big)
\cdot
\big(1 + e^{\beta} e^{\gamma R_{ij}} x_i x_j\big)
-
\big(e^{\beta} e^{\gamma R_{ij}} x_i x_j\big)
\cdot
\big(e^{\beta} e^{\gamma R_{ij}} x_i x_j\big)
}{
\big(1 + e^{\beta} e^{\gamma R_{ij}} x_i x_j\big)^2
}\\[6pt]
&=
\frac{
e^{\beta} e^{\gamma R_{ij}} x_i x_j
\Big[
(1 + e^{\beta} e^{\gamma R_{ij}} x_i x_j)
-
(e^{\beta} e^{\gamma R_{ij}} x_i x_j)
\Big]
}{
\big(1 + e^{\beta} e^{\gamma R_{ij}} x_i x_j\big)^2
}\\[6pt]
&=
\frac{
e^{\beta} e^{\gamma R_{ij}} x_i x_j
}{
\big(1 + e^{\beta} e^{\gamma R_{ij}} x_i x_j\big)^2
}
\end{align*}
Now write $p_{ij}$ and $1-p_{ij}$ in original form:
\[
p_{ij}
=
\frac{e^{\beta} e^{\gamma R_{ij}} x_i x_j}
{1+e^{\beta} e^{\gamma R_{ij}} x_i x_j}  \\ ,
\qquad
1-p_{ij}
=
\frac{1}{1+e^{\beta} e^{\gamma R_{ij}} x_i x_j}
\]
Hence
\[
p_{ij}\big(1-p_{ij}\big)
=
\frac{e^{\beta} e^{\gamma R_{ij}} x_i x_j}{\big(1+e^{\beta} e^{\gamma R_{ij}} x_i x_j\big)^{2}}.
\]
Therefore the derivative simplifies to the original logistic identity
\begin{equation}
\label{eq:dpdbeta}
\frac{\partial p_{ij}}{\partial\beta}
=
p_{ij}\big(1-p_{ij}\big)
\end{equation}
Again from \eqref{eq:pij},
\[
p_{ij}(\beta,\gamma)
=
\frac{e^{\beta} e^{\gamma R_{ij}} x_i x_j}
{1+e^{\beta} e^{\gamma R_{ij}} x_i x_j}
\]
Apply the quotient rule in original form:
\begin{align*}
\frac{\partial}{\partial \gamma}
\left(
  \frac{e^{\beta} e^{\gamma R_{ij}} x_i x_j}
       {1 + e^{\beta} e^{\gamma R_{ij}} x_i x_j}
\right)
&=
\frac{
  \left(\dfrac{\partial}{\partial \gamma}\big[e^{\beta} e^{\gamma R_{ij}} x_i x_j\big]\right)
  \left(1 + e^{\beta} e^{\gamma R_{ij}} x_i x_j\right)
  -
  \left(e^{\beta} e^{\gamma R_{ij}} x_i x_j\right)
  \left(\dfrac{\partial}{\partial \gamma}\big[1 + e^{\beta} e^{\gamma R_{ij}} x_i x_j\big]\right)
}{
  \left(1 + e^{\beta} e^{\gamma R_{ij}} x_i x_j\right)^2
}
\\[8pt]
&=
\frac{
  \left(e^{\beta} e^{\gamma R_{ij}} x_i x_j R_{ij}\right)
  \left(1 + e^{\beta} e^{\gamma R_{ij}} x_i x_j\right)
  -
  \left(e^{\beta} e^{\gamma R_{ij}} x_i x_j\right)
  \left(e^{\beta} e^{\gamma R_{ij}} x_i x_j R_{ij}\right)
}{
  \left(1 + e^{\beta} e^{\gamma R_{ij}} x_i x_j\right)^2
}
\\[8pt]
&=
\frac{
  e^{\beta} e^{\gamma R_{ij}} x_i x_j R_{ij}
  \left[
    \left(1 + e^{\beta} e^{\gamma R_{ij}} x_i x_j\right)
    -
    \left(e^{\beta} e^{\gamma R_{ij}} x_i x_j\right)
  \right]
}{
  \left(1 + e^{\beta} e^{\gamma R_{ij}} x_i x_j\right)^2
}
\\[8pt]
&=
\frac{
  e^{\beta} e^{\gamma R_{ij}} x_i x_j R_{ij}
}{
  \left(1 + e^{\beta} e^{\gamma R_{ij}} x_i x_j\right)^2
}
\end{align*}
Compute the two partial derivatives:
\[
\frac{\partial}{\partial\gamma}\big[e^{\beta} e^{\gamma R_{ij}} x_i x_j\big]
=
e^{\beta} e^{\gamma R_{ij}} x_i x_j\,R_{ij},
\]
\[
\frac{\partial}{\partial\gamma}\big[1+e^{\beta} e^{\gamma R_{ij}} x_i x_j\big]
=
e^{\beta} e^{\gamma R_{ij}} x_i x_j\,R_{ij}
\]
Therefore,
\begin{align*}
\frac{\partial p_{ij}}{\partial \gamma}
&=
\frac{
\big(e^{\beta} e^{\gamma R_{ij}} x_i x_j R_{ij}\big)
\cdot
\big(1 + e^{\beta} e^{\gamma R_{ij}} x_i x_j\big)
-
\big(e^{\beta} e^{\gamma R_{ij}} x_i x_j\big)
\cdot
\big(e^{\beta} e^{\gamma R_{ij}} x_i x_j R_{ij}\big)
}{
\big(1 + e^{\beta} e^{\gamma R_{ij}} x_i x_j\big)^{2}
}
\\[6pt]
&=
\frac{
e^{\beta} e^{\gamma R_{ij}} x_i x_j R_{ij}
\Big[
\big(1 + e^{\beta} e^{\gamma R_{ij}} x_i x_j\big)
-
\big(e^{\beta} e^{\gamma R_{ij}} x_i x_j\big)
\Big]
}{
\big(1 + e^{\beta} e^{\gamma R_{ij}} x_i x_j\big)^{2}
}
\\[6pt]
&=
\frac{
e^{\beta} e^{\gamma R_{ij}} x_i x_j R_{ij}
}{
\big(1 + e^{\beta} e^{\gamma R_{ij}} x_i x_j\big)^{2}
}.
\\[10pt]
\text{}\quad
\frac{\partial p_{ij}}{\partial \gamma}
&=
R_{ij}\,
\left(
\frac{e^{\beta} e^{\gamma R_{ij}} x_i x_j}
     {1 + e^{\beta} e^{\gamma R_{ij}} x_i x_j}
\right)
\left(
\frac{1}
     {1 + e^{\beta} e^{\gamma R_{ij}} x_i x_j}
\right)
\\[6pt]
&= R_{ij}\,p_{ij}\,(1 - p_{ij})
\end{align*}
Using the same expressions for $p_{ij}$ and $1-p_{ij}$ as above,
\[
\frac{\partial p_{ij}}{\partial\gamma}
=
p_{ij}\big(1-p_{ij}\big)\,R_{ij}
\]
Let $a_{ij}\in\{0,1\}$ be the observed indicator of the edge between $i$ and $j$. The log-likelihood is
\begin{equation}
\label{eq:loglik}
\ell(\beta,\gamma)
=
\sum_{i<j}
\left[
a_{ij}\,\log\!\left(\frac{e^{\beta} e^{\gamma R_{ij}} x_i x_j}{1+e^{\beta} e^{\gamma R_{ij}} x_i x_j}\right)
+
\big(1-a_{ij}\big)\,\log\!\left(\frac{1}{1+e^{\beta} e^{\gamma R_{ij}} x_i x_j}\right)
\right]
\end{equation}
Differentiate \eqref{eq:loglik} with respect to $p_{ij}$ 
\[
\frac{\partial\ell}{\partial p_{ij}}
=
\frac{a_{ij}}{p_{ij}}
-
\frac{1-a_{ij}}{1-p_{ij}}
=
\frac{a_{ij}-p_{ij}}{p_{ij}(1-p_{ij})}
\]
Then apply the chain rule with the explicit derivatives obtained above:
\[
\frac{\partial\ell}{\partial\beta}
=
\sum_{i<j}
\left(
\frac{a_{ij}-p_{ij}}{p_{ij}(1-p_{ij})}
\right)
\left(
p_{ij}(1-p_{ij})
\right)
=
\sum_{i<j}(a_{ij}-p_{ij}),
\]
\[
\frac{\partial\ell}{\partial\gamma}
=
\sum_{i<j}
\left(
\frac{a_{ij}-p_{ij}}{p_{ij}(1-p_{ij})}
\right)
\left(
p_{ij}(1-p_{ij})\,R_{ij}
\right)
=
\sum_{i<j}(a_{ij}-p_{ij})\,R_{ij}
\]
For each pair $(i,j)$,
\[
\frac{\partial^{2}\ell}{\partial p_{ij}^{2}}
=
-\frac{a_{ij}}{p_{ij}^{2}}
-
\frac{1-a_{ij}}{(1-p_{ij})^{2}}
\]
Taking expectation under the Bernoulli model where $\mathbb{E}[a_{ij}]=p_{ij}$ gives
\[
\mathbb{E}\!\left[-\frac{\partial^{2}\ell}{\partial p_{ij}^{2}}\right]
=
\frac{1}{p_{ij}}
+
\frac{1}{1-p_{ij}}
=
\frac{1}{p_{ij}(1-p_{ij})}.
\]
Using
\(
\frac{\partial p_{ij}}{\partial\beta}=p_{ij}(1-p_{ij})
\)
and
\(
\frac{\partial p_{ij}}{\partial\gamma}=p_{ij}(1-p_{ij})\,R_{ij}
\),
we compute each entry:

\noindent\textbf{Fisher information for $\beta$ and $\beta$:}
\[
I_{\beta\beta}
=
\sum_{i<j}
\left(
\mathbb{E}\!\left[-\frac{\partial^{2}\ell}{\partial p_{ij}^{2}}\right]
\right)
\left(
\frac{\partial p_{ij}}{\partial\beta}
\right)^{2}
=
\sum_{i<j}
\left(
\frac{1}{p_{ij}(1-p_{ij})}
\right)
\left(
p_{ij}(1-p_{ij})
\right)^{2}
=
\sum_{i<j}
p_{ij}(1-p_{ij})
\]
\noindent\textbf{Fisher information for $\beta$ and $\gamma$:}
\begin{align*}
I_{\beta\gamma}
&=
\sum_{i<j}
\left(
\mathbb{E}\!\left[-\frac{\partial^{2}\ell}{\partial p_{ij}^{2}}\right]
\right)
\left(
\frac{\partial p_{ij}}{\partial \beta}
\right)
\left(
\frac{\partial p_{ij}}{\partial \gamma}
\right)
\\[6pt]
&=
\sum_{i<j}
\left(
\frac{1}{p_{ij}(1-p_{ij})}
\right)
\big(p_{ij}(1-p_{ij})\big)
\big(R_{ij} p_{ij}(1-p_{ij})\big)
\\[6pt]
&=
\sum_{i<j}
p_{ij}(1-p_{ij})\,R_{ij}
\end{align*}
\noindent\textbf{Fisher information for $\gamma$ and $\gamma$:}
\begin{align*}
I_{\gamma\gamma}
&=
\sum_{i<j}
\left(
\mathbb{E}\!\left[-\frac{\partial^{2}\ell}{\partial p_{ij}^{2}}\right]
\right)
\left(
\frac{\partial p_{ij}}{\partial \gamma}
\right)^{2}
\\[6pt]
&=
\sum_{i<j}
\left(
\frac{1}{p_{ij}(1-p_{ij})}
\right)
\big(R_{ij} p_{ij}(1-p_{ij})\big)^{2}
\\[6pt]
&=
\sum_{i<j}
p_{ij}(1-p_{ij})\,R_{ij}^{2}
\end{align*}
These three sums correspond respectively to $Q_0(\beta,\gamma)$, $Q_1(\beta,\gamma)$, and $Q_2(\beta,\gamma)$ as defined in the main text.

\noindent\textbf{Thus, the Fisher information matrix is:}
\[
I(\beta,\gamma)
=
\begin{bmatrix}
\displaystyle \sum_{i<j} p_{ij}(1-p_{ij})
&
\displaystyle \sum_{i<j} p_{ij}(1-p_{ij})R_{ij}
\\[8pt]
\displaystyle \sum_{i<j} p_{ij}(1-p_{ij})R_{ij}
&
\displaystyle \sum_{i<j} p_{ij}(1-p_{ij})R_{ij}^{2}
\end{bmatrix}
\]
When $R_{ij}$ takes values in $\{0,1\}$, one has $R_{ij}^{2}=R_{ij}$.
Impose that the expected total number of edges equals the observed total $L_{\mathrm{total}}$:
\[
C(\beta,\gamma)
=
\sum_{i<j}
\frac{e^{\beta} e^{\gamma R_{ij}} x_i x_j}
{1+e^{\beta} e^{\gamma R_{ij}} x_i x_j}
-
L_{\mathrm{total}}
=
0
\]
Using the explicit derivatives:
\[
\frac{\partial C}{\partial\beta}
=
\sum_{i<j}
\frac{e^{\beta} e^{\gamma R_{ij}} x_i x_j}
{\big(1+e^{\beta} e^{\gamma R_{ij}} x_i x_j\big)^{2}}
=
\sum_{i<j}
p_{ij}(1-p_{ij}),
\]
\[
\frac{\partial C}{\partial\gamma}
=
\sum_{i<j}
\frac{e^{\beta} e^{\gamma R_{ij}} x_i x_j\,R_{ij}}
{\big(1+e^{\beta} e^{\gamma R_{ij}} x_i x_j\big)^{2}}
=
\sum_{i<j}
p_{ij}(1-p_{ij})\,R_{ij}
\]
Along the feasible curve defined by $C(\beta,\gamma)=0$, the total derivative with respect to $\beta$ satisfies
\[
\frac{dC}{d\beta}
=
\frac{\partial C}{\partial\beta}
+
\frac{\partial C}{\partial\gamma}\,\frac{d\gamma}{d\beta}
=
0
\]
Therefore the tangent slope is
\begin{equation}
\label{eq:slope}
\frac{d\gamma}{d\beta}
=
-
\frac{
\displaystyle\sum_{i<j}
\frac{e^{\beta} e^{\gamma R_{ij}} x_i x_j}
{\big(1+e^{\beta} e^{\gamma R_{ij}} x_i x_j\big)^{2}}
}{
\displaystyle\sum_{i<j}
\frac{e^{\beta} e^{\gamma R_{ij}} x_i x_j\,R_{ij}}
{\big(1+e^{\beta} e^{\gamma R_{ij}} x_i x_j\big)^{2}}
}
=
-
\frac{\displaystyle\sum_{i<j}p_{ij}(1-p_{ij})}
{\displaystyle\sum_{i<j}p_{ij}(1-p_{ij})\,R_{ij}}
\end{equation}
The determinant of Fisher matrix is
\[
\det I(\beta,\gamma)
=
\left(
\sum_{i<j}p_{ij}(1-p_{ij})
\right)
\left(
\sum_{i<j}p_{ij}(1-p_{ij})\,R_{ij}^{2}
\right)
-
\left(
\sum_{i<j}p_{ij}(1-p_{ij})\,R_{ij}
\right)^{2}
\]
Hence the two-dimensional Jeffreys volume density of the unconstrained family on $(\beta,\gamma)$ is
\begin{equation}
\pi_{\text{Jeffreys}}(\beta,\gamma)
=
\sqrt{
\left(
\sum_{i<j}p_{ij}(1-p_{ij})
\right)
\left(
\sum_{i<j}p_{ij}(1-p_{ij})\,R_{ij}^{2}
\right)
-
\left(
\sum_{i<j}p_{ij}(1-p_{ij})\,R_{ij}
\right)^{2}
}
\end{equation}
The effective one-dimensional Fisher information is the Schur complement of $I_{\beta\beta}$:
\[
I_{\mathcal C}
=
\left(
\sum_{i<j}p_{ij}(1-p_{ij})\,R_{ij}^{2}
\right)
-
\frac{
\left(
\sum_{i<j}p_{ij}(1-p_{ij})\,R_{ij}
\right)^{2}
}{
\left(
\sum_{i<j}p_{ij}(1-p_{ij})
\right)
}
\]
Therefore the one-dimensional Jeffreys density along the feasible curve, when parameterized by $\gamma$, is
\begin{equation}
J_{\gamma}(\gamma)
=
\sqrt{
\left(
\sum_{i<j}p_{ij}(1-p_{ij})\,R_{ij}^{2}
\right)
-
\frac{
\left(
\sum_{i<j}p_{ij}(1-p_{ij})\,R_{ij}
\right)^{2}
}{
\left(
\sum_{i<j}p_{ij}(1-p_{ij})
\right)
}
}
\end{equation}
The tangent direction along the feasible curve is the vector with components $\big(1,\frac{d\gamma}{d\beta}\big)$. The Fisher information along this direction is the quadratic form
\[
J_{\beta}(\beta)^2
=
\begin{bmatrix}
1 & \dfrac{d\gamma}{d\beta}
\end{bmatrix}
\,
\begin{bmatrix}
\displaystyle\sum_{i<j}p_{ij}(1-p_{ij})
&
\displaystyle\sum_{i<j}p_{ij}(1-p_{ij})\,R_{ij}
\\[10pt]
\displaystyle\sum_{i<j}p_{ij}(1-p_{ij})\,R_{ij}
&
\displaystyle\sum_{i<j}p_{ij}(1-p_{ij})\,R_{ij}^{2}
\end{bmatrix}
\,
\begin{bmatrix}
1\\[6pt]
\dfrac{d\gamma}{d\beta}
\end{bmatrix}
\]
This equals
\[
J_{\beta}(\beta)^2
=
\left(
\sum_{i<j}p_{ij}(1-p_{ij})
\right)
+
2\left(
\sum_{i<j}p_{ij}(1-p_{ij})\,R_{ij}
\right)\frac{d\gamma}{d\beta}
+
\left(
\sum_{i<j}p_{ij}(1-p_{ij})\,R_{ij}^{2}
\right)
\left(\frac{d\gamma}{d\beta}\right)^{2}
\]
Substitute the slope from \eqref{eq:slope} in its original sum form to obtain the same square-root term as $J_{\gamma}(\gamma)$, multiplied by the Jacobian factor
\[
\left|\frac{d\gamma}{d\beta}\right|
=
\left|
-
\frac{\displaystyle\sum_{i<j}p_{ij}(1-p_{ij})}
{\displaystyle\sum_{i<j}p_{ij}(1-p_{ij})\,R_{ij}}
\right|
=
\frac{\displaystyle\sum_{i<j}p_{ij}(1-p_{ij})}
{\displaystyle\left|\sum_{i<j}p_{ij}(1-p_{ij})\,R_{ij}\right|}
\]
Hence the one-dimensional Jeffreys density along the feasible curve, when parameterized by $\beta$, is
\begin{equation}
J_{\beta}(\beta)
=
\frac{
\displaystyle\sum_{i<j}p_{ij}(1-p_{ij})
}{
\displaystyle\left|\sum_{i<j}p_{ij}(1-p_{ij})\,R_{ij}\right|
}
\,
\sqrt{
\left(
\sum_{i<j}p_{ij}(1-p_{ij})\,R_{ij}^{2}
\right)
-
\frac{
\left(
\sum_{i<j}p_{ij}(1-p_{ij})\,R_{ij}
\right)^{2}
}{
\left(
\sum_{i<j}p_{ij}(1-p_{ij})
\right)
}
}
\end{equation}
The induced Jeffreys measure is invariant in the sense that
\[
J_{\beta}(\beta)\,|d\beta|
=
J_{\gamma}(\gamma)\,|d\gamma|
\]
For each fixed $\beta$, define
\[
F(\gamma;\beta)
=
\sum_{i<j}
\frac{e^{\beta} e^{\gamma R_{ij}} x_i x_j}
{1+e^{\beta} e^{\gamma R_{ij}} x_i x_j}
\]
Differentiate this function with respect to $\gamma$ using the explicit derivative obtained earlier:
\[
\frac{\partial F}{\partial\gamma}
=
\sum_{i<j}
\frac{e^{\beta} e^{\gamma R_{ij}} x_i x_j\,R_{ij}}
{\big(1+e^{\beta} e^{\gamma R_{ij}} x_i x_j\big)^{2}}
=
\sum_{i<j}
p_{ij}(1-p_{ij})\,R_{ij}
\ge 0
\]
Therefore $F(\gamma;\beta)$ is monotone non-decreasing in $\gamma$. Where $Q_1(\beta,\gamma)>0$, the equation $F(\gamma;\beta)=L_{\mathrm{total}}$ has a unique finite solution $\gamma(\beta)$ if $L_{\mathrm{total}}$ lies strictly between the two limits obtained as $\gamma$ tends to negative or positive infinity:
\[
\lim_{\gamma\to -\infty}
F(\gamma;\beta)
=
\sum_{R_{ij}=0}
\frac{e^{\beta}\,x_i\,x_j}{1+e^{\beta}\,x_i\,x_j},
\]
\[
\lim_{\gamma\to +\infty}
F(\gamma;\beta)
=
\sum_{R_{ij}=0}
\frac{e^{\beta}\,x_i\,x_j}{1+e^{\beta}\,x_i\,x_j}
+
\#\big\{(i,j):R_{ij}=1\big\}.
\]
These limits establish the feasibility interval for finite solutions $\gamma(\beta)$; equality at a limit corresponds to an asymptotic endpoint rather than a finite root. Numerically, Algorithm~\ref{alg:build-curve} brackets each finite solution and uses bisection in $\gamma$ on the $\beta$ grid.
Define the intrinsic arclength element along the feasible curve by
\[
ds
=
J_{\beta}(\beta)\,|d\beta|
=
J_{\gamma}(\gamma)\,|d\gamma|,
\]
which is invariant to the choice of parameter along the curve.
Let the total Jeffreys length of a feasible segment in $\beta$ be
\[
S_{\mathrm{tot}}
=
\int_{\beta_{\min}}^{\beta_{\max}}
J_{\beta}(\beta)\,d\beta
\]
Choose an equal-length partition
\[
0=s_{0}<s_{1}<\cdots<s_{M}=S_{\mathrm{tot}},
\qquad
\Delta s
=
\frac{S_{\mathrm{tot}}}{M}
\]
Select points $(\beta_{m},\gamma_{m})$ on the feasible curve by solving
\[
\int_{\beta_{\min}}^{\beta_{m}}
J_{\beta}(\beta)\,d\beta
=
s_{m},
\qquad
m=1,\ldots,M,
\]
with each $\gamma_{m}$ obtained from the constraint equation $C(\beta_{m},\gamma_{m})=0$.
All $(\beta_{m},\gamma_{m})$ lie on the feasible curve.
Numerically, Algorithm~\ref{alg:scan-by-s} approximates these points by linear interpolation between neighboring grid points, so the constraint is satisfied up to interpolation accuracy (Sec.~\ref{sec:jeffreys-discretization}).
The collection $\{(\beta_{m},\gamma_{m})\}$ forms a discretization
of the feasible set that is uniform in Jeffreys arclength. When the
curve is parameterized by $\gamma$, its induced Fisher information
coefficient is
\[
I_{\mathcal C}(\gamma)
=Q_2-\frac{Q_1^2}{Q_0}
=\sum_{i<j}p_{ij}(1-p_{ij})R_{ij}^2
-\frac{\left[\sum_{i<j}p_{ij}(1-p_{ij})R_{ij}\right]^2}
{\sum_{i<j}p_{ij}(1-p_{ij})}.
\]
The local Jeffreys line element is
$ds=\sqrt{I_{\mathcal C}(\gamma)}\,|d\gamma|$.
Thus, equal steps in Jeffreys arclength place more samples per unit
$\gamma$ where the induced Fisher information is larger. This
information-geometric weighting should not be confused with the
Euclidean curvature or apparent flatness of the plotted curve; it is
invariant under smooth changes of parameter coordinate.

\subsection{Two-Dimensional Projections of the Feasible Curves}
\label{app:2d-projections}
For completeness, Figs.~\ref{fig:appendix-entropy-beta}--\ref{fig:appendix-loglik-beta} show the remaining two-dimensional projections of the three-dimensional curves in Figs.~\ref{fig:Jeffreys} and~\ref{fig:Jeffreys_loglikelihood_3d}: entropy versus $e^{\beta}$, entropy versus $\gamma$, and log-likelihood versus $e^{\beta}$. These are alternative views of the same feasible families rather than additional reconstructions. The complementary log-likelihood projection versus $\gamma$ appears in Fig.~\ref{fig:Jeffreys_loglikelihood} of the main text. The $e^{\beta}$ horizontal coordinate is a positive reparameterization of $\beta$; it does not change the numerical selections.

\renewcommand{\subcaptext}[1]{\par\vspace{0pt}{\scriptsize #1\par}}
\setcounter{figure}{0}
\renewcommand{\thefigure}{A\arabic{figure}}
\renewcommand{\theHfigure}{A\arabic{figure}}
\begin{figure}[H]
\centering
\includegraphics[width=0.92\textwidth]{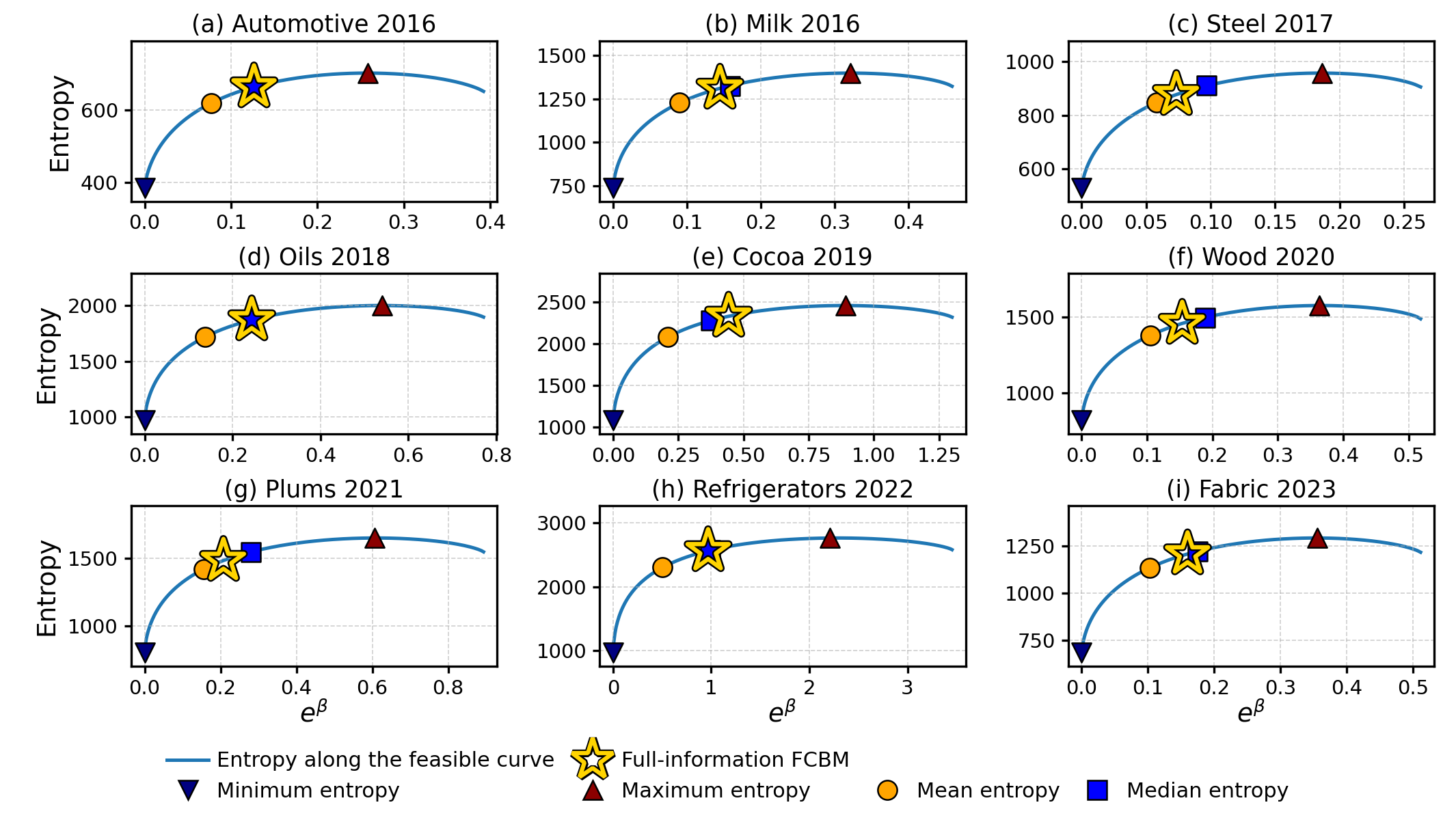}


\caption{Empirical trade networks: Shannon entropy along the feasible curve versus $e^{\beta}$. Markers indicate the full-information FCBM (\FCBMOpenSym), minimum (\MinEntropySym), maximum (\MaxEntropySym), mean (\MeanEntropySym), and median (\MedianEntropySym) entropy points. Panels (a)--(i) correspond to Fig.~\ref{fig:trade_networks}.}

\label{fig:appendix-entropy-beta}
\end{figure}

\begin{figure}[H]
\centering
\includegraphics[width=0.92\textwidth]{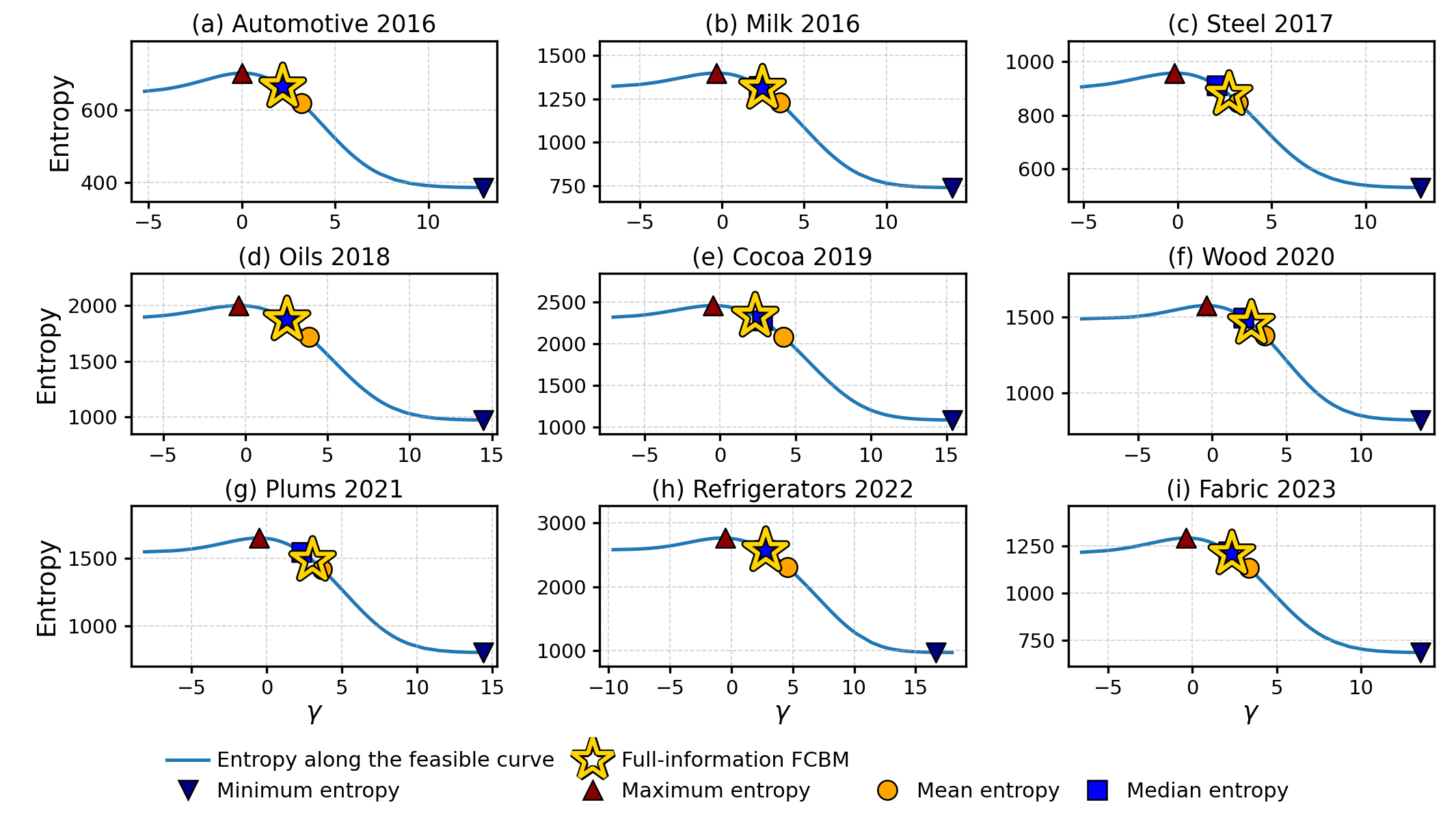}

\caption{Empirical trade networks: Shannon entropy along the feasible curve versus $\gamma$. Markers indicate the full-information FCBM (\FCBMOpenSym), minimum (\MinEntropySym), maximum (\MaxEntropySym), mean (\MeanEntropySym), and median (\MedianEntropySym) entropy points. Panels (a)--(i) correspond to Fig.~\ref{fig:trade_networks}.}

\label{fig:appendix-entropy-gamma}
\end{figure}

\begin{figure}[H]
\centering
\includegraphics[width=0.92\textwidth]{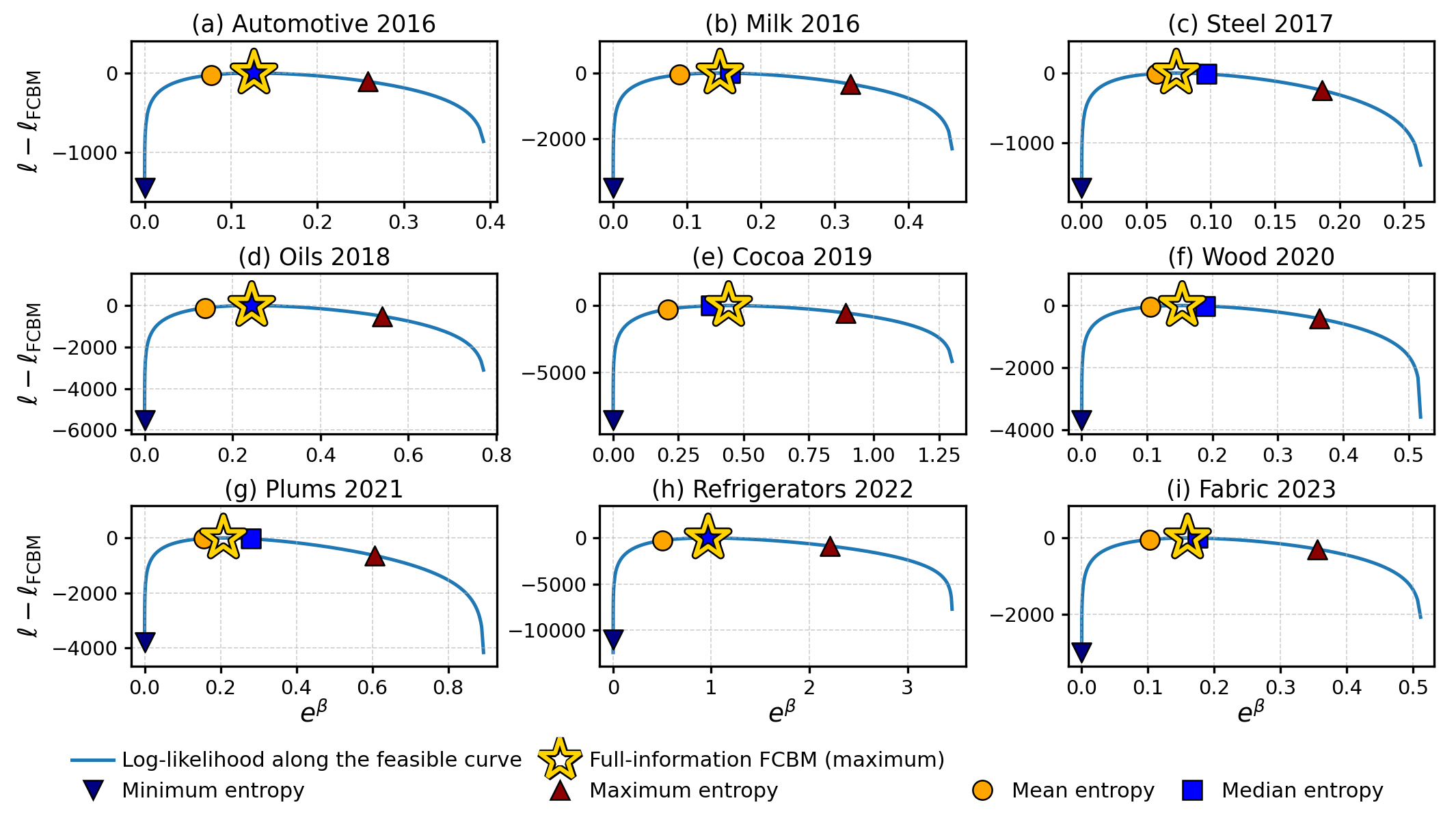}


\caption{Empirical trade networks: observed-network log-likelihood along the feasible curve versus $e^{\beta}$, shown relative to the full-information FCBM maximum as in Fig.~\ref{fig:Jeffreys_loglikelihood}. Markers indicate the full-information FCBM (\FCBMOpenSym), minimum (\MinEntropySym), maximum (\MaxEntropySym), mean (\MeanEntropySym), and median (\MedianEntropySym) entropy points. Panels (a)--(i) correspond to Fig.~\ref{fig:trade_networks}.}

\label{fig:appendix-loglik-beta}

\end{figure}

\FloatBarrier
\subsection{Numerical Values of the Model-Performance Summaries}
\label{app:perf-table}
Table~\ref{tab:model_perf} records the originally submitted FM ROC and PR AUC results, together with the reduced-information median and full-information FCBM results and the corrected undirected AIC/BIC-form scores. The historical FM AUCs were evaluated on all $n^2$ entries of the adjacency matrix, including the diagonal and both orientations; the block-aware AUCs use the unordered country pairs $i<j$. These AUC values are reported transparently as historical results and must not be used to infer paired gains across evaluation domains. Accordingly, Fig.~\ref{fig:model_perf}(a)--(b) compares only the two block-aware methods on the shared domain.
\setcounter{table}{0}
\renewcommand{\thetable}{A\arabic{table}}
\renewcommand{\theHtable}{A\arabic{table}}
\begin{table*}[!htbp]
\centering
\caption{Empirical trade networks: reported ROC AUC, PR AUC and corrected undirected penalized-likelihood scores. The FM ROC/PR AUCs reproduce the original submitted values based on all $n^2$ matrix entries, whereas the median and FCBM AUCs use unordered pairs $i<j$; cross-domain AUC gains are therefore not inferred. All AIC/BIC-form scores are calculated with undirected pair log-likelihoods. ROC AUC and PR AUC are rounded to four decimal places. Figure~\ref{fig:model_perf}(a)--(b) compares only the median and FCBM AUCs on their common sample.}
\label{tab:model_perf}
\renewcommand{\arraystretch}{1.15}

\definecolor{jpbg}{RGB}{232,245,233}   
\definecolor{fcbmbg}{RGB}{255,248,225} 
\definecolor{headbg}{RGB}{245,245,245}

\scriptsize
\setlength{\tabcolsep}{5pt}

\begingroup
\setlength{\arrayrulewidth}{0.45pt}
\arrayrulecolor{black!55}

\begin{tabular}{|l|l|c|c|c|c|}
\hline
\cellcolor{headbg}\textbf{Data} &
\cellcolor{headbg}\textbf{Model} &
\cellcolor{headbg}\textbf{ROC AUC} &
\cellcolor{headbg}\textbf{PR AUC} &
\cellcolor{headbg}\textbf{AIC}$^{\diamond}$ &
\cellcolor{headbg}\textbf{BIC}$^{\diamond}$ \\
\hline

\multirow{3}{*}{\shortstack[l]{ELEnet\\2016 Automotive\\Figure (3a)}}
& Block-Agnostic FM & 0.8914 & 0.4756 & 1600.49 & 1606.95 \\
\cline{2-6}
& \cellcolor{jpbg}Jeffreys Measure \& Median Entropy
& \cellcolor{jpbg}0.9318$^{\star}$
& \cellcolor{jpbg}0.5777$^{\star}$
& \cellcolor{jpbg}1384.61$^{\star}$
& \cellcolor{jpbg}1391.08$^{\star}$ \\
\cline{2-6}
& \cellcolor{fcbmbg}FCBM Planted Partition$^{\ddagger}$
& \cellcolor{fcbmbg}0.9318$^{\star}$
& \cellcolor{fcbmbg}0.5775
& \cellcolor{fcbmbg}1386.58
& \cellcolor{fcbmbg}1399.51 \\
\hline

\multirow{3}{*}{\shortstack[l]{BACI\\2016 Milk\\Figure (3b)}}
& Block-Agnostic FM & 0.8055 & 0.1889 & 6432.09 & 6439.70 \\
\cline{2-6}
& \cellcolor{jpbg}Jeffreys Measure \& Median Entropy
& \cellcolor{jpbg}0.8296
& \cellcolor{jpbg}0.2528
& \cellcolor{jpbg}5924.04
& \cellcolor{jpbg}5931.64$^{\star}$ \\
\cline{2-6}
& \cellcolor{fcbmbg}FCBM Planted Partition$^{\ddagger}$
& \cellcolor{fcbmbg}0.8309$^{\star}$
& \cellcolor{fcbmbg}0.2585$^{\star}$
& \cellcolor{fcbmbg}5921.25$^{\star}$
& \cellcolor{fcbmbg}5936.47 \\
\hline

\multirow{3}{*}{\shortstack[l]{UN Comtrade\\2017 Steel\\Figure (3c)}}
& Block-Agnostic FM & 0.9008 & 0.4152 & 2561.79 & 2568.98 \\
\cline{2-6}
& \cellcolor{jpbg}Jeffreys Measure \& Median Entropy
& \cellcolor{jpbg}0.9352
& \cellcolor{jpbg}0.5131
& \cellcolor{jpbg}2141.57
& \cellcolor{jpbg}2148.77 \\
\cline{2-6}
& \cellcolor{fcbmbg}FCBM Planted Partition$^{\ddagger}$
& \cellcolor{fcbmbg}0.9389$^{\star}$
& \cellcolor{fcbmbg}0.5168$^{\star}$
& \cellcolor{fcbmbg}2117.81$^{\star}$
& \cellcolor{fcbmbg}2132.20$^{\star}$ \\
\hline

\multirow{3}{*}{\shortstack[l]{UN Comtrade\\2018 Oils\\Figure (3d)}}
& Block-Agnostic FM & 0.8829 & 0.4568 & 6172.25 & 6179.96 \\
\cline{2-6}
& \cellcolor{jpbg}Jeffreys Measure \& Median Entropy
& \cellcolor{jpbg}0.9094$^{\star}$
& \cellcolor{jpbg}0.5322$^{\star}$
& \cellcolor{jpbg}5405.42$^{\star}$
& \cellcolor{jpbg}5413.12$^{\star}$ \\
\cline{2-6}
& \cellcolor{fcbmbg}FCBM Planted Partition$^{\ddagger}$
& \cellcolor{fcbmbg}0.9094$^{\star}$
& \cellcolor{fcbmbg}0.5322$^{\star}$
& \cellcolor{fcbmbg}5407.41
& \cellcolor{fcbmbg}5422.83 \\
\hline

\multirow{3}{*}{\shortstack[l]{BACI\\2019 Cocoa\\Figure (3e)}}
& Block-Agnostic FM & 0.8653 & 0.4070 & 8377.36 & 8385.15 \\
\cline{2-6}
& \cellcolor{jpbg}Jeffreys Measure \& Median Entropy
& \cellcolor{jpbg}0.8894$^{\star}$
& \cellcolor{jpbg}0.5003$^{\star}$
& \cellcolor{jpbg}7640.12
& \cellcolor{jpbg}7647.91 \\
\cline{2-6}
& \cellcolor{fcbmbg}FCBM Planted Partition$^{\ddagger}$
& \cellcolor{fcbmbg}0.8886
& \cellcolor{fcbmbg}0.4956
& \cellcolor{fcbmbg}7606.52$^{\star}$
& \cellcolor{fcbmbg}7622.12$^{\star}$ \\
\hline

\multirow{3}{*}{\shortstack[l]{BACI\\2020 Wood\\Figure (3f)}}
& Block-Agnostic FM & 0.8785 & 0.3796 & 5000.01 & 5007.64 \\
\cline{2-6}
& \cellcolor{jpbg}Jeffreys Measure \& Median Entropy
& \cellcolor{jpbg}0.9077
& \cellcolor{jpbg}0.4773
& \cellcolor{jpbg}4366.52
& \cellcolor{jpbg}4374.15 \\
\cline{2-6}
& \cellcolor{fcbmbg}FCBM Planted Partition$^{\ddagger}$
& \cellcolor{fcbmbg}0.9103$^{\star}$
& \cellcolor{fcbmbg}0.4853$^{\star}$
& \cellcolor{fcbmbg}4340.52$^{\star}$
& \cellcolor{fcbmbg}4355.78$^{\star}$ \\
\hline

\multirow{3}{*}{\shortstack[l]{UN Comtrade\\2021 Plums\\Figure (3g)}}
& Block-Agnostic FM & 0.8096 & 0.2463 & 6949.57 & 6957.09 \\
\cline{2-6}
& \cellcolor{jpbg}Jeffreys Measure \& Median Entropy
& \cellcolor{jpbg}0.8586
& \cellcolor{jpbg}0.3554
& \cellcolor{jpbg}6049.73
& \cellcolor{jpbg}6057.24 \\
\cline{2-6}
& \cellcolor{fcbmbg}FCBM Planted Partition$^{\ddagger}$
& \cellcolor{fcbmbg}0.8648$^{\star}$
& \cellcolor{fcbmbg}0.3751$^{\star}$
& \cellcolor{fcbmbg}5993.36$^{\star}$
& \cellcolor{fcbmbg}6008.39$^{\star}$ \\
\hline

\multirow{3}{*}{\shortstack[l]{BACI\\2022 Refrigerators\\Figure (3h)}}
& Block-Agnostic FM & 0.8398 & 0.4616 & 10224.93 & 10232.58 \\
\cline{2-6}
& \cellcolor{jpbg}Jeffreys Measure \& Median Entropy
& \cellcolor{jpbg}0.8750
& \cellcolor{jpbg}0.5548
& \cellcolor{jpbg}8968.49$^{\star}$
& \cellcolor{jpbg}8976.14$^{\star}$ \\
\cline{2-6}
& \cellcolor{fcbmbg}FCBM Planted Partition$^{\ddagger}$
& \cellcolor{fcbmbg}0.8752$^{\star}$
& \cellcolor{fcbmbg}0.5557$^{\star}$
& \cellcolor{fcbmbg}8969.67
& \cellcolor{fcbmbg}8984.97 \\
\hline

\multirow{3}{*}{\shortstack[l]{UN Comtrade\\2023 Fabric\\Figure (3i)}}
& Block-Agnostic FM & 0.9043 & 0.4690 & 3347.57 & 3354.93 \\
\cline{2-6}
& \cellcolor{jpbg}Jeffreys Measure \& Median Entropy
& \cellcolor{jpbg}0.9318
& \cellcolor{jpbg}0.5830
& \cellcolor{jpbg}2915.10
& \cellcolor{jpbg}2922.46$^{\star}$ \\
\cline{2-6}
& \cellcolor{fcbmbg}FCBM Planted Partition$^{\ddagger}$
& \cellcolor{fcbmbg}0.9324$^{\star}$
& \cellcolor{fcbmbg}0.5843$^{\star}$
& \cellcolor{fcbmbg}2912.98$^{\star}$
& \cellcolor{fcbmbg}2927.70 \\
\hline
\end{tabular}

\endgroup

\vspace{2mm}
\begin{flushleft}
\footnotesize
$^{\star}$ Best numerical value for ROC/PR AUC among the two block-aware methods (the same pair domain), and lowest reported AIC/BIC-form score among the three models. Historical FM AUCs are not directly comparable. \\
$^{\diamond}$ For FM and full-information FCBM these are standard AIC/BIC values. For the median-entropy rule, the same algebraic forms are descriptive penalized-likelihood scores because the selected point is not an MLE; $k=1$ is the dimension of the constrained feasible family. \\
$^{\ddagger}$ Uses additional information besides the total number of links, i.e., the number of links between countries within the same economic region.
\end{flushleft}
\end{table*}

\FloatBarrier
\renewcommand{\thetable}{A\arabic{table}}
\renewcommand{\theHtable}{A\arabic{table}}
\subsection{Controlled Synthetic Validation: Replications and Stress Test}
\label{app:synthetic-robustness}
For the empirically calibrated $3\times3$ grid of Sec.~\ref{sec:synthetic-validation}, we generated 100 independent networks per cell. The median entropy rule had the smallest probability RMSE of the four entropy representatives in 893 of the 900 realizations. Across the nine cells, per-cell median $\mathrm{RMSE}_P$ ranged from $0.0039$ to $0.0171$ for median entropy, $0.025$--$0.053$ for mean entropy, $0.079$--$0.103$ for maximum entropy, and $0.138$--$0.183$ for minimum entropy. The corresponding full-information FCBM range was $0.0032$--$0.0041$; the median rule outperformed it in 171 of the 900 individual realizations.

To assess the limits of this behavior outside the empirically calibrated range, we performed a broader stress test with $\gamma_0\in\{0.5,2,3.5\}$ and target density $\rho_0\in\{0.03,0.07,0.15\}$, using 100 independent networks per cell. For each pair $(\gamma_0,\rho_0)$, we solved for $\beta_0$ before sampling so that the expected density equals $\rho_0$. This design separates changes in the generating regional effect from changes in overall link density. Table~\ref{tab:synthetic-stress} reports how often each of the four reduced-information representatives has the lowest probability RMSE relative to the known generating probability matrix. The full-information FCBM is a separate benchmark and is not included in the winner counts. This stress-test batch of 900 networks is separate from the 900 realizations in the empirically calibrated experiment reported above.

\begin{table}[H]
\centering
\caption{Synthetic FCBM stress test outside the empirically calibrated parameter range (not an evaluation on observed trade networks). Each cell names the entropy-based representative with the lowest $\mathrm{RMSE}_P$ relative to the known generating probability matrix and gives its number of wins among 100 independently sampled synthetic networks. Where two rules win, both counts are shown. The full-information FCBM benchmark is not included in the winner counts. These 900 stress-test realizations are independent of the 900 calibrated realizations summarized in Sec.~\ref{sec:synthetic-validation}.}
\label{tab:synthetic-stress}
\renewcommand{\arraystretch}{1.2}
\setlength{\tabcolsep}{11pt}
\begin{tabular}{@{}c c c c@{}}
\toprule
 & \multicolumn{3}{c}{Target density $\rho_0$} \\
\cmidrule(lr){2-4}
$\gamma_0$ & $0.03$ & $0.07$ & $0.15$ \\
\midrule
$0.5$ & Maximum (100) & Maximum (100) & Maximum (100) \\
$2.0$ & Median (100) & Median (100) & \begin{tabular}[t]{@{}c@{}}Maximum (75)\\Median (25)\end{tabular} \\
$3.5$ & Mean (100) & Mean (100) & Median (100) \\
\bottomrule
\end{tabular}
\end{table}

The table reveals a systematic dependence on the generating regime. For weak regional effects ($\gamma_0=0.5$), maximum entropy wins in all 300 realizations across the three densities, consistent with its comparatively weak regional differentiation. For intermediate effects ($\gamma_0=2$), the median-entropy point wins all 100 realizations at $\rho_0=0.03$ and at $\rho_0=0.07$, whereas at $\rho_0=0.15$ maximum entropy wins 75 of 100 (the median wins the remaining 25). For strong effects ($\gamma_0=3.5$), the mean-entropy point wins all 100 realizations at $\rho_0=0.03$ and at $\rho_0=0.07$, while the median wins all 100 at $\rho_0=0.15$. The preferred rule therefore depends on both the regional effect and network density; it cannot be inferred from $\gamma_0$ alone.

The differences are substantial in some regimes. At $(\gamma_0,\rho_0)=(0.5,0.07)$, the per-cell median $\mathrm{RMSE}_P$ is $0.0287$ for the maximum-entropy point versus $0.0654$ for the median-entropy point; at $(3.5,0.03)$, it is $0.0114$ for the mean-entropy point versus $0.0383$ for the median-entropy point. Importantly, all representatives on a given feasible curve reproduce the same observed total number of links in expectation. Their different probability errors therefore reflect how the constrained link-probability mass is distributed across country pairs, not different expected network sparsities. These results delimit the empirical success of the Jeffreys median: it provides a central representative of the unidentified family, not a rule that universally recovers the generating regional effect. The minimum-entropy solution wins none of the tested cells and remains sensitive to the numerical truncation; lowering the lower end of the $e^{\beta}$ grid from $10^{-5}$ to $10^{-7}$ left the winning representative of every cell unchanged.

\end{document}